\documentclass[oldversion]{aa}      %not compiling otherwise
\usepackage{authblk}
\usepackage{graphicx,amsmath}      %amsmath required for split
\usepackage{amssymb}
\usepackage{natbib}             %for A&A : reimplements the LaTeX \cite command 
\usepackage{url}      %call to dace url not working otherwise
\usepackage{threeparttable}   %follow the A&A style
\usepackage{mathabx}
\usepackage{tikz}
\bibpunct{(}{)}{;}{a}{}{,} % to f

\def\approxsup{%
  \def\p{%
    \setbox0=\vbox{\hbox{$>$}}%
    \ht0=0.6ex \box0 }%
  \def\s{%
    \vbox{\hbox{$\sim$}}%
  }%
  \mathrel{\raisebox{0.7ex}{%
      \mbox{$\underset{\s}{\p}$}%
    }}%
}

\def\wb{\overline{\lambda}}
\def\wsun{\lambda_\mathrm{B_{\odot}}}

\def\wbsun{\overline{\lambda}_\mathrm{B_{\odot}}}
\def\nubsun{\overline{\nu}_\mathrm{B_{\odot}}}
\def\wstar{\lambda_{\star}}

\def\wbstar{\overline{\lambda}_{\star}}
\def\wsurf{\lambda_\mathrm{surf}}

\def\aRs{\frac{a}{R_{\star}}}
\def\iocc{i_\mathrm{occ}}

\def\bindensity{\textsc{bindensity}}
\newcommand\supsymb{\mathord{>}}
\newcommand\infsymb{\mathord{<}}

\usepackage{txfonts}
\begin{document}

\title{The ANTARESS workflow \\ I.\ Optimal extraction of spatially resolved stellar spectra with high-resolution transit spectroscopy}

%NEW VERSION  
%\author{{V.~Bourrier}\and
%{J.-B.~Delisle} \and  
%{C.~Lovis}   \and 
%{H.~M.~Cegla}\and 
%{M.~Cretignier}\and 
%{R.~Allart} \and
%{K.~Al~Moulla} \and
%{S.~Tavella} \and
%{O.~Attia} \and
%{D.~Mounzer} \and
%{V.~Vaulato} \and
%{M.~Steiner}\and 
%{T.~Vrignaud} \and
%{S.~Mercier} \and
%{X.~Dumusque} \and
%{D.~Ehrenreich} \and
%{J.~V.~Seidel} \and
%{A.~Wyttenbach} \and
%{W.~Dethier} \and
%{F.~Pepe} 
%}

%% OLDER VERSION
\author[1]{V.~Bourrier}
\author[1]{J.-B.~Delisle}   
\author[1]{C.~Lovis}    
\author[1,2]{H.~M.~Cegla} 
\author[1,3]{M.~Cretignier} 
\author[4,1]{R.~Allart} 
\author[1]{K.~Al~Moulla} 
\author[1]{S.~Tavella} 
\author[1]{O.~Attia} 
\author[1]{D.~Mounzer} 
\author[1]{V.~Vaulato} 
\author[1]{M.~Steiner} 
\author[1,5]{T.~Vrignaud} 
\author[1]{S.~Mercier} 
\author[1]{X.~Dumusque} 
\author[1]{D.~Ehrenreich} 
\author[1,6]{J.~V.~Seidel} 
\author[1]{A.~Wyttenbach} 
\author[7]{W.~Dethier} 
\author[1]{F.~Pepe}

\authorrunning{V.~Bourrier et al.}
\titlerunning{ANTARESS I}

\offprints{Vincent Bourrier (\email{vincent.bourrier@unige.ch})}

%NEW VERSION
%\institute{\affil[1]{Observatoire Astronomique de l'Universit\'e de Gen\`eve, Chemin Pegasi 51b, CH-1290 Versoix, Switzerland}
%\affil[2]{Physics Department, University of Warwick, Coventry CV4 7AL, UK}
%\affil[3]{Centre for Exoplanets and Habitability, University of Warwick, Coventry CV4 7AL, UK}
%\affil[3]{Department of Physics, University of Oxford, Oxford OX13RH, UK}
%\affil[4]{D\'epartement de Physique, Institut Trottier de Recherche sur les Exoplan\`etes, Universit\'e de Montr\'eal, Montr\'eal, Qu\'ebec H3T 1J4, Canada\thanks{Trottier Postdoctoral Fellow}}
%\affil[5]{Institut d'astrophysique de Paris, CNRS, UMR 7095, Sorbonne Universit\'e, 98 bis bd Arago, 75014, Paris, France}
%\affil[6]{European Southern Observatory, Alonso de C\'ordova 3107, Vitacura, Regi\'on Metropolitana, Chile}
%\affil[7]{Univ. Grenoble Alpes, CNRS, IPAG, 38000 Grenoble, France}}
%
%\institute{}

%% OLDER VERSION
\affil[1]{Observatoire Astronomique de l'Universit\'e de Gen\`eve, Chemin Pegasi 51b, CH-1290 Versoix, Switzerland}
\affil[2]{Physics Department, University of Warwick, Coventry CV4 7AL, UK}
\affil[3]{Centre for Exoplanets and Habitability, University of Warwick, Coventry CV4 7AL, UK}
\affil[3]{Department of Physics, University of Oxford, Oxford OX13RH, UK}
\affil[4]{D\'epartement de Physique, Institut Trottier de Recherche sur les Exoplan\`etes, Universit\'e de Montr\'eal, Montr\'eal, Qu\'ebec H3T 1J4, Canada\thanks{Trottier Postdoctoral Fellow}}
\affil[5]{Institut d'astrophysique de Paris, CNRS, UMR 7095, Sorbonne Universit\'e, 98 bis bd Arago, 75014, Paris, France}
\affil[6]{European Southern Observatory, Alonso de C\'ordova 3107, Vitacura, Regi\'on Metropolitana, Chile}
\affil[7]{Univ. Grenoble Alpes, CNRS, IPAG, 38000 Grenoble, France}
\institute{}

   \date{Received 12/01/2024; accepted 13/07/2024}
 
  \abstract
{High-resolution spectrographs open a detailed window onto the atmospheres of stars and planets. As the number of systems observed with different instruments grows, it is crucial to develop a standard in analyzing spectral time series of exoplanet transits and occultations, for the benefit of reproducibility. Here, we introduce the \textsc{antaress} workflow, a set of methods aimed at processing high-resolution spectroscopy datasets in a robust way and extracting accurate exoplanetary and stellar spectra. While a fast preliminary analysis can be run on order-merged 1D spectra and cross-correlation functions (CCFs), the workflow was optimally designed for extracted 2D echelle spectra to remain close to the original detector counts, limit the spectral resampling, and propagate the correlated noise. Input data from multiple instruments and epochs were corrected for relevant environmental and instrumental effects, processed homogeneously, and analyzed independently or jointly. In this first paper, we show how planet-occulted stellar spectra extracted along the transit chord and cleaned from planetary contamination provide a direct comparison with theoretical stellar models and enable a spectral and spatial mapping of the photosphere. We illustrate this application of the workflow to archival ESPRESSO data, using the Rossiter-McLaughlin effect Revolutions (RMR) technique to confirm the spin-orbit alignment of HD\,209458b and unveil biases in WASP-76b's published orbital architecture. Because the workflow is modular and its concepts are general, it can support new methods and be extended to additional spectrographs to find a range of applications beyond the proposed scope.  In a companion paper, we will present how planet-occulted spectra can be  processed further to extract and analyze planetary spectra decontaminated from the star, providing clean and direct measurements of atmospheric properties. }


   \maketitle

%%%%%%%%%%%%%%%%%%%%%%%%%%%%%%%%%%%%%%%%%%%%%%%%%%%%%%%%%%%%%%%%%%%%%%%%%%%%%%%%%

\section{Introduction}
\label{sec:intro}

The last two decades have seen a stupendous rise in the development of ground-based spectrographs, fuelled by their application in both the planetary and stellar communities (\citealt{Pepe2014}). The efficiency, spectral resolution, and wavelength coverage of available instrumentation have increased to the point where several spectrographs across the world now allow for observations in the optical and near-infrared (NIR) domains with resolving powers up to more than 100\,000 (e.g., ESPRESSO, \citealt{Pepe2021}; NIRPS, \citealt{Bouchy2017}). Another key point that qualifies these spectrographs to characterize exoplanetary systems (besides their high spectral resolution) is that they are fiber-fed and extremely stabilised in temperature and pressure, which ensures a high stability for their point-spread function. This allows for the lines from species in stellar and planetary atmospheres to be resolved with
a high precision, thus determining their chemical composition (by identifying which transition the line corresponds to), physical properties (by analyzing the line profile shape), and dynamical motions (by analyzing the Doppler shifts of the lines with respect to their expected rest position). While spectrographs in the exoplanet field were originally designed to search for planets through the Keplerian motion of their star (\citealt{Mayor1995,baranne1996}), transiting planets further unlocked their potential. Spectra measured during transit do indeed mix information from the unocculted photosphere, from the local photosphere fully occulted by the opaque planetary layers, and from the planet atmospheric limb filtering the local photospheric light.

%-----------------------------

The first exploitation of an exoplanet spectroscopic transit (\citealt{Queloz2000}) repurposed a velocimetric technique originally applied to eclipsing binaries. As a transiting body occults a local region of the photosphere, it removes its light from the observed stellar spectrum. This distortion of the disk-integrated stellar lines, known as the Rossiter-McLaughlin (RM) effect (\citealt{Rossiter1924,McLaughlin1924}), traces the trajectory of the body across the stellar disk, thus constraining the velocity field of the photosphere and the orbital architecture of the system. The latter is an important tracer for exoplanets, as their spin-orbit angle is inherited from the protoplanetary disk or shaped by late dynamical migration (see \citealt{Albrecht2022} and \citealt{Bourrier2023} for reviews of these processes). Measurements of the RM effect evolved over the years, from analyzing the anomalous radial velocity (hearafter, $rv$) deviation from the Keplerian motion of the star (e.g., \citealt{Ohta2005,Gimenez2006,Hirano2011,Boue2013}) to modeling the distorted disk-integrated stellar lines using Doppler tomography (e.g., \citealt{cameron2010a,bourrier2015_koi12,Temple2019}), and, finally, extracting the planet-occulted stellar lines to analyze their $rv$ centroids via the reloaded RM effect (\citealt{Cegla2016}) or model their profiles via the RM effect Revolutions (\citealt{Bourrier2021}). The possibility to analyze the local stellar lines occulted by a transiting planet translates, effectively, into a spatial mapping of the photosphere (e.g., \citealt{Dravins2017,Dravins2021}).

%-----------------------------

At wavelengths where the atmospheric components of an exoplanet absorb the stellar light, the transit depth appears larger than in white light (\citealt{seager2000}). This is the conceptual idea that \citet{Charbonneau2002} used to derive the transmission spectrum of an exoplanet atmosphere through its spectroscopic transit. While the medium spectral resolution ($R\sim$ 5500) of the \textit{Hubble} Space Telescope Imaging Spectrograph (HST/STIS) did not allow them to resolve the narrow planetary absorption lines, they measured a larger transit depth in the region of the sodium doublet that probably arises from the line wings (e.g., \citealt{Carteret2023}). Transit spectroscopy then expanded from space-borne, medium-resolution (e.g., \citealt{VM2003,Sing2008a,Deming2013,Stevenson2014,Sing2015}) to ground-based, high-resolution (\citealt{Redfield2008,Snellen2008}). While absolute transit depth is lost from the ground, the higher spectral resolution ($R\approxsup$ 60~000--100~000) allows for the planetary atmospheric lines to be resolved, while avoiding their blurring by the planet orbital motion, so that their core can be detected through comparisons with the surrounding continuum. High-resolution spectroscopy with stabilized spectrographs gives access to atmospheric signatures from both non-transiting and transiting planets, since the ability to perform precise $rv$ measurements allows us to track the Doppler shift of individual spectral lines along the orbital motion. Molecular bands imprinted in the bright dayside spectrum of the planet, for example from H$_2$O and CO, could thus be detected in the NIR (\citealt{Brogi2012,Birkby2013,Lockwood2014,snellen2014}), while iron emission lines were recently detected in the optical spectrum of ultra-hot gas giants (\citealt{Pino2020,Nugroho2020,Yan2020}). During transit, measurements of absorption lines blueshifted with respect to the planet orbital motion -- again, from molecules in the NIR first (\citealt{Snellen2010}) and then from atoms in the optical (\citealt{Wyttenbach2015,Louden2015}) were attributed to winds in the planetary atmosphere. Spectrally resolving  the planetary lines provides further constraints on the thermal and dynamical broadening mechanisms in the atmosphere (\citealt{Seidel2019,Seidel2020c,Wyttenbach2020}). Combined with the possibility to resolve absorption lines temporally during transit, high-resolution spectroscopy thus allows mapping spatially and dynamically the planetary atmosphere (\citealt{Louden2015,Allart2018,Ehrenreich2020,Gandhi2022,Seidel2023a}). 

%%%%%%%%%%%%%%%%

Characterizing exoplanetary atmospheres from the ground, which was long thought to be unfeasible due to time-variable telluric contamination, became possible thanks to high spectral resolution. The fast orbital motion of close-in planets allows us to disentangle in $rv$ space the fast-moving planetary lines from telluric lines and, outside of the transit, from the unocculted stellar lines. Analyses remain complicated during transit, when disk-integrated stellar lines may be distorted by both planet-occulted stellar lines and planetary atmospheric lines. On the one hand, atoms in the atmosphere of the hottest gas giants may absorb the same transitions as the stellar atmosphere and contaminate stellar lines traditionally used for RM analysis (e.g., \citealt{Bourrier2020_HEARTSIII,Ehrenreich2020}). More critically, transmission spectra of planetary atmospheres are usually corrected for the stellar lines they filter using the out-of-transit stellar spectrum. Because the disk-integrated spectrum is often not a good estimate of local stellar lines (e.g., \citealt{Czesla2015,Yan2017,Casasayas2020}), planet-occulted line distortions (POLDs) are introduced at every transition absorbed in the stellar spectrum (\citealt{Dethier2023}). The $rv$ tracks of the planetary orbital motion and of the occulted stellar surface generally overlap over a substantial fraction of the transit (e.g., \citealt{CasasayasBarris2022}), so that planetary lines are contaminated by POLDs in all but the hotter stars, which have few absorption lines. Various approaches have been taken to correct transmission spectra for POLDs a posteriori (\citealt{Yan2017,Casasayas2020,CasasayasBarris2021,Borsa2018,Borsa2021,Mounzer2022}), or to model them together with planetary signatures (\citealt{Dethier2023}). As a complementary approach, we propose building data-driven estimates of the planet-occulted stellar lines, by cleanly separating spectral contributions from the planet-occulted stellar regions and from the planetary atmosphere. 

%%%%%%%%%%%%%%%%

This requires spectra to be corrected for instrumental and environmental effects, and made as close as possible to how they would be measured directly out of the planetary system. Instrumental pipelines, such as the Data Reduction Software (DRS) of the ESPRESSO and HARPS spectrographs, provide a first layer of calibrations (in wavelength and flux) and corrections (for the grating response, cosmic rays, etc). This reduction, however, is generally performed independently on individual exposures, making it difficult to separate systematic noise from stellar and planetary signals. Processing  multiple exposures from a given star together offers the advantage of using the stellar spectrum as a common reference, to better identify and clean spurious features (e.g., \textsc{yarara}, \citealt{Cretignier2021}; \textsc{apero}, \citealt{Cook2022}). Transit observations offer the additional advantage of providing time series of consecutive exposures, typically within a given night. In that case, the stability of the stellar spectrum and observing conditions over the course of a few hours can be exploited to perform more advanced corrections (e.g., \citealt{Snellen2010}; \textsc{SLOPpy}, \citealt{Sicilia2022}). However, a variety of methods are employed in the community to reduce and analyze high-resolution transit spectroscopy datasets (e.g., \citealt{Wyttenbach2017,Casasayas2019,Hoeijmakers2020}), making comparisons between the results of these analysis difficult.

To make the most out of transit observations and ensure the reproducibility of results, we developed a novel workflow, the Advanced and Neat Techniques for the Accurate Retrieval of Exoplanetary and Stellar Spectra (\textsc{antaress}), which is described in two main papers focusing on planet-occulted and planetary spectra, respectively. In Sect.~\ref{sec:scope}, we introduce the general scope and concept of the workflow. Section~\ref{sec:pl_sys_des} presents our description of the planetary system, in particular, the stellar surface. In Sect.~\ref{sec:mock}, we explain how the workflow can be used to generate realistic mock datasets. The methods devised to correct and process high-resolution spectroscopy datasets are introduced in Sects.~\ref{sec:init_corr} and \ref{sec:proc_mod}, while Sect.~\ref{sec:gen_mod} describes generic methods used across the workflow. In Sect.~\ref{sec:ana_mod}, we present the methods dedicated to the analysis of the processed datasets and their application to archival data of two planets representative of high-resolution transit studies. Following a discussion of the performance of the workflow in Sect.~\ref{sec:perf}, we present our conclusions in Sect.~\ref{sec:conclu}.

%%%%%%%%%%%%%%%%%%%%%%%%%%%%%%%%%%%%%%%%%%%%%%%%%%%%%%%%%%%%%%%%%%%%%%%%%%%%%

\section{ ANTARESS workflow}
\label{sec:scope}

\begin{figure*}
\begin{minipage}[tbh!]{\textwidth}
\includegraphics[trim=0cm 0cm 0cm 0cm,clip=true,width=0.9\columnwidth]{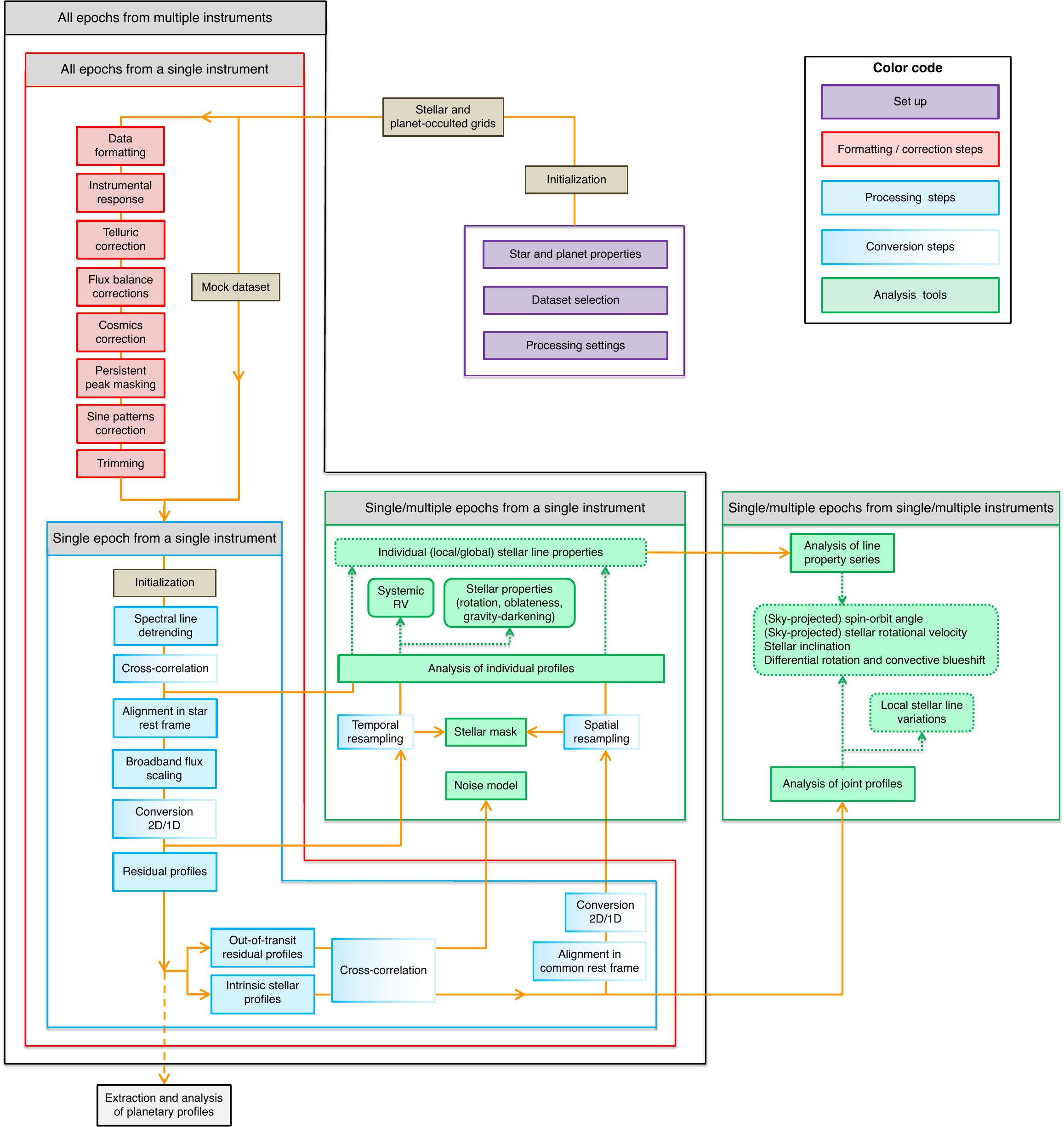}
\centering
\end{minipage}
\caption[]{Chart of the \textsc{antaress} process flow. 
}
\label{fig:flowchart}
\end{figure*}

\subsection{Concept}
\label{sec:workflow}

The \textsc{antaress} workflow consists in a set of methods to process spectral time series from an exoplanetary system, obtained consecutively in one or several epochs. The main goals are to retrieve spectra from local regions of the stellar surface or the transmission and emission spectra from the planet atmosphere; however, the workflow can also be used more generally to clean and format stellar spectra. Depending on the objectives, the time series can thus overlap with the transit or eclipse of exoplanets, cover various parts of their orbit, or even monitor a star individually. The workflow was designed to process the spectral time series from various instruments homogeneously, to remain as close as possible to the original data, and to pay particular attention to the propagation of uncertainties and spectro-temporal resampling. We are presenting \textsc{antaress} in several papers, describing the theoretical concepts behind the methods and their application to archival datasets to assess their performance and showcase new results. The present study describes the corrections that must be applied to the time series and the steps needed to extract the spectra of stellar regions occulted by a transiting planet. A companion paper will describe the subsequent methods developed to extract and analyze planetary spectra. Once the full \textsc{antaress} workflow is described, we will release the implementation we are developing in an open-source, user-friendly Python pipeline.\\

Multiple epochs and instruments can be processed with the \textsc{antaress} workflow to compare their outputs, analyze them jointly, or combine them and increase the detectability of extracted signals. Methods must be applied according to the sequences illustrated in Fig.~\ref{fig:flowchart}, which depend on the input type of the data (e.g., spectral corrections are not relevant for cross-correlated spectra) and on the goal of the study (e.g., stellar spectra need to be cross-correlated for the analysis of the RM effect). Methods are grouped in three main categories, represented by colored boxes in Fig.~\ref{fig:flowchart}. Input data are first corrected for instrumental and environmental effects (red boxes), then processed in a way that allows for the extraction of the stellar surface and planetary spectra (blue boxes). Finally, the data are analyzed to derive the orbital architecture of the system and properties of the stellar and planetary atmospheres (green boxes). 

\textsc{Corrections} methods are applied to all epochs from a given instrument together, as some corrections are better constrained using multiple spectral time series. They are applied first in the sequence so that datasets from various instruments and epochs are made comparable, and can be processed and analyzed in the same way. The existing set of corrections was developped for ground-based, stabilized high-resolution spectrographs, and some corrections are specific to a given instrument (e.g., to deal with ESPRESSO interference patterns). The current workflow is thus applicable to the spectrographs listed in Table~\ref{tab:anta_inst}. The concepts behind the \textsc{Processing} and \textsc{Analysis} methods, however, are generic and applicable to time series from any medium- or high-resolution spectrograph. The \textsc{antaress} workflow could thus be applied to additional instruments (Table~\ref{tab:anta_inst}), provided that the necessary correction methods are developed. \textsc{Processing} methods aim at extracting specific types of spectral profiles and converting them in the format required for the various \textsc{Analysis} methods available in the workflow. Profiles from a given epoch or from several epochs of a common instrument can be binned temporally and spatially at different steps of the processing, to create new series of profiles that can be analyzed at higher signal-to-noise ratio (S/N), or a single master profile representative of the star or planetary atmosphere. Most \textsc{Analysis} methods are either aimed at fitting the extracted spectral profiles or their derived properties, and can be applied to various outputs of the \textsc{Processing} workflow. Profiles and properties can be analyzed jointly over multiple epochs of different instruments to get the highest possible precision from their fit. \\

\begin{table}[tbh]
\caption{Spectrographs suitable to the \textsc{antaress} workflow}\centering
\begin{threeparttable}
\begin{tabular}{lcccl}
\hline
\noalign{\smallskip}  
\textbf{Spectrograph}   &       \textbf{Reference}                                                                                      & \textbf{Status}         \\   
\noalign{\smallskip}
\hline
\hline
CARMENES                                        &       \citealt{Quirrenbach2016}                                                                       &        Applied           \\     %,Quirrenbach2018,Quirrenbach2020
CORALIE                                         &       \citealt{Queloz2001}                                                                                    &        Applied           \\  
ESPRESSO 1/4UT         &        \citealt{Pepe2021}                                                                              &          Applied          \\  
EXPRES                                  &       \citealt{Jurgenson2016}                                                                          &          Applied          \\  
GIANO                                   &   \citealt{Claudi2018}                                                                                        &          Applicable      \\  
HARPS                                   &       \citealt{Mayor2003}                                                                             &          Applied          \\  
HARPS-N                                 &       \citealt{Cosentino2012}                                                                          &          Applied          \\  
HST/STIS                                &                                                                                                                               &    Applicable    \\  
JWST/NIRSPEC                    &                                                                                                                               &    Applicable    \\  
MAROON-X                    &   \citealt{Seifahrt2018}                                                                          &    Applicable    \\  
NIRPS                                   &       \citealt{Bouchy2017}                                                                                    &          Applied          \\  
SOPHIE                                  &       \citealt{Perruchot2008}                                                                         &          Applied          \\  
SPIRou                                  &       \citealt{Donati2020}                                                                                    &          Applicable       \\  
UVES                                            &   \citealt{Dekker2000}                                                                                        &          Applicable      \\  
\hline
\end{tabular}
\begin{tablenotes}[para,flushleft]
The list of spectrographs that could be applicable to the workflow is not exhaustive.
  \end{tablenotes}
  \end{threeparttable}
\label{tab:anta_inst}
\end{table}

While the \textsc{antaress} workflow eventually aims at retrieving the orbital architecture of planetary systems as well as properties of the stellar and planetary atmospheres, it requires a number of bulk and orbital properties for planets and their host stars to be fixed. The main properties that need to be defined to apply the workflow can be found in Table~\ref{tab:sys_prop}.

\begin{table*}[tbh]
\caption{Properties of the HD\,209458 and WASP-76 system used in our analysis}\centering
\begin{threeparttable}
\begin{tabular}{lcccl}
\hline
\noalign{\smallskip}  
\textbf{Parameter}      & \textbf{Symbol}                       &        \multicolumn{2}{c}{\bf Value}  & \textbf{Unit}          \\   
                               &                                             &   \textbf{HD\,209458}   &  \textbf{WASP-76}  &    \\  
\noalign{\smallskip}
\hline
\hline
Star &                                       &     &   &    \\  
\hline
Spectral type                               &                                                           &    F7             &       F9            & \\
Radius                                                          & $R_{\star}$                                 &  1.160 [1]   &  1.77 [4]   & $R_{\Sun}$   \\
Microturbulence velocity                & $\xi$                                         &  1.03 [1]   &  1.38 [4]   & km\,s$^{-1}$                                                                                                 \\
Effective temperature                   & $T_{\rm eff}$                                 &  6069 [1]   &  6316 [4]   & \,K \\
Surface gravity                                                         & $log\,g $                                               &  4.41 [1]      &   4.13 [4]    & cgs \\
Metallicity                                             &  [Fe/H]                                               &  0.02 [1]      &   0.34 [4]    &  \\
Projected rotational velocity   & $v$\,sin\,$i_\mathrm{\star}$  &  4.272 [0]  &   0.86 [0]  & km\,s$^{-1}$                                                                                               \\
Limb-darkening coefficients             & $u_\mathrm{1}$ &  0.380 [1] & 0.393 [5] / Chromatic [6]$^{\ddagger}$       &     \\
                                        & $u_\mathrm{2}$ &  0.234 [1]  & 0.219 [5] / Chromatic [6]$^{\ddagger}$       &     \\                                                                                                    
Stellar reflex motion amplitude         & $K_{\star}$ &  84.27 [2]  &  116.02 [5]  & m\,s$^{-1}$                                                                                               \\
\hline
Planet &                                             &     &   &    \\  
\hline
Orbital period                          & $P$      &   3.52474859 [2]  & 1.80988198 [5]   & days  \\
Eccentricity                            & $e$        &   0 [2]  & 0 [5]  &  \\
Argument of periastron                  & $\omega$    &   90 [2]  &  90 [5]   & deg  \\  
Mid-transit time                                & $T_\mathrm{0}$-2~450~000       &  4560.80676 [1]$^{\dagger}$    &   8080.626165 [5]     & BJD$_{\rm TDB}$  \\
Orbital inclination                             & $i$    &  86.71 [1]$^{\dagger}$    &   89.623 [5]    & deg   \\
Scaled semi-major axis                  & $a_p/R_{\star}$     &   8.90 [1]$^{\dagger}$    &   4.08 [5]   & \\  
Projected spin-orbit angle              & $\lambda$   &  1.06 [0]  &   -37.1 [0]   & deg \\ 
Planet-to-star radius ratio     & $R_\mathrm{p}/R_{\star}$      &  0.12086  [3]    &  0.10852 [5] / Chromatic [6]$^{\ddagger}$     &  \\
\hline
\end{tabular}
\begin{tablenotes}[para,flushleft]
References. [0] Our work; [1] \citealt{CasasayasBarris2021}; [2] \citealt{Bonomo2017}; [3] \citealt{Torres2008}; [4] \citealt{Tabernero2021}; [5] \citealt{Ehrenreich2020}; [6] \citealt{Fu2021}\\
Notes. $^{\dagger}$ We use the values adjusted from the RM analysis in [1]. $^{\ddagger}$ We use the values measured at low spectral resolution by [6].
  \end{tablenotes}
  \end{threeparttable}
\label{tab:sys_prop}
\end{table*}

\subsection{Data format and notations}
\label{ref:data_form}

The \textsc{antaress} workflow can be applied to extracted echelle order (S2D) or order-merged (S1D) spectra, as well as cross-correlated spectra (CCFs). However, our approach is most relevant when applied to S2D spectra, which are closest to the measurements made with instrument detectors. We recommend that the S1D spectra and CCFs only be used as input for preliminary analyses, as we will show that the properties they yield for the stellar surface (Sect.~\ref{sec:perf}) and planetary atmosphere can be biased compared to processing S2D spectra. Hereafter we use the generic term profile to refer both to spectra, defined as a function of wavelength, and CCFs, defined as a function of $rv$. Our methods are applicable to input spectral data defined in the air (typically from optical ground-based spectrographs like ESPRESSO) or vacuum (typically from space-borne and IR ground-based spectrographs but also, e.g., for CARMENES and EXPRES). 

Throughout the paper, the symbol $E$ refers to an epoch and the symbol $t$ to the time at which a given exposure was taken; $F$ refers to the flux emitted by the star or a region of the star in the direction of the observer, measured as a function of wavelength $\lambda$ and at a distance $D$ (Table~\ref{tab:not_prof}). For the sake of clarity we write the spectral bin width $d\lambda(\lambda,E,t)$ as $d\lambda$. We assume that spectra input in the workflow are defined as a function of wavelength $\lambda_\mathrm{B_{\odot}}$ in the solar system barycentric rest frame. The symbol $\wb$ refers to wavelengths bins with a much larger spectral width than the instrumental pixel size. Some methods Doppler-shift profiles between two rest frames using the relevant $rv$, defined as negative when an object at rest in one of the frames is moving toward the other. Finally, the true flux that would be measured without any perturbations can be written as :
\begin{equation}
\rm F(\lambda,E,t) = F_{\star}(\lambda,E) \, \delta_{\rm p}(\lambda,E,t)
\label{eq:Fp}
,\end{equation}
where $F_{\star}$ is the stellar spectrum, and $\delta_{\rm p}$ the sum of all contributions from the planet, namely, the ratio of the in-transit flux to the stellar flux, $\delta_{\rm p}^{\rm tr}$, the fraction of starlight reflected by the planet, $\delta_{\rm p}^{\rm refl}$, and the ratio of the thermal planetary flux to the stellar flux, $\delta_{\rm p}^{\rm th}$. The flux coming from the planet can thus also be written as $F_{\rm p}(\lambda,E,t) = F_{\star}(\lambda,E) \, (\delta_{\rm p}^{\rm refl}(\lambda,E,t) + \delta_{\rm p}^{\rm th}(\lambda,E,t) ) $.

\begin{table*}[tbh]
\caption{Nomenclature of the main types of profiles across the workflow        }\centering
%\begin{threeparttable}
\begin{tabular}{lccc}
\hline
\noalign{\smallskip}  
\textbf{Symbol}      & \textbf{Type}                       &        \textbf{Frame}  & \textbf{Description}          \\   
\noalign{\smallskip}
\hline
\hline
$F^\mathrm{meas}$ &     Disk-integrated  & Heliocentric & Profiles integrated over the stellar disk, initially provided               \\  
    &     &    &   by a DRS, then corrected by the workflow.           \\
$F^\mathrm{sc}$ & Disk-integrated & Stellar  & $F^\mathrm{meas}$ profiles aligned and scaled to   \\
    &     &    &   their correct relative flux levels.  \\
$F^\mathrm{res}$ & Residual & Stellar  & Difference profiles between a reference for           \\
    &     &    & the unocculted star and each $F^\mathrm{sc}$.    \\
$F^\mathrm{intr}$ & Intrinsic & Stellar, photosphere & In-transit $F^\mathrm{res}$ profiles, \textit{i.e.}, planet-occulted stellar  \\
    &     &    &   profiles, corrected for planetary contamination.  \\
\hline
\end{tabular}
%\begin{tablenotes}[para,flushleft]
%  \end{tablenotes}
%  \end{threeparttable}
\label{tab:not_prof}
\end{table*}

\subsection{Resampling, error propagation, and fitting}
\label{sec:resamp}

The standard approach in the literature of transmission spectroscopy is to resample transit spectra on the same wavelength table during the reduction. However, several steps require to shift spectra between different rest frames (typically the star and planet frames), with shifts specific to each exposure. This combination of shifts and resampling has two main consequences: loss of spectral resolution and the introduction of correlated uncertainties. We thus propose two improvements to the processing of transit spectra. First, each spectrum is processed on its original spectral table as long as possible, namely, its spectral table is shifted without resampling. Second, when resampling has to be applied (typically to combine spectra together) we calculate the resulting correlations and propagate them afterwards.

We represent a spectrum as an array of flux densities over pixels, to which we associate a covariance matrix. Diagonal values of the matrix, noted as $\sigma^2(\lambda,E,t)$, correspond to the variance of the signal measured in the pixel at wavelength $\lambda$, for the exposure at time $t$ of epoch $e$. 
While the \textsc{antaress} workflow can be applied to input data affected by correlated noise, spectra provided by the DRS of standard spectrographs are usually uncorrelated and associated with diagonal covariance matrices. If uncertainties are not provided with measured spectra, a reasonable assumption for high-S/N data ($\approxsup$50) consists in defining a photon-noise variance proportional to the number of photoelectrons received during an exposure.

The resampling of a spectrum on a different spectral table is performed through the application of an interpolation algorithm, which accounts for error propagation (Sect.~\ref{apn:resamp_error}). To limit the spreading of correlation between the resampled pixels we propose a linear and a cubic interpolator,  both of which preserve the integral of the flux on each original pixel (Fig.~\ref{fig:bindensity}). Fast processing and analysis can be carried out using linear interpolation, but the preferred approach to limit blurring is cubic interpolation. 

Models can be fitted to the data products of the \textsc{antaress} workflow using various \textsc{Analysis} methods (Fig.~\ref{fig:flowchart}), following the method described in Sect.~\ref{apn:fit_cov}. Preliminary fits can be performed using Levenberg-Marquardt minimization and the variance of the data alone, but final fits should be performed with Markov chain Monte Carlo (MCMC) sampling and account for covariances to fully benefit from our approach and avoid biases.

\subsection{Application}
\label{sec:scope_app}

To illustrate the capabilities of our method we apply it throughout this paper to archival ESPRESSO transit observations of HD\,209458b and WASP-76b. The different properties of these two systems make the retrieval of the spatially resolved stellar spectra of the sodium doublet (for HD\,209458b), and of the stellar CCFs of neutral iron (for WASP-76b), particularly interesting. HD\,209458b orbits a G-type star on an aligned orbit and has no atmospheric signature in the core of the sodium doublet (\citealt{CasasayasBarris2021,Dethier2023}). WASP-76b orbits an F-type star on a highly misaligned orbit and the stellar light is absorbed by neutral iron in its atmosphere (\citealt{Ehrenreich2020}). General properties of the systems required for our analysis are given in Table~\ref{tab:sys_prop}.

For each planet we exploit two transit datasets obtained with ESPRESSO on 20 July 2019 (epoch 1) and 11 November 2019 (epoch 2) for HD\,209458b and on 03 September 2018 (epoch 1) and 31 October 2018 (epoch 2) for WASP-76b. Information about the scheduling, instrumental settings, and night conditions for these epochs can be found in \citet{CasasayasBarris2021} for HD\,209458b and in \citet{Ehrenreich2020} for WASP-76b. For an optimal application of the \textsc{antaress} workflow to transit datasets we emphasize the importance of securing observations both before and after the transit. This allows building a reference spectrum for the unocculted star, and characterizing possible short-term variations due to the star, the Earth atmosphere, and the instrument that can then be corrected for over the entire dataset. One of the interests of using ESPRESSO data as a case in point is that it is representative of a series of stabilized, high-resolution echelle spectrographs used to observe stars and transiting planets to which our methods can be applied (Table~\ref{tab:anta_inst}). ESPRESSO is installed at the Very Large Telescope (VLT) at ESO's Paranal site. The light is dispersed on 85 spectral orders\footnote{Each order is covered by two independent slices on the detectors, which are hereafter treated as individual orders.} from 380 to 788\,nm. Our present analysis starts with S2D spectra extracted from the detector images, corrected for, and calibrated by version 3.0.0 of the DRS pipeline (\citealt{Pepe2021}).

%%%%%%%%%%%%%%%%%%%%%%%%%%%%%%%%%%%%%%%%%%%%%%%%%%%%%%%%%%%%%%%%%%%%%%%%%%%%%

\section{Description of planetary systems}
\label{sec:pl_sys_des}

\subsection{Coordinates}
\label{sec:pl_coord}

\begin{figure}
\includegraphics[trim=0cm 0cm 0cm 0cm,clip=true,width=0.9\columnwidth]{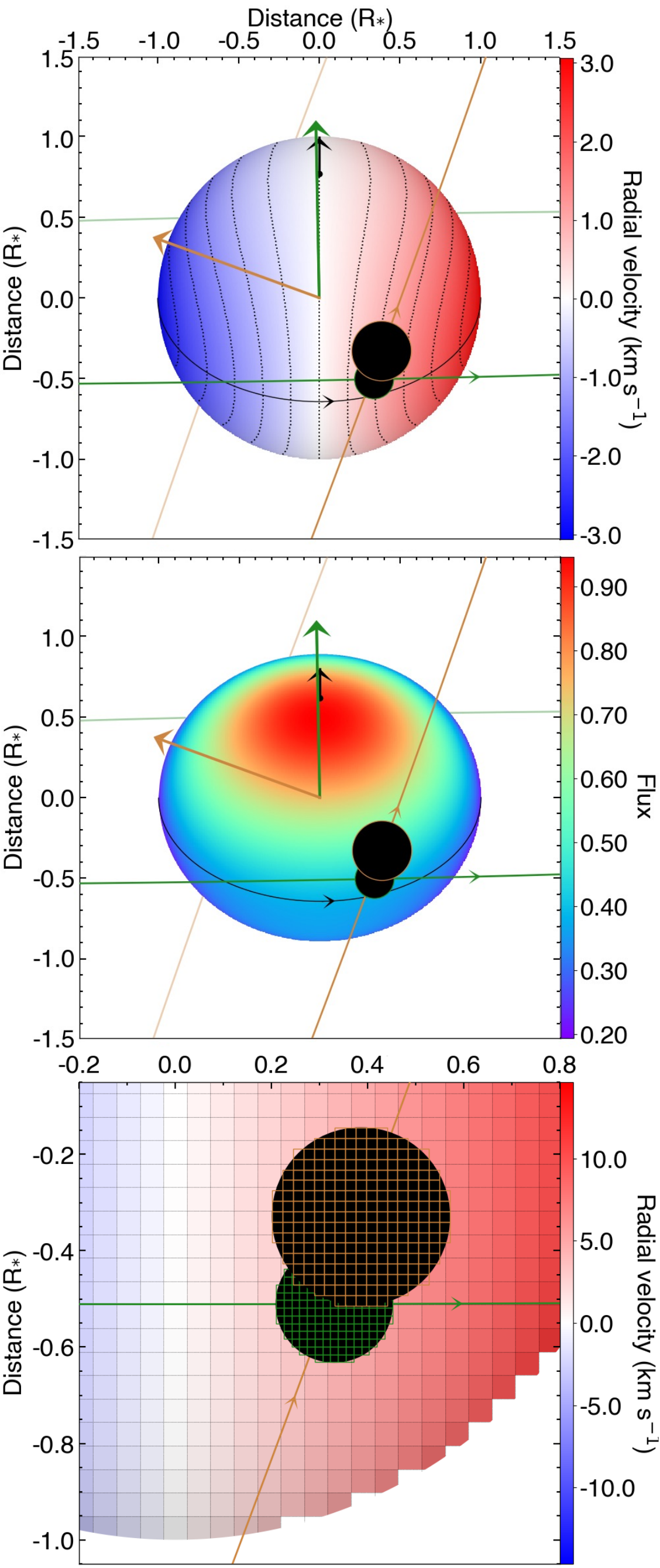}
\centering
\caption[]{Views of \textsc{antaress} sky-projected stellar grid for an imaginary system in various configurations highlighting the workflow possibilities. The plots X and Y axis correspond to $x_{\star \rm sky}$ and $y_{\star \rm sky}$ (see text). The stellar equator is shown as a solid black line, and the projected stellar spin axis as a black arrow extending from the north pole. Two planets (black disks) move along their orbital trajectory, whose 3D orientation is indicated by the normal to the orbital plane (green and golden arrows). \textit{Top}: Star shown rotating differentially between equator and poles and is colored as a function of its photospheric $rv$ field (dotted black lines show iso-$rv$ curves). \textit{Middle}: Star is fast-rotating and oblate, colored as a function of its intensity field (darkened at the limbs and brightened at the poles by gravity-darkening). \textit{Bottom}: Zoom-in on the planet-planet occultation. The grids discretizing the star and planets are displayed with a low resolution for clarity purpose. }
\label{fig:System_view}
\end{figure}

We use specific coordinates to describe the position of the planets and stellar surface regions in the \textsc{antaress} methods. The two main reference frames (Fig.~\ref{fig:Frames}), with origin at the star center, are the sky-projected stellar rest frame ($x$ is the node line of the stellar equator, $y$ the sky-projected stellar spin, $z$ the line-of-sight (LOS); Fig.~\ref{fig:System_view}) and the star rest frame ($x$ is the stellar equator, $y$ the stellar spin, $z$ the LOS). We refer to App.~\ref{apn:pl_coord} for details about our other coordinate systems. 

Particular care must be taken about the definition of transit windows. We consider that an exposure is in-transit if the planetary disk overlaps with the stellar disk during at least a fraction of the exposure. Most exoplanet host stars rotate slowly enough that they can be considered spherical, in which case the condition for overlap is $r_\mathrm{sky} < R_{\star}$+$R_{p}$, where $r_\mathrm{sky}$ is the farthest distance between the planet and star centers during the exposure, projected onto the plane of sky. The workflow can account for host stars that rotate fast enough to be oblate. In that case the transit condition is best solved numerically. We assess whether the planet limb intersects with the sky-projected photosphere, namely, whether the $x_{\star \rm sky}$ and $y_{\star \rm sky}$ coordinates of the discretized planet limb provide a solution to the quadratic describing the oblate photosphere (Eq. 13 and 14 in \citealt{Barnes2009}\footnote{We caution that all instances of $(1-f^2)$ in Eq. 14 of \citet{Barnes2009} must be replaced with $(1-f)^2$.}). \\

\subsection{Planetary atmospheric masking}
\label{sec:atm_mask}

If the transiting planet possesses an atmosphere the absorption lines of its constituent species may be present in the spectra processed from the full stellar disk. While extracting those lines is one of the end goals of the workflow, in earlier steps they act as contaminants of the stellar lines. We thus propose a procedure to exclude spectral ranges contaminated by narrow absorption lines from the planetary atmosphere. This requires a list of wavelengths for the electronic transitions to be masked and a $rv$ window corresponding to the absorbed range in the planet rest frame (assumed common to all lines). First, we calculate the $z'_{\rm sky}$ component of the planet velocity vector in the sky-projected orbital reference frame, by applying the changes from Eqs.~
\ref{eq:orb2los} and 
\ref{eq:los2sky} to the orbital velocity coordinates (Eq.~\ref{eq:der_orb}). Then, we calculate an effective planetary orbital $rv$ for each exposure, $rv_{\rm p/\star}$, by oversampling $z'_{\rm sky}$ to account for its possibly asymmetrical variations around the exposure center. The excluded $rv$ window, shifted to the star rest frame using the $rv_{\rm p/\star}$ values, is used directly with CCF profiles or converted into spectral ranges for each electronic transition to be masked. Pixels can be masked at various steps of the workflow, and are propagated afterward through our resampling algorithm.

We illustrate the interest of atmospheric masking throughout the processing of the WASP-76b dataset, excluding conservatively the range [-25 ; 10]\,km\,s$^{-1}$ contaminated by planetary iron lines (\citealt{Ehrenreich2020}). These lines are present in the F9 stellar mask used by \citet{Ehrenreich2020}, through the ESPRESSO DRS, to compute the CCFs of WASP-76. To be comparable with their analysis, we use the F9 mask linelist to set the excluded transitions.

%%%%%%%%%%%%%%%%%%%%%%%%%%%%%%%%%%%%%%%%%%%%%%%%%%%%%%%%%%%%%%%%%%%%%%%%%%%%%

\subsection{Description of the stellar surface}
\label{sec:st_surf}

The \textsc{Processing} and \textsc{Analysis} methods require the calculation of various properties associated with the planet-occulted and disk-integrated stellar surfaces, in particular their broadband emission (e.g., Sect.~\ref{sec:flux_scaling}) and spectral profiles (e.g., Sects.~\ref{sec:mock} and~\ref{sec:ana_mod}). Since photospheric properties are not spatially uniform and the regions occulted at the stellar limbs or by multiple planets may have complex shapes, we use a numerical approach to calculate accurately these properties.

\subsubsection{Local profiles}
\label{sec:st_grid}

Spectral lines emitted by an ``elementary'' region of the photosphere are defined by their intrinsic profile, the $rv$ of the region (to Doppler-shift the profile) and its broadband intensity (to scale the profile in flux).

The $rv$ of a region at position ($x_{\star \rm sky}$,$y_{\star}$) (see Sect.~ \ref{sec:pl_coord}) is defined as:
\begin{equation}
rv_\mathrm{surf/\star} = x_{\star \rm sky} \, v_{\rm eq}\sin\,i_\mathrm{\star} \frac{\Omega(y_{\star})}{\Omega_{\rm eq}} +  \sum_{j} c_\mathrm{cb,j} \mu^{j}
\label{eq:rv_surf}
,\end{equation}
with $\Omega(y_{\star}) = \Omega_{\rm eq}\,(1-\alpha_\mathrm{dr} \, y_{\star}^2-\beta_\mathrm{dr} \, y_{\star}^4)$ the stellar rotation rate as a function of latitude (equal to $\Omega_{\rm eq}$ at the stellar equator), described by a Sun-like differential rotation law, and $c_\mathrm{cb,j}$ the coefficients of a polynomial law describing convective blueshift variations (see \citealt{Cegla2016} for more details). 

We define white-light and chromatic intensities for each region. The former, based on average stellar properties over the full spectral band covered by the processed data, are used to scale cross-correlated profiles. The latter, which account for variations in the broadband stellar emission, are used to scale spectral profiles in a given band. Intensities are modulated by limb-darkening, described in our formalism through uniform, linear, quadratic, nonlinear, power-2, or solar laws with coefficients specific to the spectral band considered. They can further be modulated by gravity darkening, which has the spectral balance of a black body at photospheric temperatures calculated with the corrected formalism of \citet{Barnes2009}. We caution the importance of calculating $\mu$ = $\cos\theta$ accurately in the case of an oblate photosphere, as $\theta$ is the angle between the LOS and the local normal to the photosphere, which differs in this case from the stellar radius. We thus calculate $\mu$ numerically without approximations, using the projection of the local normal\footnote{Defined as $(-\partial p/\partial x_{\star \rm sky},-\partial p/\partial y_{\star \rm sky},1)$, where the photosphere is described by $z_{\star \rm sky} = p(x_{\star \rm sky},y_{\star \rm sky})$ following Eq. 13 in \citet{Barnes2009}.} on the LOS. 

We propose three possibilities to define intrinsic line profiles:
\begin{itemize}
\item Analytical: calculated for properties that vary across the photosphere according to parametric laws of a chosen spatial coordinate (e.g., a linear variation of contrast with $\mu$). These variations and the model profile can be informed by measured intrinsic profiles (Sect.~\ref{sec:lprof_var} and \ref{sec:fit_series}). Analytical profiles can be broadened by macroturbulence, using a radial-tangential (\citealt{Gray1975,Gray2021}) or anisotropic Gaussian (\citealt{Takeda2017}) kernel. 
\item Data-driven: intrinsic profiles extracted from transit time series are aligned into a common rest frame (Sect.~\ref{sec:intr_prof}) and spatially resampled at higher S/N (Sect.~\ref{sec:temp_resamp}) into a single master profile, or into a series function of the resampling coordinate that can then be interpolated in the chosen region.  
\item Theoretical: using \textsc{pySME} (\citealt{Wehrhahn2023}), informed by a 1D stellar atmospheric model matching the type of the host star and its microturbulence velocity, and by a VALD linelist (\citealt{Piskunov1995,Kupka2000,Ryabchikova2015}) generated with the host properties. Solar abundances are used by default (\citealt{Asplund2009}) but can be varied globally or for specific species to better match the lines of interest. The spectral lines of chosen species can further be calculated using pre-computed grids of non local thermal equilibrium (NLTE) departure coefficients (see the \textsc{pySME} website and \citealt{Amarsi2020,Amarsi2022}). Theoretical profiles are calculated over a grid in $\mu$ to sample center-to-limb variations and be interpolated in the chosen region. This approach could be extended to 3D stellar atmospheric models in future versions of the workflow.
\end{itemize}
Analytical models are typically good proxies for simple profiles, as in CCFs, but only allow defining a single line. Data-driven and theoretical profiles are better options to define complex spectral lines or a full spectrum, and by construction the former directly traces the line profiles of the target star. Using observed data to provide direct estimates of the spatially resolved stellar profiles, with no assumption on their shape, is one of the main interests of the \textsc{antaress} workflow.

%-----------------------------------------------------------

\subsubsection{Disk-integrated and planet-occulted profiles}
\label{sec:pl_occ_grid}

The model disk-integrated stellar profile outside of the transit is defined by integrating the elementary profiles over a cartesian grid discretizing the entire spherical or oblate sky-projected photosphere (Fig.~\ref{fig:System_view}). Planet-occulted profiles are integrated over the region transited during the full duration of an exposure. Disk-integrated profiles within the transit are then calculated directly as the difference between the out-of-transit disk-integrated profile and the planet-occulted profiles in each exposure. If relevant for comparison with observations (Sects.~\ref{sec:mock} and~\ref{sec:ana_mod}), those various profiles are further convolved with the response of the corresponding instrument. \\

The extraction and interpretation of planet-occulted stellar profiles is central to the \textsc{antaress} workflow. The corresponding models are thus calculated over a finer grid than the one used for the star, each occulting planetary disk being discretized with a resolution adjusted to its size in the processed spectral band. This spatial oversampling allows better resolving the partial occultation of the stellar limb and accounting for spatial variations of photospheric properties across occulted areas (Sect.~\ref{fig:System_view}). A temporal oversampling can further be applied when the chord transited during an exposure is longer than the spatial scale of stellar surface variations, by positioning the planetary grid at regular intervals along the chord to account for the blurring introduced by the planet motion.

\begin{figure}
\includegraphics[trim=0cm 0cm 0cm 0cm,clip=true,width=\columnwidth]{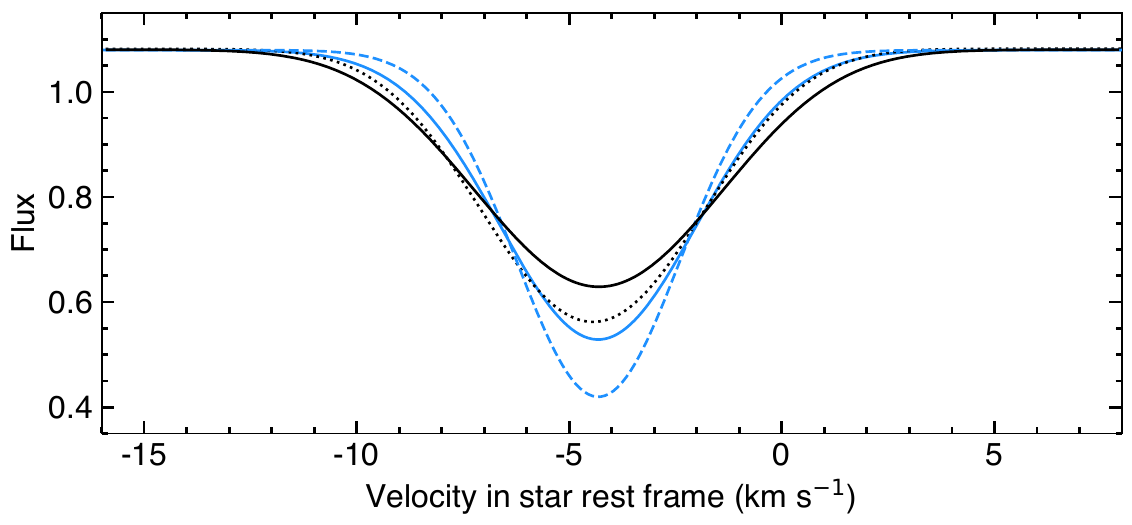}
\centering
\caption[]{Model line profile from the region occulted during a 15\,min exposure ($R_{p}$ = $R_{J}$, $R_{\star}$ = $R_{\Sun}$, $v$\,sin\,$i_\mathrm{\star}$ = 25\,km\,s$^{-1}$). The reference black profile (high precision, 1.5\,min oversampling) can be compared with the blue (low precision, rotational broadening accounted for, 1.5\,min oversampling), blue dashed (low precision, rotational broadening not accounted for, 1.5\,min oversampling), and dotted black (high precision, no oversampling) profiles.}
\label{fig:Precisions}
\end{figure}

Stellar line profiles from planet-occulted regions can be calculated at either low or high precision (Fig.~\ref{fig:Precisions}). In the first case, brightness-averaged properties are calculated over each planet-occulted grid along the chord, and their mean over all grids define the exposure planet-occulted line profile. In the second case, elementary line profiles are summed over the cells of each planet-occulted grid, and then averaged over all grids. Our tests suggest that models at low precision without oversampling approximate well enough the planet-occulted line profile for slow rotators, planets with small transit depth, and short exposure times whereas models at high precision with oversampling are required for giant planets transiting fast-rotating or bright stars (when blurring due the planet size and motion is strong and detectable), and for multiple transiting planets. Details can be found in App.~\ref{apn:pl_occ_grid}. 

%%%%%%%%%%%%%%%%%%%%%%%%%%%%%%%%%%%%%%%%%%%%%%%%%%%%%%%%%%%%%%%%%%%%%%%%%%%%%

\section{Mock datasets}
\label{sec:mock}

\begin{figure*}
\begin{minipage}[tbh!]{\textwidth}
\includegraphics[trim=0cm 0cm 0cm 0cm,clip=true,width=\columnwidth]{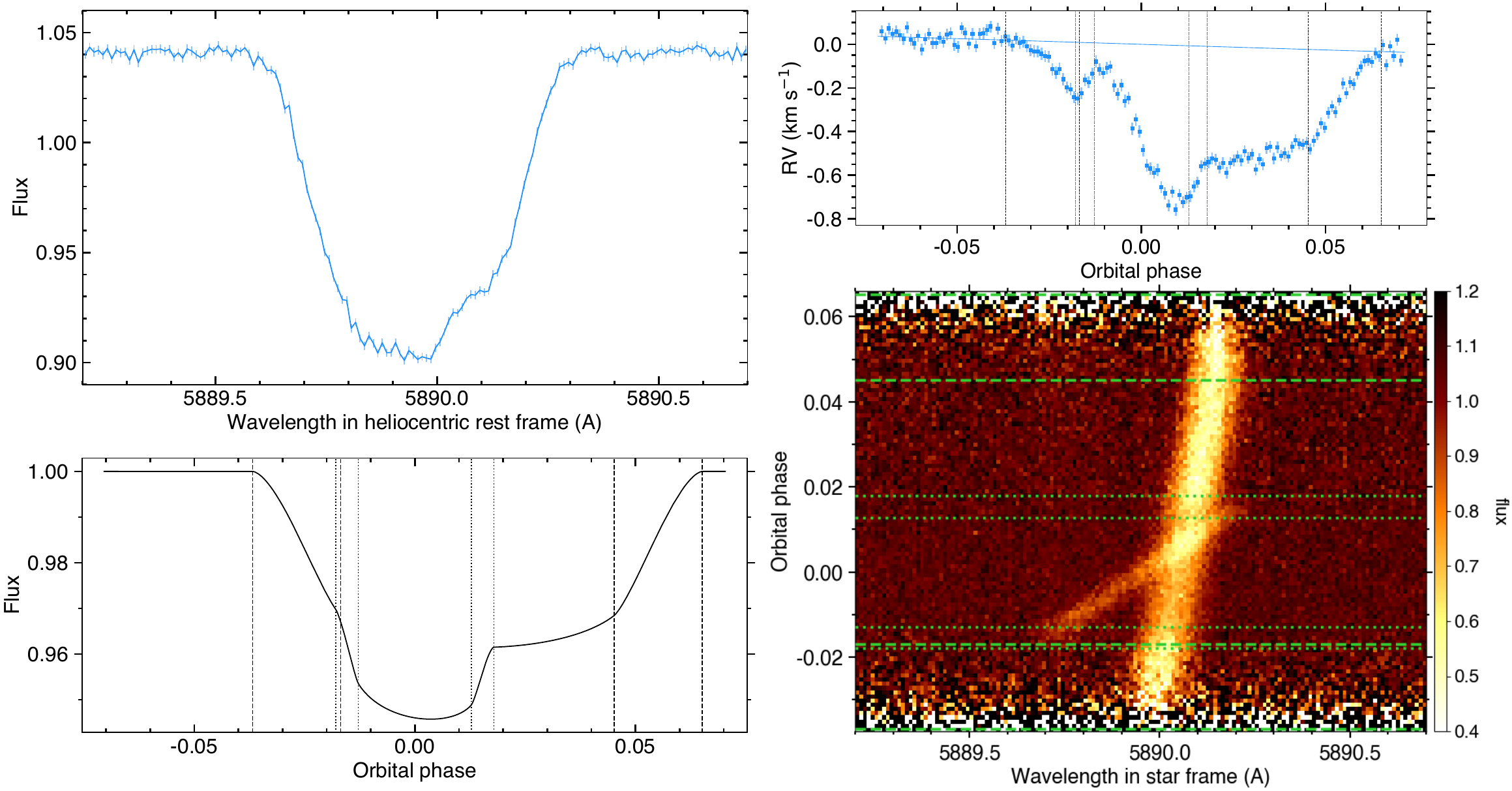}
\centering
\end{minipage}
\caption[]{Mock ESPRESSO-like time series in the sodium D2 line, for HD\,209458b (R$_\mathrm{p}$ = 1.36\,R$_\mathrm{J}$ ; P = 3.5\,d) and a mock companion (R$_\mathrm{p}$ = 2\,R$_\mathrm{J}$ ; P = 7\,d) transiting simultaneously HD\,209458 (1.16\,$R_\Sun$) with $v\sin\,i_\mathrm{\star}$ = 15\,km\,s$^{-1}$. The transit configuration is illustrated in Fig.~\ref{fig:System_view}. Orbital phases are relative to the outer planet. Dotted and dashed lines show the contacts of the inner and outer planets, respectively. \textit{Top left}: In-transit disk-integrated profile distorted by the occultation of the two planets. \textit{Bottom left}: Double-transit light curve. We note that photometric light curves are typically not an observable from ground-based spectroscopic datasets. \textit{Top right}: RV derived from a Gaussian fit to the disk-integrated profile series (similar to the output of most spectrographs DRS), blending the well-known RM anomalies from the larger planet on a highly misaligned orbit ($\sim$flat deviation) and from the smaller planet on its aligned orbit (S-shaped deviation). The solid blue line show the mock Keplerian stellar reflex motion. \textit{Bottom right}: 2D map of intrinsic profiles (see Sect.~\ref{sec:intr_prof}), showing the tracks of local stellar lines occulted by the planets along their respective transit cores.}
\label{fig:mock}
\end{figure*}

The description of the stellar surface (Sect.~\ref{sec:st_surf}) and noise properties (Sect.~\ref{sec:resamp}) in the \textsc{antaress} methods can be used to generate realistic datasets for the planetary system of interest, which can then be processed like real observational datasets for tests and predictions. Mock datasets can be used in particular to evaluate the detectability of RM signatures, as illustrated with the simultaneous transits of two exoplanets in Fig.~\ref{fig:mock}.

The method is generic and allows creating a series of disk-integrated profiles, defined by their start and end BJD times, as they would be observed with one of the spectrographs implemented in Table~\ref{tab:anta_inst}. Local profiles are defined in spectral or CCF format (Sect.~\ref{sec:st_surf}) and integrated over the stellar disk numerically (Sect.~\ref{sec:pl_occ_grid}), accounting for planetary occultation during in-transit exposures.

These disk-integrated profiles are generated in the stellar rest frame over a spectral grid at the chosen sampling. They are then Doppler-shifted to the heliocentric rest frame, using the stellar Keplerian motion induced by all planets in the system and a chosen systemic velocity (Sect.~\ref{sec:align_star}), before being convolved with the spectrograph response. If the spectral sampling of the mock dataset is low, the disk-integrated profiles are first calculated at high resolution and resampled after instrumental convolution to avoid resolution loss.

Profiles are defined in spectral flux density $F_\mathrm{mock}$, namely, the number of photoelectrons per spectral bin measured during an exposure, multiplied by the exposure time, by a mean flux density over the epoch, and by a constant instrumental flux calibration factor, $g_\mathrm{mock}$. The last two fields allow introducing variations in the mock stellar emission measured with different instruments or epochs. The final profiles comparable to observations $F^\mathrm{meas}_\mathrm{mock}$ are drawn from a Poisson distribution with number of events set by $F_\mathrm{mock}$. Uncertainties on $F^\mathrm{meas}_\mathrm{mock}$ are then attributed as $\sigma^\mathrm{meas}_\mathrm{mock} = \sqrt{g_\mathrm{mock} F^\mathrm{meas}_\mathrm{mock}}$ (Sect.~\ref{sec:count_sc}).

%%%%%%%%%%%%%%%%%%%%%%%%%%%%%%%%%%%%%%%%%%%%%%%%%%%%%%%%%%%%%%%%%%%%%%%%%%%%%

\section{Corrections methods}
\label{sec:init_corr}

Input spectra are assumed to be corrected for standard instrumental effects (flat field, blaze, background) and are put into a common format at the start of the workflow (Sect.~\ref{ref:data_form}). Our methods exploit housekeeping and complementary data\footnote{Provided by the DRS of standard spectrographs or, for activity indexes, by the Data \& Analysis Center for Exoplanets (DACE) hosted by the University of Geneva at \url{https://dace.unige.ch/}.} to monitor the quality of the observations and inform the corrections. If available, spectra can be replaced by their sky-corrected values in the relevant spectral orders. Exposures and orders considered too noisy to be exploited are removed, and spurious spectral features (e.g., bad detector pixels or poorly-corrected telluric lines) are permanently masked, with thresholds depending on the dataset and objectives.

The current \textsc{antaress} workflow (Fig.~\ref{fig:flowchart}) can correct for telluric contamination, spectro-temporal flux balance variations, cosmic rays and spurious peaks, as well as ESPRESSO interference patterns. Additional methods could be added to deal with other instrument-specific corrections (e.g., fringing in near-IR spectrographs, breathing in HST/STIS) and stellar noise (e.g., spots, pulsations in early-type stars). In the following we use $F^{\rm meas}$ and $F^{\rm corr}$ to refer to spectra prior to, and after, a given correction. We note that most of the proposed correction methods rely on a fixed model to correct $F^{\rm meas}$, and therefore do not propagate possible uncertainties associated with the model. Corrected spectra are made comparable between epochs of a given instrument, but still differ between instruments because of instrumental responses. This is taken into account when jointly interpreting datasets from different spectrographs together (Sect.~\ref{sec:ana_mod}).  \\

%%%%%%%%%%%%%%%%%%%%%%%%%%%%%%%%%%%%%%%%%%%%%%%%%%%%%%%%%%%%%%%%%%%%%%%%%%%%%

\subsection{Spectral flux calibration}
\label{sec:count_sc}

Input spectra $F^{\rm meas}$ and their associated uncertainties $\sigma^{\rm meas}$ are generally provided in flux units, related to the number of photoelectrons $N^{\rm meas}$ cumulated during an exposure per unit of wavelength as:
\begin{equation}
\begin{split}
F^{\rm meas}(\lambda,E,t) &= g_{\rm cal}(\lambda,E,t) N^{\rm meas}(\lambda,E,t),  \\
\sigma^{\rm meas}(\lambda,E,t) &= g_{\rm cal}(\lambda,E,t) \sqrt{N^{\rm meas}(\lambda,E,t)}  \\
\end{split}
\label{eq:gdet}
\end{equation}
At this stage, no resampling has been applied to the spectra and there is no covariance between their bins. The $g_{\rm cal}$ profiles represent calibrations applied by the instrument's DRS to the raw photoelectrons count, typically to correct for the blaze function (i.e.,\textit{} the transmission of the spectrograph grating). While the application of $g_{\rm cal}$ in DRS sets spectra closer to their true spectral flux balance, it artificially changes the relative noise level between different spectral regions. 

\begin{figure}
\includegraphics[trim=0cm 0cm 0cm 0cm,clip=true,width=\columnwidth]{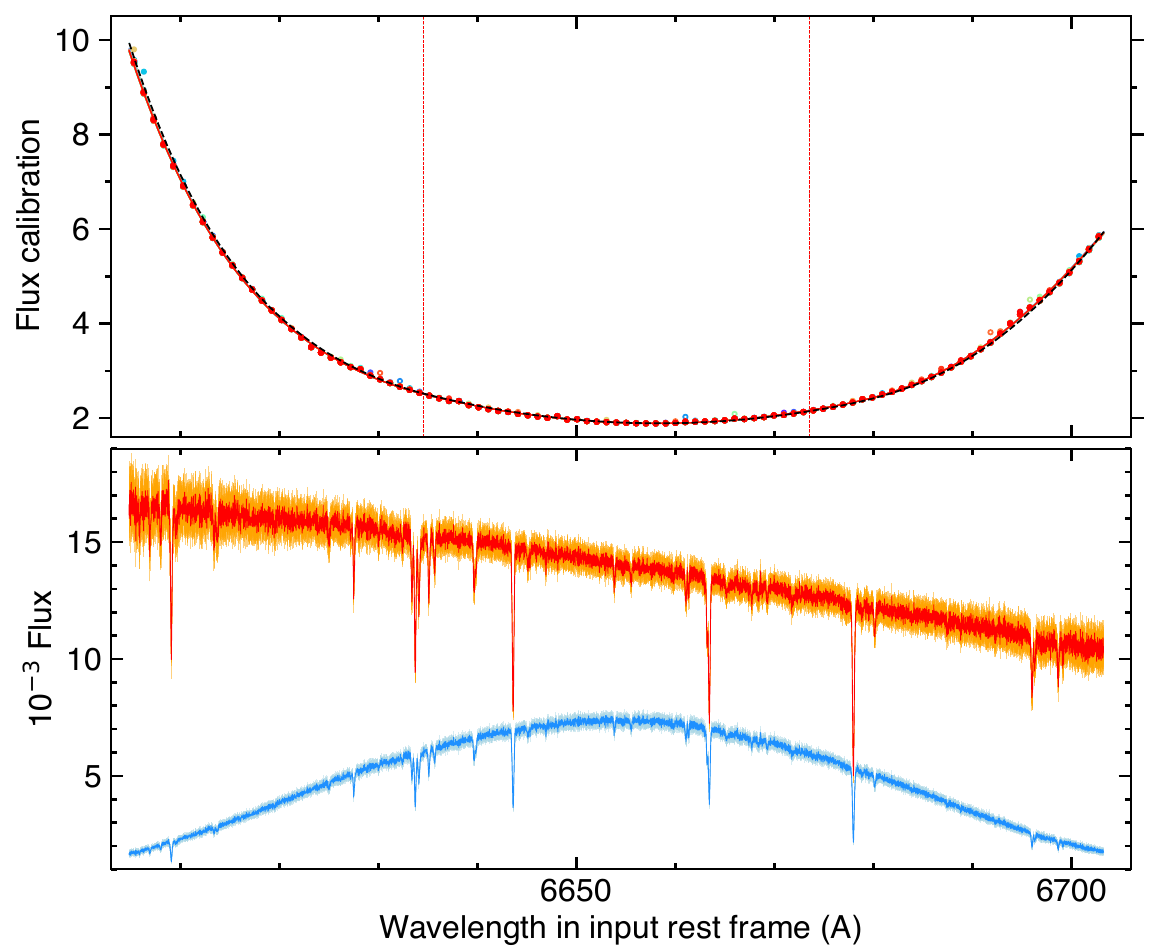}
\centering
\caption[]{Flux calibration. \textit{Top panel:} Disks (colored over the rainbow scale as a function of orbital phase, but mostly identical) show the $g_{\rm cal}(\wb,t)$ values measured at low spectral resolution in a given order, for all exposures of WASP-76 epoch 2. The black dashed line shows the median profile $\infsymb g \supsymb_{\rm cal}$ over both epochs. Dotted vertical lines indicate where the three piece-wise polynomials join. \textit{Bottom panel:} Spectrum measured during one of WASP-76 epoch 2 exposure, scaled back from the DRS calibrated counts (red, with orange uncertainties) to \textsc{antaress} equivalent raw counts (blue, with light blue uncertainties). Plotted uncertainties are amplified by three to visualize the difference in relative errors between the order edge and center. }
\label{fig:Flux2Count}
\end{figure}

Let us consider a flat spectrum, measured with lower count levels at the edges of a CCD (e.g., $N/10$, with photon noise uncertainty $\sqrt{N/10}$) compared to its middle ($N$, $\sqrt{N}$) due to the blaze function. Correcting for the blaze yields similar flux levels ($\propto N$) across the detector, but uncertainties at the edge and middle become proportional to $\sqrt{10 N}$ and $\sqrt{N}$, respectively, so that the error ratio between edge and middle is 10 times larger after flux calibration (see Fig.~\ref{fig:Flux2Count}). This illustrates how combining different spectral regions in calibrated flux units, in particular when binning spectra over wide bands (Sect.~\ref{sec:col_bal}) or cross-correlating them (Sect.~\ref{sec:CCFs}), artificially inflates uncertainties compared to operating in raw flux units, decreasing the precision on derived properties.

To circumvent this issue, we estimate a flux calibration profile $\infsymb g \supsymb_{\rm cal}$ using the spectra themselves (Sect.~\ref{apn:count_sc}). This profile (Fig.~\ref{fig:Flux2Count}) only approximates the original calibration from the DRS (in particular it is overestimated in regions of low S/N because we neglect readout noise), but allows temporarily scaling back spectra into raw flux units in the relevant methods (Sect.~\ref{sec:spec_weights}), while maintaining the relative flux balance between exposures to avoid introducing biases.

%%%%%%%%%%%%%%%%%%%%%%%%%%%%%%%%%%%%%%%%%%%%%%%%%%%%%%%%%%%%%%%%%%%%%%%%%%%%%

\subsection{Telluric correction}
\label{sec:tell_corr}

In spectra obtained from the ground, the stellar or planetary lines of interest can overlap with absorption (or emission) lines from Earth's atmosphere, especially broad molecular absorption bands. The \textsc{antaress} workflow does not yet correct for telluric emission lines, whose amplitude and profile vary in complex ways during a night. Fiber-fed spectrographs often monitor the sky simultaneously with the target, so that emission lines can potentially be removed in sky-corrected spectra.

Different methods exist to correct for telluric absorption in the literature: contemporaneous spectral measurements of spectrophotometric standard stars close to the target (e.g., \citealt{VidalMadjar1986}), model-based algorithms (e.g., \textsc{molecfit}, \citealt{Smette2015, Kausch2015}; \textsc{tapas}, \citealt{Bertaux2014}; \textsc{telfit}, \citealt{Gullikson2014}), or empirical methods (e.g., \textsc{sysrem}, \citealt{Mazeh2007}; the PCA approach from \citealt{Artigau2014}; \textsc{yarara}, \citealt{Cretignier2021}). We adapted in the \textsc{antaress} workflow the Automatic Telluric Correction (\textsc{atc}, \citealt{Allart2022}), used for the DRS of state-of-the-art spectrographs (ESPRESSO, NIRPS, HARPS). This model-based approach is a simple line-by-line radiative transfer code, which calculates the telluric spectrum of a single atmospheric layer based on the sky conditions and the physical properties of molecular lines (\textsc{hitran} database, \citealt{Rothman2021}). The model accounts for the most relevant telluric molecules (H$_2$O, O$_2$, CH$_4$, CO$_2$), and could process additional species if needed. The \textsc{atc} provides similar performances as other established methods for exoplanet transmission spectroscopy (\textsc{sysrem}, \citealt{Snellen2010}; \textsc{molecfit}, e.g. \citealt{Allart2017,Seidel2019,Hoeijmakers2020,Sedaghati2021}) but can be applied automatically to individual spectra, which better suits the \textsc{antaress} approach. 

\begin{figure}
\includegraphics[trim=0cm 0cm 0cm 0cm,clip=true,width=\columnwidth]{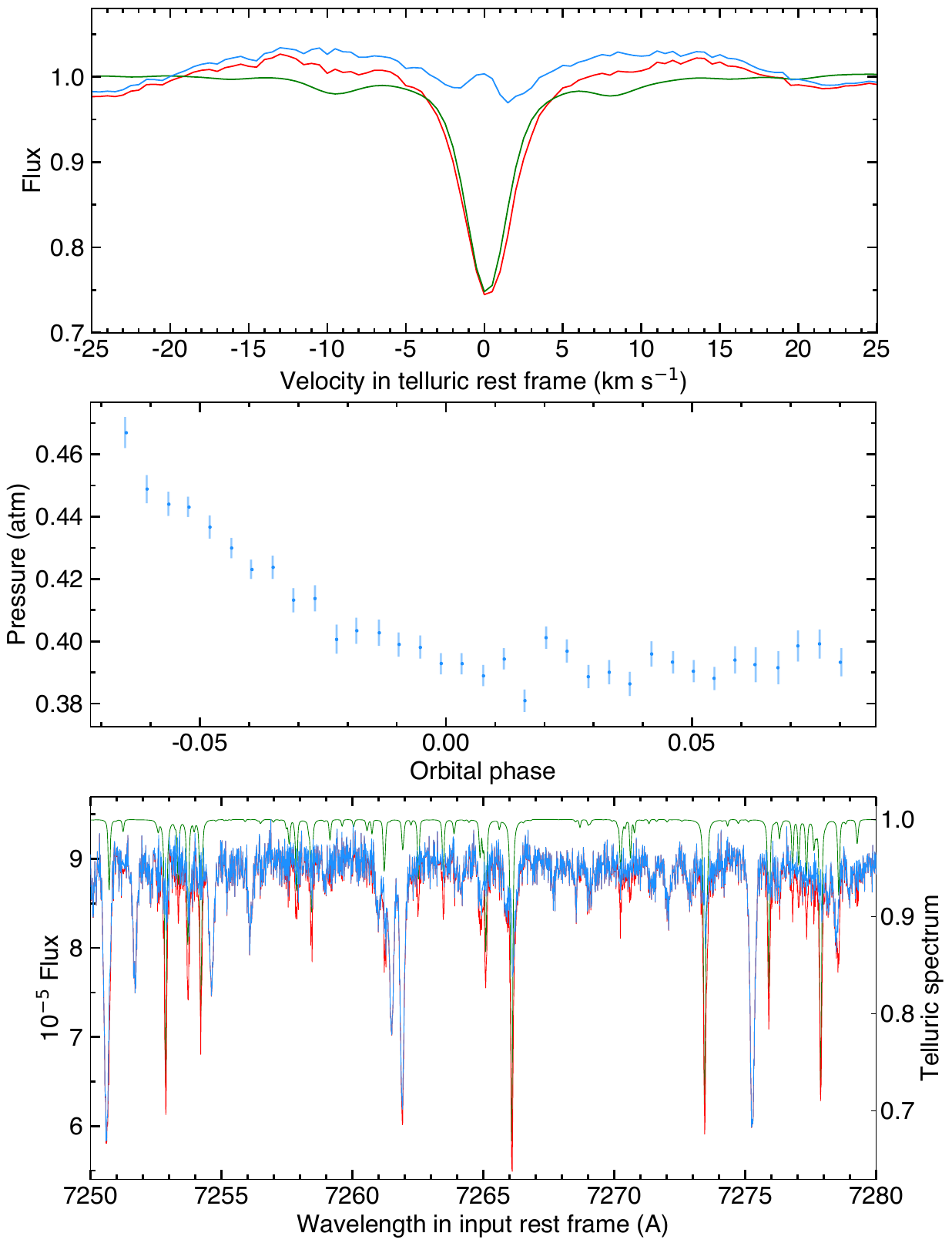}
\centering
\caption[]{Telluric correction for WASP-76 epoch 1. \textit{Top panel:} CCF of H$_{2}$O over the telluric lines used in the fit, from an observed spectrum before (red) and after (blue) correction, and from its best-fit telluric spectrum (green). \textit{Middle panel:} Average atmospheric pressures derived for H$_{2}$O over the epoch. \textit{Bottom panel:} Spectrum of the exposure in the top panel, before (red) and after (blue) correction, with its best-fit telluric model (green).}
\label{fig:tell_corr}
\end{figure}

Properties of the theoretical atmospheric layer are derived for each molecule by fitting a CCF of the model telluric spectrum to the equivalent CCF of an exposure spectrum, using housekeeping data from the spectrograph facility to inform the fit. Once atmospheric properties are derived for all molecules, we generate a global telluric spectrum $\delta_{\Earth}$ at the resolution and sampling of the observed spectrum, corrected as:
\begin{equation}
F^{\rm corr}(\wsun,E,t) = \frac{\rm F^{\rm meas}(\wsun,E,t)}{\rm \delta_{\Earth}(\wsun,E,t)}
.\end{equation} 
The method is detailed in Sect.~\ref{apn:tell_corr} and illustrated in Fig.~\ref{fig:tell_corr}. The WASP-76b and HD\,209458b datasets were corrected for H$_{2}$O and O$_{2}$ telluric absorption. We note that the derived $\delta_{\Earth}$ are also used to exclude telluric lines from CCF masks built with the workflow (Sect.~\ref{sec:CCF_masks}).

%%%%%%%%%%%%%%%%%%%%%%%%%%%%%%%%%%%%%%%%%%%%%%%%%%%%%%%%%%%%%%%%%%%%%%%%%%%%%

\subsection{Flux balance corrections}
\label{sec:col_bal}

This subset of methods deal with changes in the relative flux balance of spectra over time, which can be described in the most general case through a coefficient $c$:
\begin{equation}
\rm F^{\rm meas}(\wsun,E,t) = c(\wbsun,E,t) \, F_{\star}(\wsun,E) \, \delta_{\rm p}(\wsun,E,t)
.\end{equation}
The planetary contribution $\delta_{\rm p}$ could be isolated if $c$ had a repeatable temporal behaviour (e.g., the breathing effect of the HST, whose achromatic flux variations are phased with the telescope orbit, \citealt{Sing2019}, \citealt{Bourrier2017_HD976}) or if they could be measured independently (e.g., using a comparison star). Since the current \textsc{antaress} workflow focuses on ground-based observations, $c$ represents at first order the ``color effect'' caused by Earth atmospheric diffusion (e.g., \citealt{bourrier2014b}). The spectral balance of $c$ varies over scales of tens to hundreds of \AA,\,, with variations more regular as a function of light frequency $\nu$. Furthermore, $c$ evolves during an epoch (as the target moves through the sky) as well as between epochs (as atmospheric conditions and the sky position of the star evolves). As a result, low-frequency variations of planetary origin cannot be separated from atmospheric diffusion and are removed by our correction. An estimate of $\delta_{\rm p}$ must be re-injected later-on in spectra corrected for flux balance (Sect.~\ref{sec:flux_scaling}). Importantly, the correction has no impact at short spectral scales. Because high-resolution spectra are spectro-photometrically accurate at high spectral frequency, the true profiles of narrow stellar and planetary lines can thus be retrieved once spectra have been scaled at low frequency (Sect.~\ref{sec:res_prof}). We note that our description for $c$ may also account for chromatic fiber losses possibly caused by a wavelength-dependent seeing profile and its projection on the spectrograph fiber.

\begin{figure}
\includegraphics[trim=0cm 0cm 0cm 0cm,clip=true,width=\columnwidth]{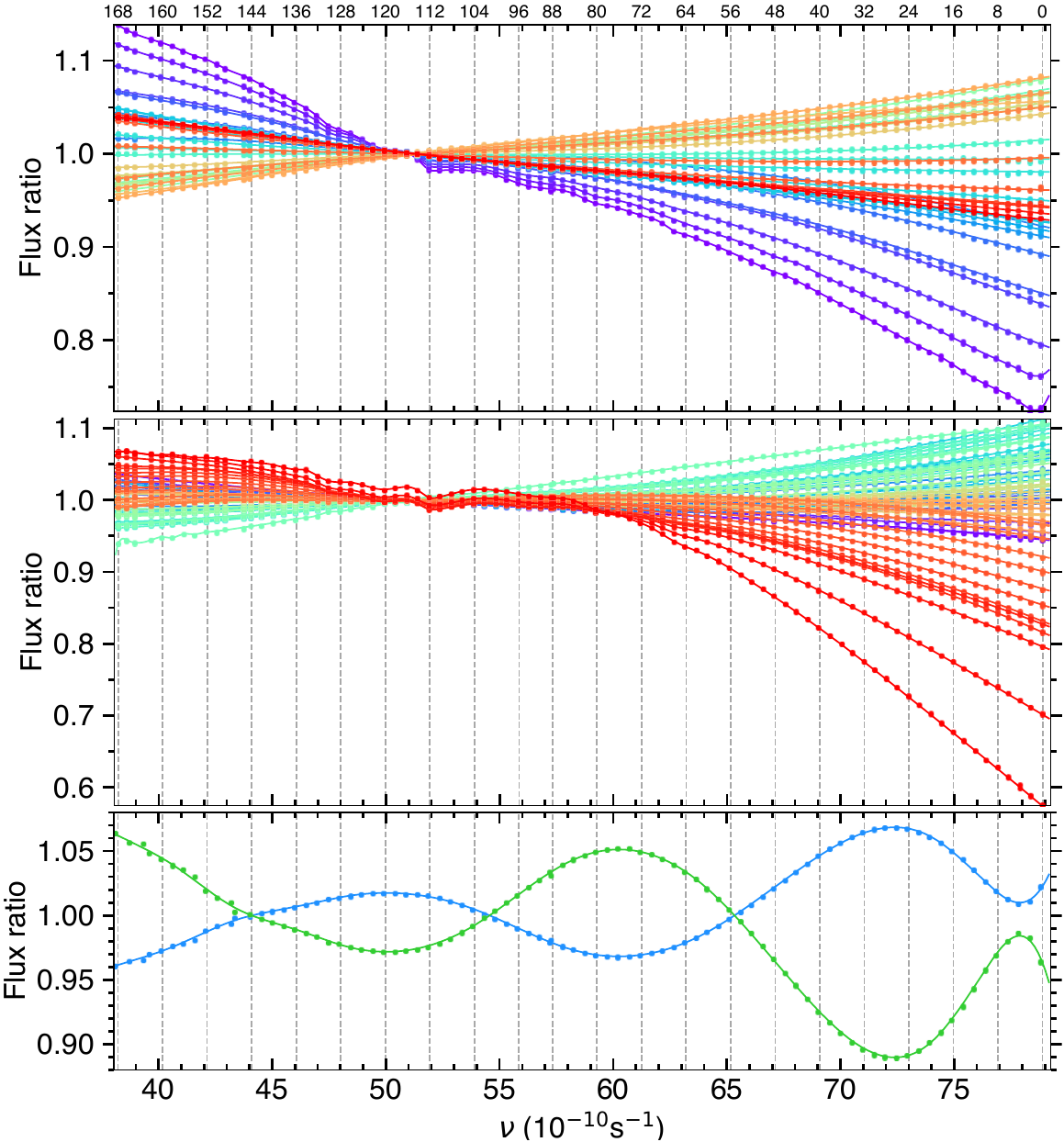}
\centering
\caption[]{Flux balance variations in WASP-76 spectra, between each exposure and their median reference in epoch 1 (top panel) and 2 (middle panel), and between the median reference and their mean over both epochs (bottom panel). Disks show the spectra-to-reference ratios measured at low spectral resolution in each exposure (top and middle panels, colored from purple to red over the rainbow scale with increasing orbital phase) or in each epoch (bottom panel, in blue for epoch 1 and green for epoch 2). Solid lines with matching colors show their best-fit models. Vertical dashed grey lines indicate the central position of one in every eight ESPRESSO slice. The medium-frequency variations captured around slice 120 are observed at high airmass and might correspond to the ozone Chappuis absorption bands at $\lambda$ = 5750 and 6030\AA\, ($\nu$ = 52.1 and 49.7$\times10^{-10}$s$^{-1}$).}
\label{fig:ColorBalanceModel}
\end{figure}

We reset the flux balance of the $\rm F^{\rm meas}$ spectra to that of a reference spectrum expressed as $\rm F_{\rm ref}(\lambda,E) = \rm C_{\rm ref}(\lambda,E) F_{\star}(\lambda,E)$, considering that this reference does not necessarily have the same balance as the true stellar spectrum. An analytical model $\rm R^{\rm theo}$ describing the flux balance variations between $\rm F^{\rm meas}$ and $\rm F_{\rm ref}$ (Fig.~\ref{fig:ColorBalanceModel}) is derived in $\nu$ space as described in Sect.~\ref{apn:col_bal}, and applied as:
\begin{equation}
\begin{split}
F^{\rm corr}(\wsun,E,t) &= F^{\rm meas}(\wsun,E,t) \frac{R^{\rm norm}(E,t)}{R^{\rm theo}(\wsun,E,t)}, \\
\rm where  \, R^{\rm norm}(E,t) &= \frac{\sum_{\wsun} F^{\rm meas}(\wsun,E,t) d\wsun}{\sum_{\wsun} F_{\rm ref}(\wsun,E) d\wsun}. \\
\end{split}
\label{eq:fbal}
\end{equation}
Corrected spectra have the same broadband profile as $F_{\rm ref}$ but keep the same global flux level as the original spectra to remain as close as possible to the measured counts (absolute flux scaling is performed in Sect.~\ref{sec:flux_scaling}). 

\begin{figure}
\includegraphics[trim=2cm 0cm 2cm 0cm,clip=true,width=\columnwidth]{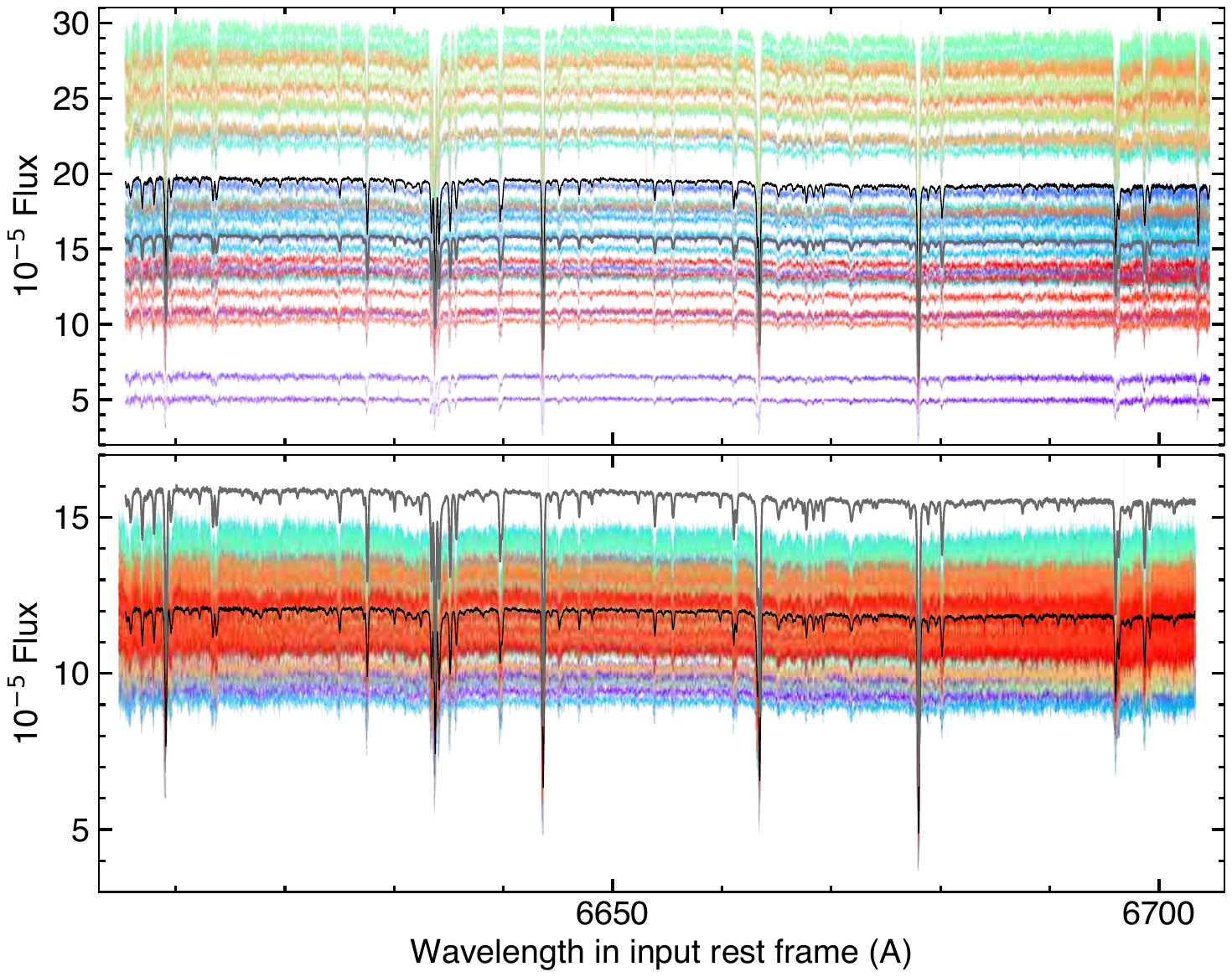}
\centering
\caption[]{Data-driven reference spectra of WASP-76 for flux balance corrections in epoch 1 (top panel) and 2 (bottom panel). The dark grey profile is the mean of the median spectra in each epoch (solid black profiles), taken as final reference for the two epochs. Spectra from individual exposures are colored over the rainbow scale with increasing orbital phase.}
\label{fig:scaling_master}
\end{figure}

We first set all spectra in a given epoch to the same flux balance, taking their median spectrum as reference (see Sect.~\ref{apn:col_bal} and Fig.~\ref{fig:scaling_master}). The correction for the dominant color effect is performed over all spectral orders together. A second-order correction can be applied in individual orders, with $c$ tracing local variations in Earth absorption or in the instrumental response (blaze function, flat field). This correction must be used with caution, as it may also capture small-scale features of instrumental, stellar, or planetary origin, such as the broad wings of planetary absorption lines or the ESPRESSO wiggle pattern (Sect.~\ref{sec:wig_mod}). Because the WASP-76 and HD\,209458 datasets are dominated by this latter effect, we do not run them through the spectral order correction.

When several epochs are processed, spectra must be set to the correct relative balance of the stellar spectrum between epochs so that line profiles in different spectral bands can be properly compared and possibly combined. We thus apply a second correction derived from the ratio between the median reference spectrum in each epoch and a final reference (Sect.~\ref{apn:col_bal}), which can be common to all epochs (if the stellar flux balance remains stable, Fig.~\ref{fig:scaling_master}) or account for variations between epochs. The final reference should further be set to a spectrum representative of the absolute stellar emission in each epoch if one wants to extract low-frequency emission spectra from the planet.

%%%%%%%%%%%%%%%%%%%%%%%%%%%%%%%%%%%%%%%%%%%%%%%%%%%%%%%%%%%%%%%%%%%%%%%%%%%%%

\subsection{Cosmics correction}
\label{sec:cosm_corr}

This method exploits the fact that an epoch is made of a series of consecutive exposures, assumed to have the same broadband spectral profile thanks to the previous steps, except for slight variations due to stellar activity and planetary absorption/emission. This makes it possible to compare a given exposure spectrum with the mean spectrum over adjacent exposures, $F_{\rm comp}$, following the procedure described in Sect.~\ref{apn:cosm_corr}, and to identify and replace spectral bins contaminated by cosmic rays. A bin is considered to be cosmics-affected if:
\begin{equation}
\begin{split}
F^{\rm meas}(\wstar,E,t) - F_{\rm comp}(\wstar,E) &>     \\
\alpha_{\rm cosm} \, \mathrm{max}(\sigma_{\rm F_{\rm comp}}&(\wstar,E),\sigma^{\rm meas}(\wstar,E,t)). \\
\end{split}
\end{equation}
Where $\alpha_{\rm cosm}$ is a chosen threshold, adjusted depending on the quality of each dataset so as not to exclude noisy pixels as cosmics. The condition on $\sigma^{\rm meas}$ accounts for the processed spectrum possibly being much noisier than adjacent exposures in a given bin - in which case $F^{\rm meas}(\wstar,E,t)$ can deviate significantly from $F_{\rm comp}(\wstar,E)$ due to photon noise alone. This can occur over pixels corrected for deep telluric lines that shift during an epoch. Because cosmic rays can generate charges across several pixels before being absorbed, due to their trajectory and local charge bleeding of the detector, we allow pixels adjacent to a cosmics-flagged bin to be replaced as well. Corrected flux values are set to $F_{\rm comp}(\wstar,E)$ and their variance\footnote{At this stage of the workflow spectra still display no covariance since they were processed on their own spectral grid.} to $\sigma_{\rm F_{\rm comp}}(\wstar,E)^2$.

\begin{figure}
\includegraphics[trim=0cm 0cm 0cm 0cm,clip=true,width=\columnwidth]{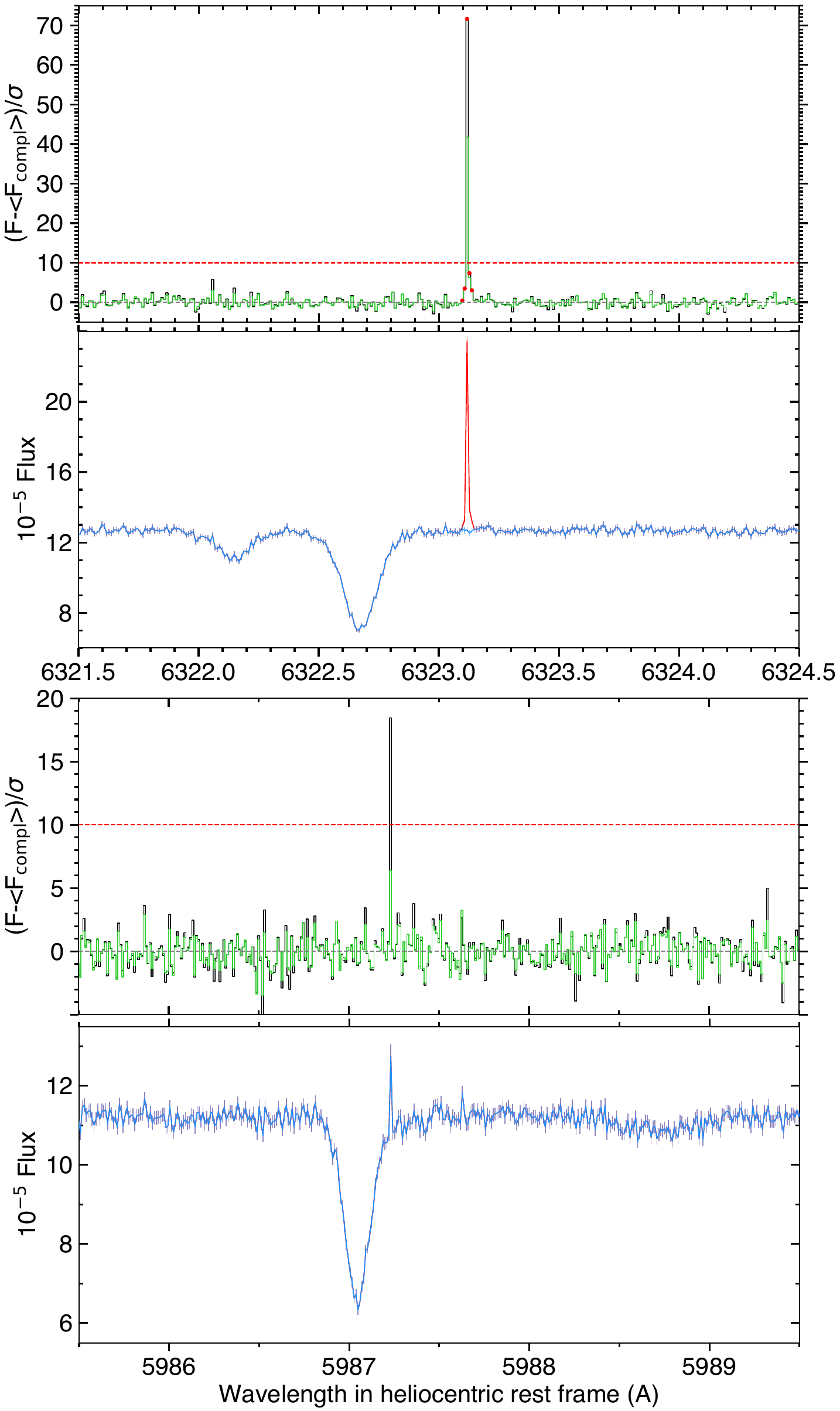}
\centering
\caption[]{Cosmics processing applied to epoch 2 of WASP-76. Top subpanels show $F(t) - F_{\rm comp}$ divided by $\sigma_{\rm F_{\rm comp}}$ (black) and $\sigma^{\rm meas}(t)$ (green), with $\alpha_{\rm cosm}$ plotted as a red line, in a sample exposure whose flux spectrum is shown in the bottom subpanels. The red pixel in the top block (together with adjacent pixels) is corrected because it deviates with respect to both its own uncertainty and to the standard deviation over adjacent exposures. The pixel in the bottom block is not corrected because it could result from photon noise in the exposure.}
\label{fig:Cosm_corr}
\end{figure}

%%%%%%%%%%%%%%%%%%%%%%%%%%%%%%%%%%%%%%%%%%%%%%%%%%%%%%%%%%%%%%%%%%%%%%%%%%%%%

\subsection{Persistent peaks masking}
\label{sec:perspeak_corr}

This method deals with flux peaks induced, for example, by hot detector pixels that were not masked by the instrument DRS, or telluric emission lines that were not removed using sky-corrected data. In contrast to transient cosmics, the affected bins show spurious flux values persisting over time. We thus identify and mask pixels that deviate from the stellar continuum by more than a threshold over all exposures (typically hot pixels) or over consecutive exposures (typically telluric emission lines, strengthening at twilight as airmass increases) in an epoch.

\begin{figure}
\includegraphics[trim=0cm 0cm 0cm 0cm,clip=true,width=\columnwidth]{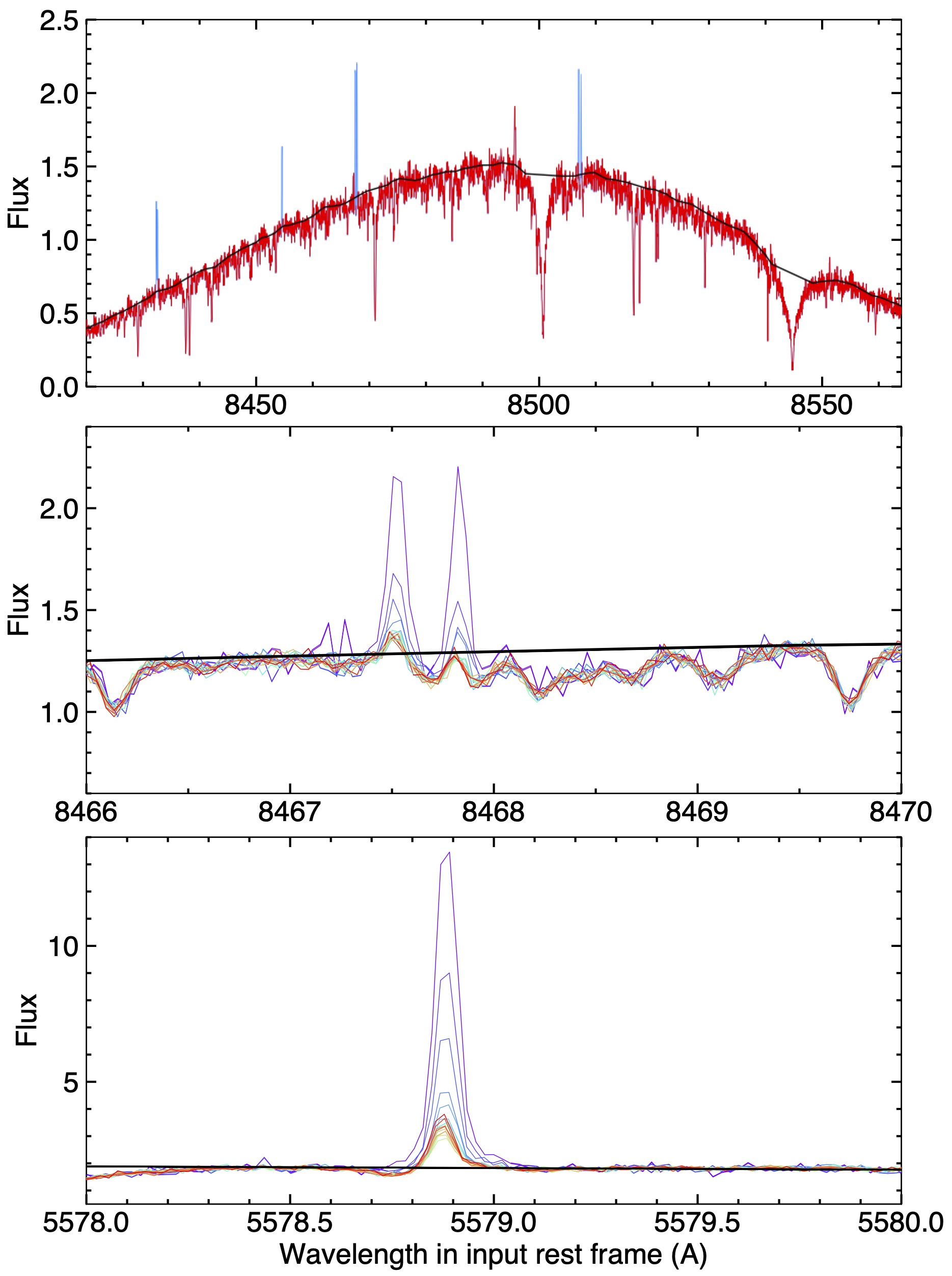}
\centering
\caption[]{Persistent peaks masking of a CARMENES time series. \textit{Top panel:} Telluric emission lines (highlighted in blue in the first exposure spectrum, shown in red) were identified through comparison with the stellar continuum (black). \textit{Middle panel:} Zoom on the strongest telluric line from the top panel, plotted over the rainbow scale with increasing orbital phase. This specific line was flagged over the first ten exposures, after which the decreasing airmass makes it negligible. \textit{Bottom panel:} Zoom on one of the strongest telluric lines in the CARMENES range, masked over the entire epoch. }
\label{fig:Perspeaks}
\end{figure}

Since no spurious peaks were found in the ESPRESSO data, we show in Fig.~\ref{fig:Perspeaks} an application to CARMENES transit spectra of WASP-156b (\citealt{Bourrier2023}) where strong telluric emission lines varying in strength but persisting over consecutive exposures could be identified.

%%%%%%%%%%%%%%%%%%%%%%%%%%%%%%%%%%%%%%%%%%%%%%%%%%%%%%%%%%%%%%%%%%%%%%%%%%%%%
%%%%%%%%%%%%%%%%%%%%%%%%%%%%%%%%%%%%%%%%%%%%%%%%%%%%%%%%%%%%%%%%%%%%%%%%%%%%%

\subsection{Wiggle correction}
\label{sec:wig_mod}

\begin{figure*}
\begin{minipage}[tbh!]{\textwidth}
\includegraphics[trim=0cm 0cm 0cm 0cm,clip=true,width=\columnwidth]{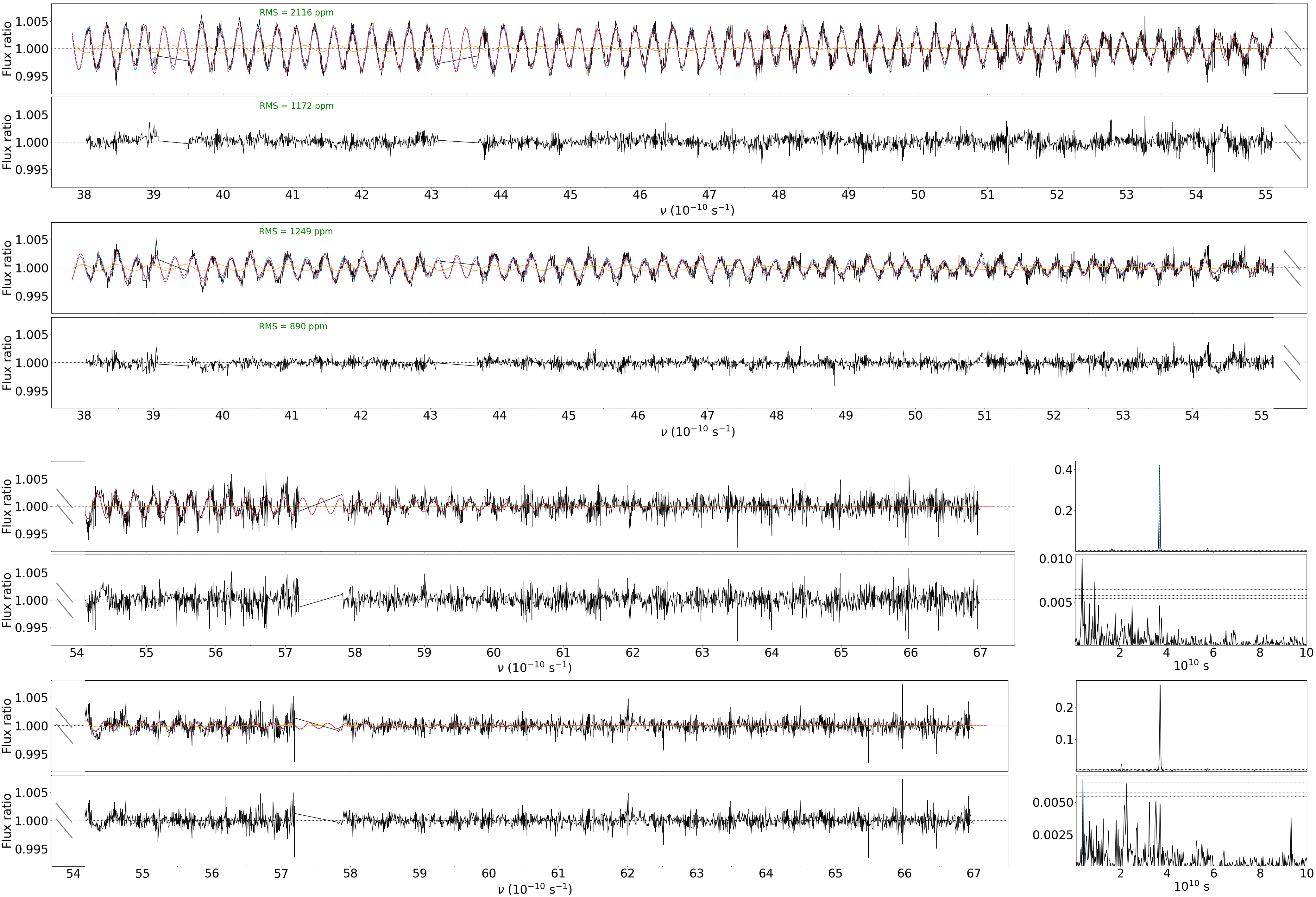}
\centering
\end{minipage}
\caption[]{Transmission spectra of HD\,209458 as a function of light frequency, highlighting different wiggle patterns at the start (exposure 14, top panel) and end (exposure 79, bottom panel) of epoch 2. Top (resp. bottom) subpanels show data before (resp. after) wiggle correction. The wiggle model, overplotted as a dashed red line, is a beat pattern between a dominant (blue) and weaker (orange) component. Right panels show periodograms of the data (horizontal dashed lines are false-alarm probability levels at 1, 5, and 10\%), highlighting the removal of these components.  }
\label{fig:wig_trans}
\end{figure*}

ESPRESSO spectra are modulated by ``wiggles'' (Fig.~\ref{fig:wig_trans}), sinusoidal patterns arising from interferences within the Coud\'e Train optics that affect the four VLT units. Most observations are affected by low-frequency wiggles (amplitudes up to several \%, periods between $\sim$10-100\,\AA\,; \citealt{Allart2020,Tabernero2021}), with the possible occasional contribution of high-frequency wiggles (amplitude $\sim$0.1\%, period $\sim$1\,\AA\,; \citealt{CasasayasBarris2021}). The wiggles' amplitude and period vary with wavelength and over timescales of minutes, making it difficult to determine their behaviour. Nonetheless, since wiggles modify the local flux balance between exposures and thus bias transmission spectra, several empirical corrections have been attempted:
\begin{itemize}
\item Cubic splines (\citealt{Tabernero2021,Azevedo2022}) fitted over a specific spectral range. 
\item Sines (\citealt{Borsa2021,CasasayasBarris2021,Benatti2021,Seidel2021,Seidel2022}) combined over different spectral ranges with specific period, phase, and amplitude. 
\item A broad Gaussian smoothing filter (\citealt{Merritt2020,Kesseli2022,Maguire2023,Zhang2022}) or Savitzky-Golay filter (\citealt{Allart2020}) to capture low-frequency wiggles, and a narrow median filter to capture high-frequency wiggles (\citealt{CasasayasBarris2022}).
\end{itemize}
While these methods can be efficient at removing wiggles over local spectral bands, they consider each exposure independently and apply blind corrections without understanding of the underlying wiggle pattern. Such separate, piecewise corrections could introduce variations between transmission spectra, and inadvertently remove stellar and planetary signals. Given the importance of ESPRESSO for the astrophysical community, we thus devised a method to describe the spectro-temporal behaviour of wiggles and correct them homogeneously. This method is generic enough that it can be adapted to similar patterns, such as HARPS-N ripples (\citealt{Thompson2020}), HARPS interference pattern (\citealt{Cretignier2021}), or fringing on NIR spectrographs (e.g., \citealt{Guilluy2023}).

Through trial and error we determined that wiggles are best described as a beat pattern between a dominant sinusoidal component with frequency $\sim3.8\times10^{10}\,s$ (periods\footnote{The period in wavelength space is $P(\lambda) = \lambda^2/(F(\lambda)c)$} between $\sim$13-54\AA\,.) and a weaker one at $\sim3.1\times10^{10}\,s$ (periods between $\sim$16-65\AA\,). The components' amplitude and frequency can be defined as polynomials of light frequency. The chromatic coefficients of these polynomials, as well as the phase reference of the components, can be expressed as a linear combination of the telescope cartesian pointing coordinates, defined by its azimuth and altitude. The full model is thus controlled by a set of pointing parameters for each component (see Sect.~\ref{apn:wig_mod_descr}). These parameters change when the LOS crosses the meridian, but the model remains continuous in value and derivative. We emphasize that the telescope guide star should not be changed during an epoch, as we found that it resets all pointing parameters (see WASP-76, Fig.~\ref{fig:apn_Pointing_ana_wig_guidshift}).

\begin{figure}
\includegraphics[trim=1.5cm 2cm 1cm 6cm,clip=true,width=\columnwidth]{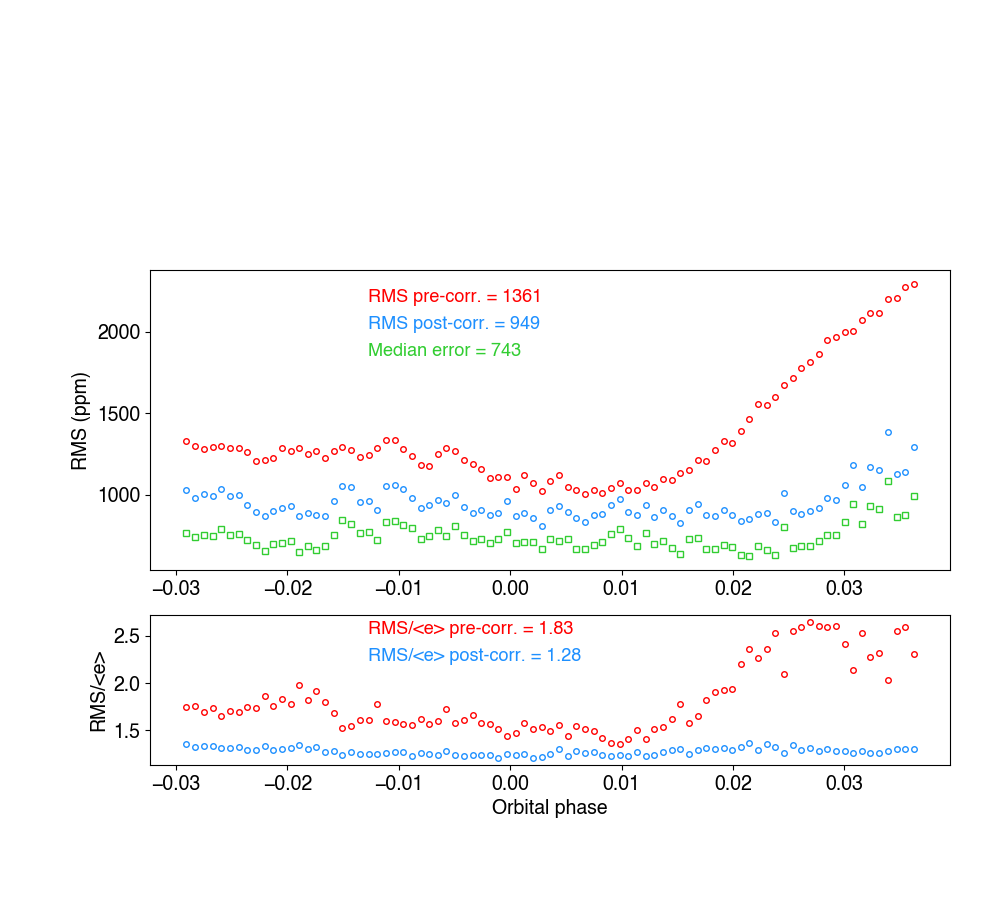}
\centering
\caption[]{RMS of HD\,209458 transmission spectra in epoch 2 before (red) and after (blue) wiggle correction. Green squares are the medians of the propagated error tables. The bottom panel shows the ratio between RMS and median error, expected to be unity for photon-dominated noise. The correction removes the structure in this ratio, which can be attributed to systematic noise from the wiggles.}
\label{fig:RMS_all_wig}
\end{figure}

Because our wiggle model still depends on many free parameters, its characterization for a given dataset has to be performed semi-automatically. Our procedure (see Sect.~\ref{apn:wig_car}) follows iterative steps that aims at initializing the model close to its best fit. First, we screen transmission spectra manually to identify the spectral ranges that constrain the wiggles and to assess the strength of the two components. Then we sample the chromatic variations of the components' frequency and amplitude in each exposure, and determine their optimal polynomial models. Those polynomials are used to initialize the spectral wiggle model fitted to each exposure. These fits yield time series of polynomial coefficients for the amplitude and frequency of the two components, as well as phase values, which are fitted in turn with the pointing coordinate function. Finally, the resulting set of pointing parameters is used to initialize the spectro-temporal wiggle model fitted to the full dataset. Original flux spectra are corrected for the wiggles through division with this global best-fit model (Fig.~\ref{fig:wig_trans}). The efficiency of the wiggle correction is illustrated in Fig.~\ref{fig:RMS_all_wig}. 

Our exploration of the wiggles hints at trends between targets, offering hope that our model can be further simplified with generic laws linked to the telescope/instrument properties or the observed sky region. Our findings that wiggles are dominated by two components with reproducible frequencies already provide information to identify the exact optical elements that are responsible for the interference pattern. A systematic exploration of the wiggle properties over all ESPRESSO transit datasets, which can be done in a homogeneous way with \textsc{antaress}, would help to further characterize their behavior (\citealt{Allart2020}). 

%%%%%%%%%%%%%%%%%%%%%%%%%%%%%%%%%%%%%%%%%%%%%%%%%%%%%%%%%%%%%%%%%%%%%%%%%%%%%
%%%%%%%%%%%%%%%%%%%%%%%%%%%%%%%%%%%%%%%%%%%%%%%%%%%%%%%%%%%%%%%%%%%%%%%%%%%%%

\section{Processing methods}
\label{sec:proc_mod}

At this stage of the workflow (Fig.~\ref{fig:flowchart}) spectral data are corrected for instrumental and environmental variations. We now describe the methods developed to extract disk-integrated and planet-occulted profiles in each epoch, and to process them into various formats. Keeping with the spirit of \textsc{antaress}, data is processed as S2D spectra until otherwise required. They can then be converted into 1D spectra (Sect.~\ref{sec:2D_1D}) or cross-correlated with the appropriate mask (Sect.~\ref{sec:CCFs}), and resampled into new time
series  or single master profiles (Sect.~\ref{sec:temp_resamp}). The processing pathway depends on the objective and corresponding \textsc{Analysis} methods (Sect.~\ref{sec:ana_mod}), such as measuring global photospheric properties from a master disk-integrated stellar spectrum, or performing a RM analysis using time
series  of planet-occulted CCFs. We will describe methods to further process S2D spectra until the extraction of planetary atmospheric profiles in a companion paper.

%%%%%%%%%%%%%%%%%%%%%%%%%%%%%%%%%%%%%%%%%%%%%%%%%%%%%%%%%%%%%%%%%%%%%%%%%%%%%

\subsection{Stellar line detrending}
\label{sec:st_line_detrend}

This method corrects time
series  of disk-integrated stellar profiles for systematic trends identified between the line properties ($rv$ residuals, FWHM, or constrast) and environmental/housekeeping parameters (e.g., time or S/N). This is done by fitting polynomials, sinusoids, or their combination, to the out-of-transit line properties (Sect.~\ref{sec:lprof_var}) and using the derived model to correct the entire time
series , in particular during transit (see details in Sect.~\ref{apn:st_line_detrend}). This approach works best when stellar baseline can be measured both before and after the transit.  

We highlight the importance of CCFs, whose properties can be measured at high precision, to devise detrending models that can then be applied to individual spectral lines or full spectra (for a direct detrending of CCFs, see \citealt{Bourrier2022,Bourrier2023}). Using this approach (not detailed here) to correct the S2D spectra of WASP-76 and HD\,209458's first epochs for correlations of their CCFs with phase and S/N reduced the out-of-transit dispersion of their $rv$ residuals from 2.8 to 1.9\,m\,s$^{-1}$ (WASP-76) and 1.25 to 1.13\,m\,s$^{-1}$ (HD\,209458), and contrast (normalized to its mean) from 383 to 258 ppm (WASP-76) and 275 to 174 ppm (HD\,209458).

%%%%%%%%%%%%%%%%%%%%%%%%%%%%%%%%%%%%%%%%%%%%%%%%%%%%%%%%%%%%%%%%%%%%%%%%%%%%%

\subsection{Alignment in star rest frame}
\label{sec:align_star}

Disk-integrated profiles must be aligned in the rest frame of the star, taking care of shifting their individual spectral grid rather than resampling them on a common grid (Sect.~\ref{sec:resamp}). Complementary spectral profiles, such as telluric spectra (Sect.~\ref{sec:tell_corr}) or flux calibration profiles (Sect.~\ref{sec:count_sc}), follow the same procedure so that they can be used for weighing (Sect.~\ref{sec:spec_weights}). Wavelengths in the star rest frame $\lambda_\star$ are calculated using the formula of the classical Doppler shift, for a stationary receiver located at the solar solar System barycenter and the star as the source moving in an arbitrary direction:
\begin{equation}
\lambda_\star(t) = \frac{\lambda_\mathrm{B_{\odot}}}{ (1+\frac{rv_{\star/B_{\star}}(t)}{c})(1+\frac{rv_{B_{\star}/B_{\odot}}}{c}) }
\end{equation}
where $\rm rv_{\star/B_{\star}}$ is the star Keplerian $rv$ relative to the planetary system barycenter and $\rm rv_{B_{\star}/B_{\odot}}$ is the systemic $rv$ between the stellar and solar System barycenters. If disk-integrated profiles are in CCF format, the sum of these two $rv$ is instead subtracted from their $rv$ grids.

The Keplerian $rv$ (Eq.~\ref{eq:kep}) should account for all bodies that induce a substantial reflex motion over the course of the epoch. This motion is small during the transit of planets on low-eccentricity orbits and remains below 500\,m\,s$^{-1}$ for most known host stars, which is lower than the best resolutions available (1.6\,km\,s$^{-1}$ for ESPRESSO in ultra high-resolution mode, \citealt{Pepe2021}). There are nonetheless a few known planets inducing larger motions, and accounting for this correction may become important with next-generation spectrographs. 

The systemic $rv$ can be known a priori from velocimetry, but it must be measured from each dataset. First, because it can vary between instruments and epochs. Second, to work in the rest frame of a specific stellar line, which may deviate from the overall stellar rest frame depending on the line shape and formation layer. The measurement of $\rm rv_{B_{\star}/B_{\odot}}$ is done by isolating a given stellar line or converting disk-integrated spectra into CCFs (Sect.~\ref{sec:CCFs}), aligning the profiles by correcting for the Keplerian motion, binning them over out-of-transit exposures (Sect.~\ref{sec:temp_resamp}), and deriving the centroid of the master stellar line (Sect.~\ref{sec:ana_indiv_star_spec}). 

We neglected the Lorentz factor $1/\sqrt{1-(v/c)^2}$ that accounts for relativistic effects in the Doppler shift formula, as the contribution from the systemic motion is constant, that from the Keplerian motion amounts to $\sim$0.04\,cm\,s$^{-1}$ for a velocity of 500\,m\,s$^{-1}$.

%%%%%%%%%%%%%%%%%%%%%%%%%%%%%%%%%%%%%%%%%%%%%%%%%%%%%%%%%%%%%%%%%%%%%%%%%%%%%

\subsection{Broadband flux scaling}
\label{sec:flux_scaling}

This method resets profile time
series  to their true relative flux level. This step if not necessary if absolute fluxes can be measured or if the planet contribution is not removed by the flux balance correction (Sect.~\ref{sec:col_bal}), but is generally critical for ground-based observations. 

\begin{figure}
\includegraphics[trim=1.5cm 0.5cm 4.5cm 6cm,clip=true,width=\columnwidth]{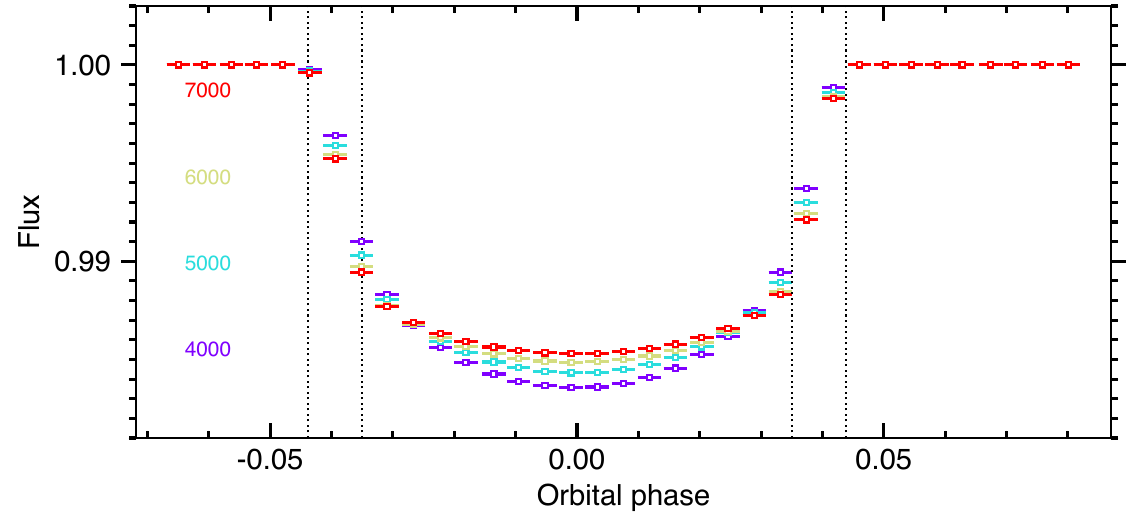}
\centering
\caption[]{Chromatic broadband scaling. Scaling values for WASP-76b exposures (epoch 1) were derived from a chromatic set of \textsc{batman} light curves (see text). }
\label{fig:LC_scaling_module_chrom}
\end{figure}

\begin{figure}
\includegraphics[trim=1.5cm 0.5cm 4.5cm 6cm,clip=true,width=\columnwidth]{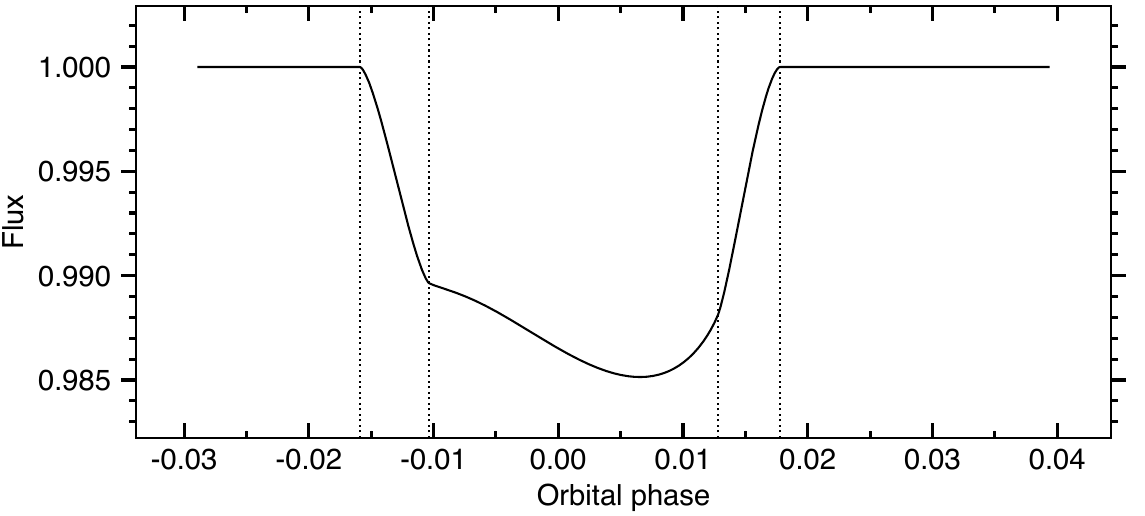}
\centering
\caption[]{Numerical broadband scaling. The light curve, generated with \textsc{antaress} stellar grid, corresponds to the mock system in Fig.~\ref{fig:System_view} (middle panel). Note the distorted shape of the light curve due to the misaligned orbit of the planet across the gravity-darkened star, and the asymmetric transit contacts due to its oblateness.  }
\label{fig:LC_scaling_module_mock}
\end{figure}

At this stage the $F^{\rm meas}$ spectra have the low-frequency profile of the chosen reference $F_{\rm ref}$ but retain their original flux level. We first normalize them to a common integrated flux $TF_{\rm ref}$, set to the median flux over the epoch or to a physically-motivated value specific to each epoch. Then, we scale the normalized profiles using light curves that match $lc$, the flux modulation  observed from the system that we calculate using the \textsc{batman} package (for simple transit light curves, Fig.~\ref{fig:LC_scaling_module_chrom}), the \textsc{antaress} stellar surface model (to account for simultaneous transits and complex photospheres, Fig.~\ref{fig:LC_scaling_module_mock}), or a measured/theoretical light curve imported into the workflow (to account for complex features, specific to a given epoch). The real light curves trace temporal variations of the broadband stellar intensity and planetary absorption/emission, and our scaling light curves can thus be calculated chromatically over $\wbstar$. Final rescaled profiles (see the full procedure in Sect.~\ref{apn:flux_scaling}) can be expressed as:
\begin{equation}
\begin{split}
F^{\rm sc}(\wstar,E,t) &= F(\wstar,E,t) C^{\rm norm}_{\rm ref}(\wbstar,E), \\
\mathrm{with}\,C^{\rm norm}_{\rm ref}(\wbstar,E) &= \frac{\mathrm{TF}_{\rm ref}}{\sum_{\wstar} F_{\rm ref}(\wstar,E) d\wstar}. 
\end{split}
\label{eq:Fsc}
\end{equation}
Where the $F$ spectral time
series  have the low-frequency profile and global flux level of $F_{\rm ref}$, and the relative flux level of the true spectra.

Both the HD\,209458b and WASP-76b datasets are scaled using \textsc{batman} light curves. While the medium-resolution sodium signature in the transmission spectrum of HD\,209458b (\citealt{Charbonneau2002}) may trace broad wings from the planetary atmospheric lines (\citealt{VM2011a}), the signatures in the line cores arise from an incorrect estimate of the planet-occulted stellar line (\citealt{CasasayasBarris2021,Dethier2023}). In-transit spectra around the sodium doublet thus mainly decrease in flux due to the planet atmospheric continuum in this range, and we apply an achromatic scaling using the constant transit depth and limb-darkening from Table~\ref{tab:sys_prop}. In the case of WASP-76b, we study spectra over the entire ESPRESSO optical range. In-transit scaling must thus account for variations in limb-darkening and planetary absorption at low spectral resolution over this range. We use the non-linear limb-darkening coefficients and transit depths derived from HST/STIS spectra (\citealt{Fu2021}), which both show strong variations between 300 and 600\,\AA\, (Fig.~\ref{fig:LC_scaling_module_chrom}).

%%%%%%%%%%%%%%%%%%%%%%%%%%%%%%%%%%%%%%%%%%%%%%%%%%%%%%%%%%%%%%%%%%%%%%%%%%%%%

\subsection{Residual profiles}
\label{sec:res_prof}

This method allows retrieving profiles that contain the starlight occulted by a planet\footnote{What we call planet-occulted starlight is actually light from the unocculted stellar profile coming from the equivalent region of the star outside of the transit.} and filtered by its atmosphere. At this stage, disk-integrated profiles are aligned in the star rest frame, have a common spectral flux balance and correct relative flux levels. They can be decomposed between regions occulted or not by the planet as:
\begin{equation}
\begin{split}
F^{\rm sc}(\wstar,E,t) = C^{\rm norm}_{\rm ref}(\wbstar,E) ( F_p(\wstar,E,t) + \\
 \sum_{i \notin \iocc} f_i(\wstar,E)\,S_i + \sum_{i \in \iocc} f_i(\wstar,E)\,(S_i - S_i^p(\wstar,E) ),
\end{split}
\label{eq:Fsc_Fres}
\end{equation}
where $F_p$ is the light emitted/reflected by the planet (Eq.~\ref{eq:Fp}) and $f_i$ the surface flux density from region $i$ of the stellar disk, which has a sky-projected surface $S_i$ along the LOS. $S_i^p$ is the portion of $S_i$ hidden by the planet, which can be written as the sum of the equivalent surfaces occulted by the opaque planetary disk $S_i^{p[thick]}(\wbstar,E)$ and by the optically thin atmospheric annulus $S_i^{p[thin]}(\wstar,E)$. The spectral dependence of these surfaces traces broadband variations of the atmospheric continuum and high-frequency absorption by the annulus. The surfaces are also time-dependent because of the partial occultation of the star during ingress and egress, of the Doppler shift of the atmospheric signal due to the planetary orbital motion, and of the different layers of planetary atmosphere that occult the star at different times of the transit (e.g., \citealt{Ehrenreich2020}).

\begin{figure}
\includegraphics[trim=0.5cm 7.5cm 0.9cm 7cm,clip=true,width=\columnwidth]{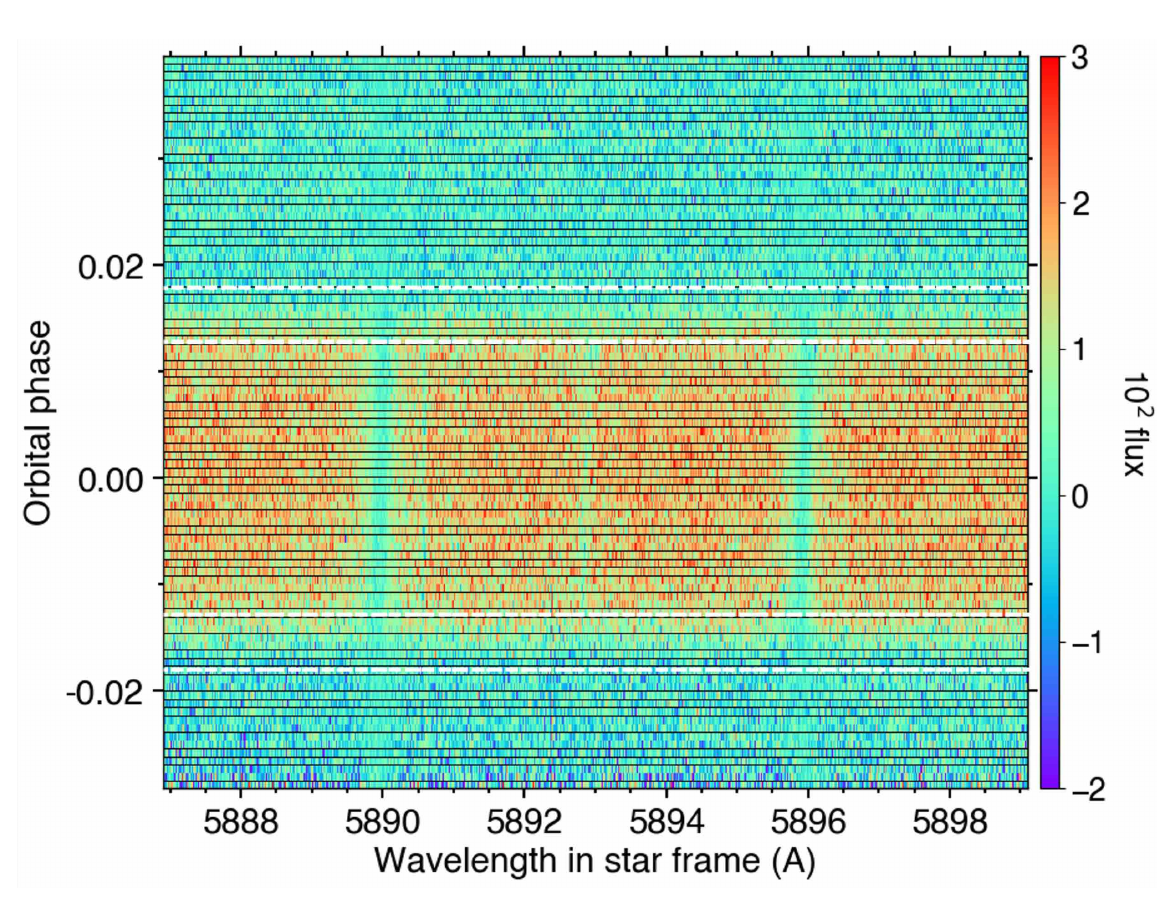}
\centering
\caption[]{Map of residual profiles during HD\,209458b epoch 1, in the region of the sodium doublet. Values are colored as a function of flux and plotted as a function of wavelengths in the star rest frame (in abscissa) and orbital phase (in ordinate). The planet-occulted stellar lines of sodium and nickel are clearly visible during transit (contacts are shown as white dashed lines). Visual inspection of out-of-transit and in-transit continuum values do not show any strong systematic features.}
\label{fig:HD209_Res_Na}
\end{figure}

Residual profiles are extracted as the difference between the unocculted star and exposure profiles (Fig.~\ref{fig:HD209_Res_Na}): 
\begin{equation}
\begin{split}
F^{\rm res}(\wstar,E,t) &= F_{\star}^{t}(\wstar,E) - F^{\rm sc}(\wstar,E,t) \\
                                           = C^{\rm norm}_{\rm ref}(\wbstar,E) (-F_p(\wstar,E,t) &+ \sum_{i \in \iocc} f_i(\wstar,E) \, S_i^p(\wstar,E) ).       \\              
\end{split}                             
\label{eq:fres} 
\end{equation}
In contrast to standard approaches, we calculate a reference for the star, $F_{\star}^{t}$, directly on the spectral grid of each exposure to avoid unnecessary resampling and subsequent resolution loss or introduction of correlated noise (see the detailed procedure in Sect.~\ref{apn:res_prof}). Residual profiles are also extracted outside of the transit to retrieve the light potentially coming from the planet, $F_p$, and to analyze the noise distribution of the data (in the absence of planetary signal and stellar variability, out-of-transit $F^{\rm res}$ values should distribute as Gaussian noise). 
 
Residual profiles can then be processed in two different ways. Either all planetary contributions are removed to convert residual profiles into intrinsic stellar profiles, from which are derived stellar surface properties and the orbital system architecture (as described in the following sections), or residual profiles can be corrected for planet-occulted stellar lines to isolate $S_i^{p[thin]}$ and $F_p$.

%%%%%%%%%%%%%%%%%%%%%%%%%%%%%%%%%%%%%%%%%%%%%%%%%%%%%%%%%%%%%%%%%%%%%%%%%%%%%

\subsection{Intrinsic profiles}
\label{sec:intr_prof}

This method converts in-transit residual profiles $F^{\rm res}$ into intrinsic stellar profiles $F^{\rm intr}$. First, residual profiles are reset to a common flux level by correcting for broadband flux variations (Sect.~\ref{sec:flux_scaling}). Then spectral ranges contaminated by the planetary atmosphere, if present, are identified and masked (Sect.~\ref{sec:atm_mask}). Finally, intrinsic spectra are corrected for chromatic deviations around the average $rv$ of the occulted regions induced by broadband variations in the planet continuum. The resulting profiles can be expressed as: 
\begin{equation}
\begin{split}
F^{\rm intr}(\wstar,E,t) = C^{\rm norm}_{\rm ref}(\wbstar,E) \times & \\
\, \frac{I_\mathrm{occ}(\wstar,E,t) \, \alpha_\mathrm{occ}(\wbstar,E,t) \, S_\mathrm{occ}^{\rm p[thick]}(\wbstar,E,t) }{1-lc(\wstar,E,t)}&                                 
\end{split}
\label{eq:intr}
.\end{equation}
Here, $I_\mathrm{occ}$ is the specific intensity emitted by the occulted region, and $\alpha_\mathrm{occ}$ its broadband intensity variations. Because the spectral flux scaling $lc$ can be expressed as a function of $\alpha_\mathrm{occ}$ and the planetary continuum, $S_\mathrm{occ}^{\rm p[thick]}$, the above expression can be simplified to (for full details see Sect.~\ref{apn:intr_prof}):
\begin{equation}
\begin{split}
F^{\rm intr}(\wstar,E,t) &= F^{\rm norm}_{\rm ref}(\wbstar,E) I_\mathrm{occ}(\wstar,E,t)
\end{split}     
\label{eq:fintr_final}                          
.\end{equation}
Here, $F^{\rm norm}_{\rm ref}$ only contains components constant in time over the transit and with low-frequency spectral variations. Intrinsic profiles are thus independent of planetary occultation (at both low- and high- spectral resolution) and broadband stellar flux variations, and only trace variations in the spectral profile $I_\mathrm{occ}$ of the occulted stellar lines. In this way, the \textsc{antaress} workflow allows mapping spatially the spectral emission from exoplanet hosts, providing direct observational constraints for theory and models of stellar atmospheres (Sect.~\ref{sec:RMR_spec}). 

\begin{figure}
\includegraphics[trim=2cm 1cm 3cm 4cm,clip=true,width=\columnwidth]{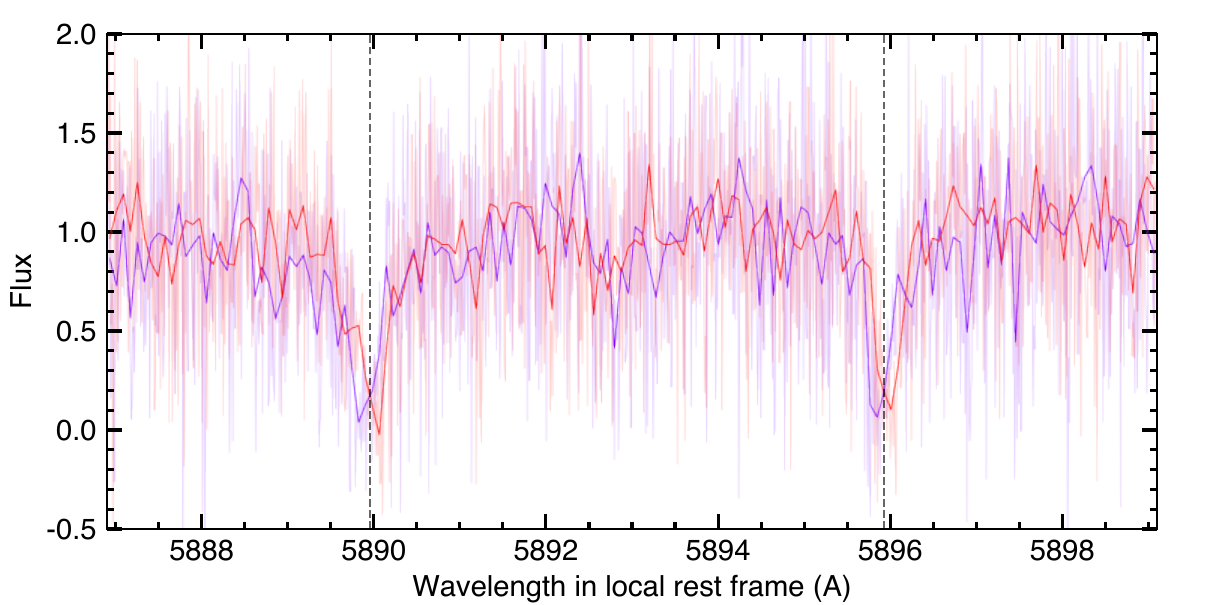}
\centering
\caption[]{Intrinsic spectra of HD\,209458 in epoch 1, showing the sodium doublet from occulted regions of the photosphere. The purple (from end of ingress) and red (from beginning of egress) spectra trace HD\,209458b moving from the blueshifted to the redshifted stellar hemisphere along its aligned orbit (spectra are downsampled for clarity, with original data shown as faded profiles). }
\label{fig:Intrinsic_profiles}
\end{figure}

%%%%%%%%%%%%%%%%%%%%%%%%%%%%%%%%%%%%%%%%%%%%%%%%%%%%%%%%%%%%%%%%%%%%%%%%%%%%%%%%%%%%%%%%%%%%%%%%%%%%%%

\subsubsection{Cross-correlated residual and intrinsic spectra}
\label{sec:CCF_intr}

\begin{figure}
\includegraphics[trim=1cm 5cm 1cm 0.5cm,clip=true,width=\columnwidth]{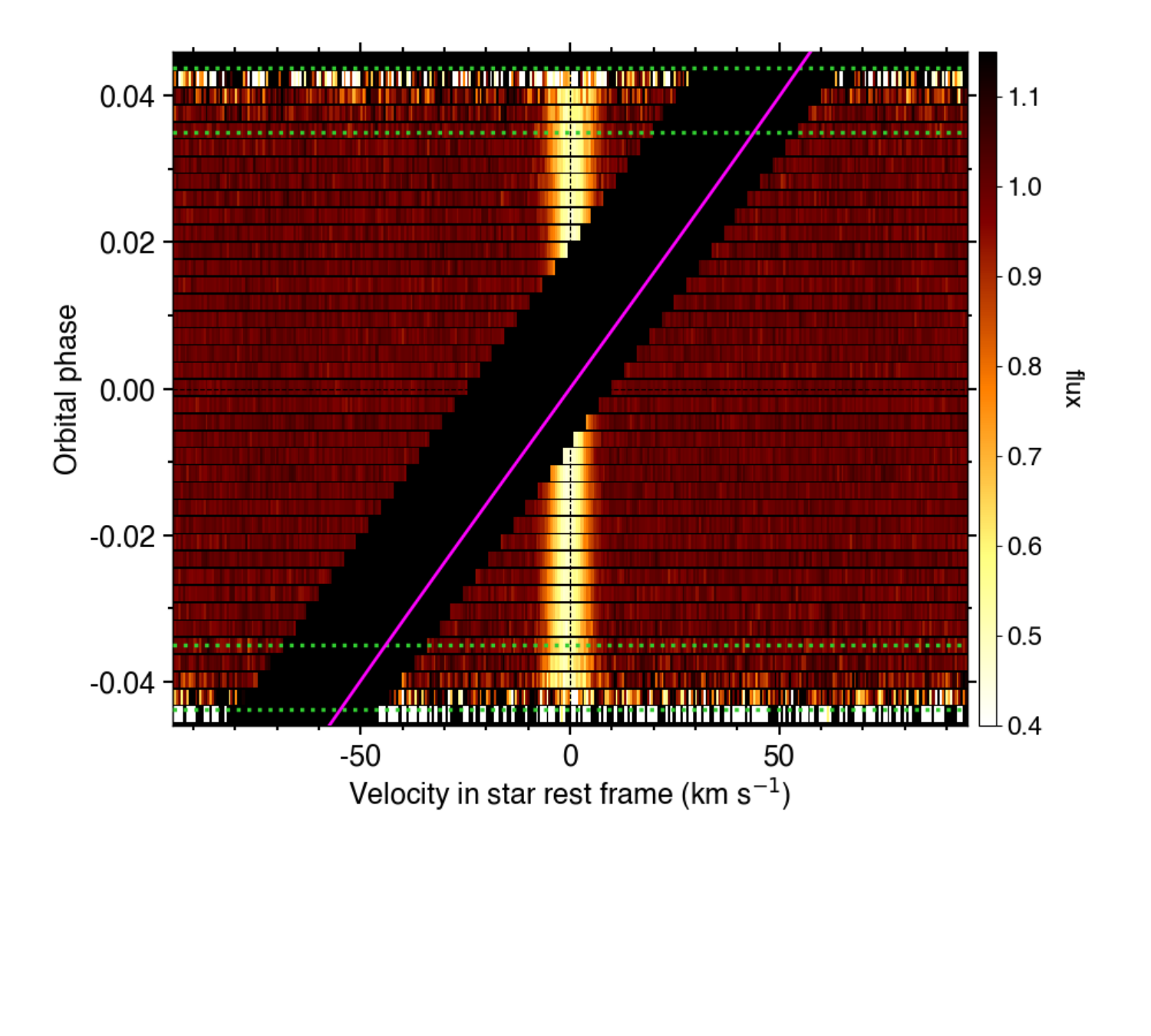}
\centering
\caption[]{Intrinsic CCF map of WASP-76 in epoch 2, plotted as a function of $rv$ in the star rest frame (in abscissa) and orbital phase (in ordinate). Values are colored as a function of normalized flux, unveiling the track of local stellar lines occulted by the planet along its transit chord (green dotted lines show transit contacts). The magenta line shows the planet orbital track, along which was excluded the range contaminated by its atmosphere. CCFs were computed with \textsc{antaress} mask built from WASP-76 data (Sect.~\ref{sec:CCF_masks}).}
\label{fig:Intr_CCF_map_WASP76}
\end{figure}

Intrinsic spectra can be cross-correlated with a weighted mask to create high-S/N CCFs (Sect.~\ref{sec:CCFs}), from which series of stellar surface properties along the transit chord can be derived. The method can be applied with masks used by standard spectrograph DRSs, or preferably with a mask generated self-consistently from the processed data (see Sect.~\ref{sec:CCF_masks}). ``White'' intrinsic CCFs built from a mask encompassing the full spectrum (Fig.~\ref{fig:Intr_CCF_map_WASP76}) can first be used to study the stellar surface velocity field and system orbital architecture (Sect.~\ref{sec:RMR}). Intrinsic CCFs built from masks specific to a given species can then be used to study how the properties of the stellar atmospheric layers populated by this species deviate from the average photospheric properties probed by the white CCFs (Sect.~\ref{sec:RMR_spec}). We emphasize that, in contrast to disk-integrated CCFs distorted by planet-occulted lines and planetary atmospheric lines, CCFs computed from intrinsic spectra only trace the stellar emission and can be directly interpreted with stellar line models with fewer risks of biases. Particular care is taken to deal with planetary contamination, as described in Sect.~\ref{apn:intr_prof}. 

When computing intrinsic CCFs we cross-correlate out-of-transit residual spectra with the same mask. In-transit residual CCFs, computed as the difference between intrinsic CCFs and their best-fit model (Sect.~\ref{sec:ana_indiv_CCF}), are then rescaled by $(1 - lc_{\rm CCF}(\mathrm{rv},E,t))/lc_{\rm CCF}(\mathrm{rv},E,t)$ (where $lc_{\rm CCF}(\mathrm{rv})$ is the CCF of $lc(\wstar)$, Sect.~\ref{sec:CCFs}) to be made comparable to out-of-transit residual CCFs. The resulting series of residual CCFs allow studying the quality of the fit to the intrinsic stellar line, and the noise distribution of the data over a full epoch (Sect.~\ref{sec:RMR}).

%%%%%%%%%%%%%%%%%%%%%%%%%%%%%%%%%%%%%%%%%%%%%%%%%%%%%%%%%%%%%%%%%%%%%%%%%%%%%%%%%%%%%%%%%%%%%%%%%%%%%%

\subsubsection{Alignment of intrinsic profiles}
\label{sec:align_intr_prof}

Intrinsic profiles are extracted in the star rest frame, with stellar lines Doppler-shifted due to the motion of photosphere. This method aligns them in a common frame, in which they can be resampled temporally/spatially (Sect.~\ref{sec:temp_resamp}) to be analyzed at higher S/N (Sect.~\ref{sec:ana_Intr}) or used to extract atmospheric profiles. We take care of shifting the individual spectral grid of each profile rather than resampling them on a common grid (Sect.~\ref{sec:resamp}). Wavelengths in the rest frame of the stellar surface $\wsurf$ are calculated for the receiver located at the star rest frame and the local photosphere as the source:
\begin{equation}
\wsurf(t) = \frac{\wstar}{ (1+\frac{\rm rv_{\rm surf}(\wstar,t)}{c}) }
,\end{equation}
where $\rm rv_{\rm surf}(\wstar,t)$ is the $rv$ of the planet-occulted region, subtracted from the $rv$ grid of intrinsic profiles if they are in CCF format. We neglect relativistic effects due to the motion of the stellar photosphere, as they are negligible. 

Values of $rv_{\rm surf}$ can be set to the shifts derived for each intrinsic line profile (Sect.~\ref{sec:ana_Intr}). This approach is accurate but may be imprecise, or even impossible if the line cannot be fitted due to a low S/N or inconvenient planet masking (Sect.~\ref{sec:atm_mask}). Alternatively, $rv_{\rm surf}$ can be calculated for all planet-occulted regions with \textsc{antaress} stellar surface model (Sect.~\ref{sec:st_surf}), accounting for $rv$ contributions relevant for the studied line (e.g., convective blueshift in the line spectral range).

%%%%%%%%%%%%%%%%%%%%%%%%%%%%%%%%%%%%%%%%%%%%%%%%%%%%%%%%%%%%%%%%%%%%%%%%%%%%%
%%%%%%%%%%%%%%%%%%%%%%%%%%%%%%%%%%%%%%%%%%%%%%%%%%%%%%%%%%%%%%%%%%%%%%%%%%%%%

\section{Generic methods}
\label{sec:gen_mod}

We describe here generic methods that are used in several steps of the \textsc{antaress} workflow.

\subsection{Stellar continuum}
\label{sec:st_cont}

The stellar continuum is required in particular by the \textsc{Peak masking} (Sect.~\ref{sec:perspeak_corr}) and \textsc{CCF masks} (Sect.~\ref{sec:CCF_masks}) methods. Section~\ref{apn:st_cont} summarizes our approach to derive the continuum from the processed data (Fig.~\ref{fig:CCF_mask_HD209_strict}), using a simplified \textsc{SNAKE} sequence (smoothing, finding local maxima, measuring $\alpha$-shape, interpolating envelop) from the \textsc{RASSINE} algorithm (see \citealt{Cretignier2020b} for details). 

\subsection{Spectral weight profiles}
\label{sec:spec_weights}

\begin{figure}
\includegraphics[trim=0cm 0cm 0cm 0cm,clip=true,width=\columnwidth]{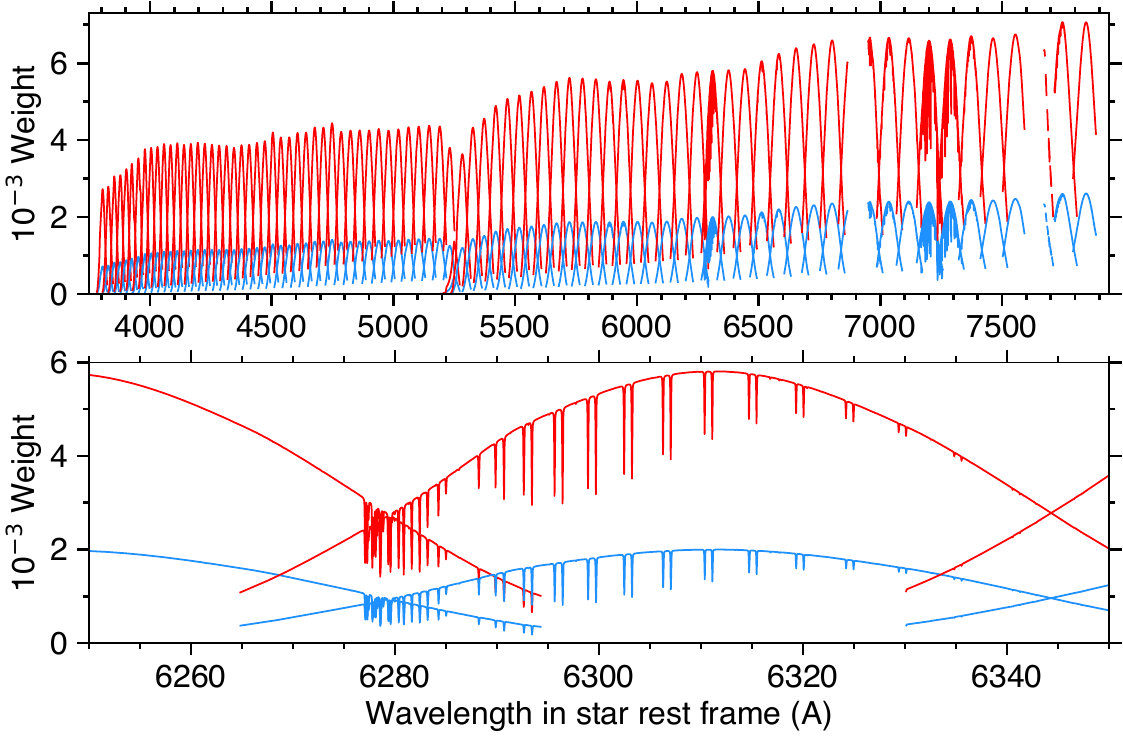}
\centering
\caption[]{Order-per-order weight profiles of disk-integrated stellar spectra in the first (blue) and last (red) exposures of WASP-76 epoch 1 (the lower panel shows a zoom-in of the upper panel). This shows how lower weights are given at the edges of detector orders, at the bottom of telluric lines, and in regions more strongly corrected for Earth's atmospheric diffusion (\textit{i.e.}, toward blue wavelengths and at the start/end of epochs). The stellar spectrum is not included, as it does not contribute to disk-integrated spectra weighing. 
}
\label{fig:weight_DI}
\end{figure}

\textsc{antaress} methods perform weighted averages of profile time
series  (Sect.~\ref{sec:temp_resamp}), or overlapping orders in a given profile (Sect.~\ref{sec:2D_1D}), using the squared inverse of the flux error in each pixel as weight. This accounts for different pixel flux precision between profiles due to low-frequency variations in their overall flux level, and to high-frequency variations in their spectral flux distribution. However, because the uncertainty associated with a measured pixel flux is a biased estimate of its true error, we devised an original approach to estimate its value (Sect.~\ref{apn:est_true}). Weight profiles on disk-integrated and intrinsic spectra measured over $\Delta t$ are then defined as (Sect.~\ref{apn:w_DI}):
\begin{equation}
\begin{split}
W^{\rm sc}(\wstar) &= \frac{1}{\sigma^{\rm sc, true}(\wstar)^2 }, \\
W^{\rm intr}(\lambda) &= \frac{(1 - lc(\lambda))^2}{ \sigma_{\rm \star}^{\rm meas}(\lambda)^2 + \sigma^{\rm sc,true}(\lambda)^2 }, \\
\rm with \, \sigma^{\rm sc,true}(\wstar) &= lc(\wstar) \sqrt{\tilde{g}_{\rm cal}(\wstar) F^{\rm meas}_{\star}(\wstar) C_{\rm corr}(\wstar)/ \Delta t}.  \\
\end{split}
\label{eq:weights_prof}
\end{equation}
The disk-integrated spectra are averaged in the star rest frame, while intrinsic spectra are averaged in the star ($\lambda$ = $\wstar$) or photosphere rest frame ($\lambda$ = $\wsurf$). The flux calibration profile $\tilde{g}_{\rm cal}$ (Sect.~\ref{sec:count_sc}), color effect (through $C_{\rm corr}$, Sect.~\ref{sec:col_bal}), and flux scaling $lc(\wstar)$ (Sect.~\ref{sec:flux_scaling}) contribute to low-frequency noise variations, weighing disk-integrated and intrinsic spectra in the same direction through $\sigma^{\rm sc,true}$. The $\tilde{g}_{\rm cal}$ profiles have a strong impact when averaging overlapping orders (Fig.~\ref{fig:weight_DI}).
 
\begin{figure}
\includegraphics[trim=0cm 0cm 0cm 0cm,clip=true,width=\columnwidth]{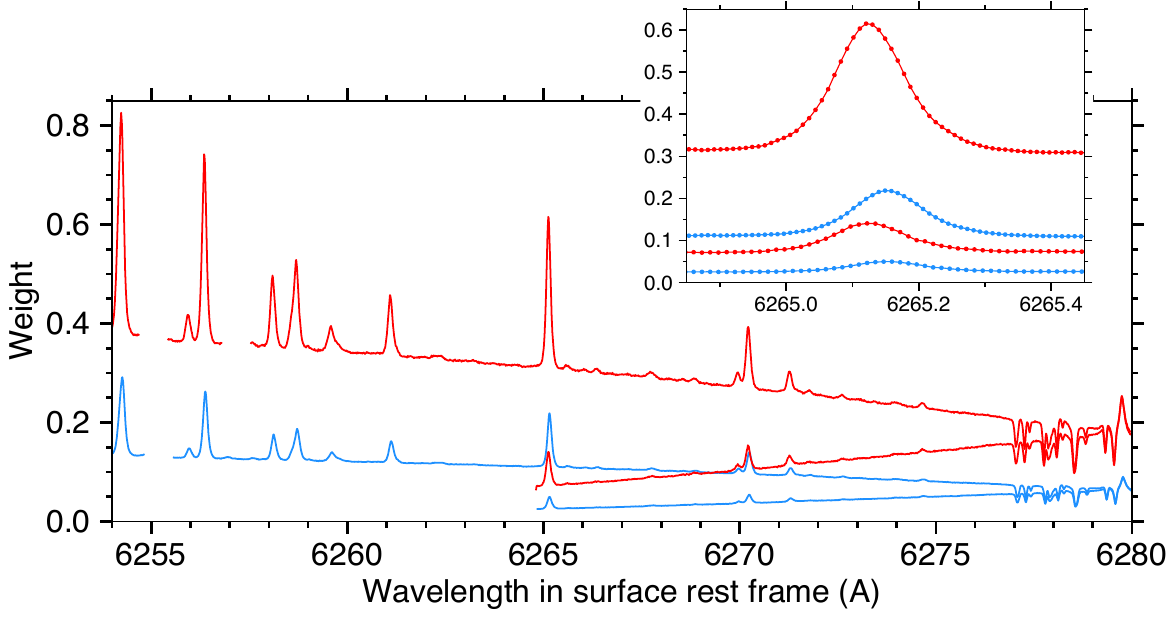}
\centering
\caption[]{Order-per-order weight profiles of intrinsic stellar spectra near the second (blue) and third (red) transit contacts of WASP-76 in epoch 1, as a function of wavelength in the surface rest frame. Besides the contributions highlighted in Fig.~\ref{fig:weight_DI}, we note the impact of masked planetary ranges (which redshift over the transit as the planet orbits faster than the star rotates) and disk-integrated stellar lines (which blueshift over the transit as the planet first occults the stellar hemisphere rotating toward, and then away, from the observer; see inset).}
\label{fig:weight_Intr}
\end{figure}

High-frequency noise variations arise from tellurics (through $C_{\rm corr}$, Sect.~\ref{sec:tell_corr}) and disk-integrated stellar lines (through $F^{\rm meas}_{\star}$). Tellurics weigh in the same direction disk-integrated and intrinsic spectra, with lower weights for deeper lines. Because tellurics shift with the Earth's motion in the star or surface rest frames, the noise and associated weight of a telluric-corrected pixel evolves over time. Stellar lines remain fixed in the star rest frame and only contribute to the weighing of intrinsic spectra in the surface rest frame. In that case, the disk-integrated lines in which intrinsic lines are imprinted shift between exposures and can induce strong noise variations in a given pixel (Fig.~\ref{fig:weight_Intr}). Counter-intuitively, $W^{\rm intr}$ is inversely proportional to $F^{\rm meas}_{\star}$ because the unocculted stellar spectrum acts as a (noisy) background light source when computing its difference with in-transit spectra. The parts of resulting intrinsic spectra that fall at the bottom of disk-integrated stellar lines are thus measured more precisely than those falling in the wings or continuum, since the flux and associated noise are lower there. When planetary atmospheric lines contaminate intrinsic spectra, the masked ranges also have a strong impact on the weighing due to their large shifts during transit (Sect.~\ref{sec:temp_resamp}). 

Low-frequency noise variations are generally accounted for in the literature by using the S/N, which approximates the true error on the disk-integrated flux over wide spectral bands. The impact of tellurics and stellar lines, however, is often ignored despite the biases it can introduce in averaged intrinsic stellar (Sect.~\ref{apn:test_wm}) and planetary spectra.

%%%%%%%%%%%%%%%%%%%%%%%%%%%%%%%%%%%%%%%%%%%%%%%%%%%%%%%%%%%%%%%%%%%%%%%%%%%%%%%%%%%%%%%%%%%%%

\subsection{Temporal and spatial resampling}
\label{sec:temp_resamp}

This method is used to calculate the weighted mean of a series of profiles, in new bins along a chosen dimension $x$:
\begin{equation}
F(\lambda,x,E) = \frac{\sum_{t} F(\lambda,x(t),E) W(\lambda,x(t),E) \, \Delta x(t)}{ \sum_{t} W(\lambda,x(t),E) \, \Delta x(t)  } 
\label{eq:av_prof}
.\end{equation}
Here, $F$ and $W$ are the flux and weight profiles associated to the exposure at time $t$ (Sect.~\ref{sec:spec_weights}). Details on the resampling procedure are given in Sect.~\ref{apn:temp_resamp}.

\begin{figure}
\includegraphics[trim=0cm 0cm 0cm 0cm,clip=true,width=\columnwidth]{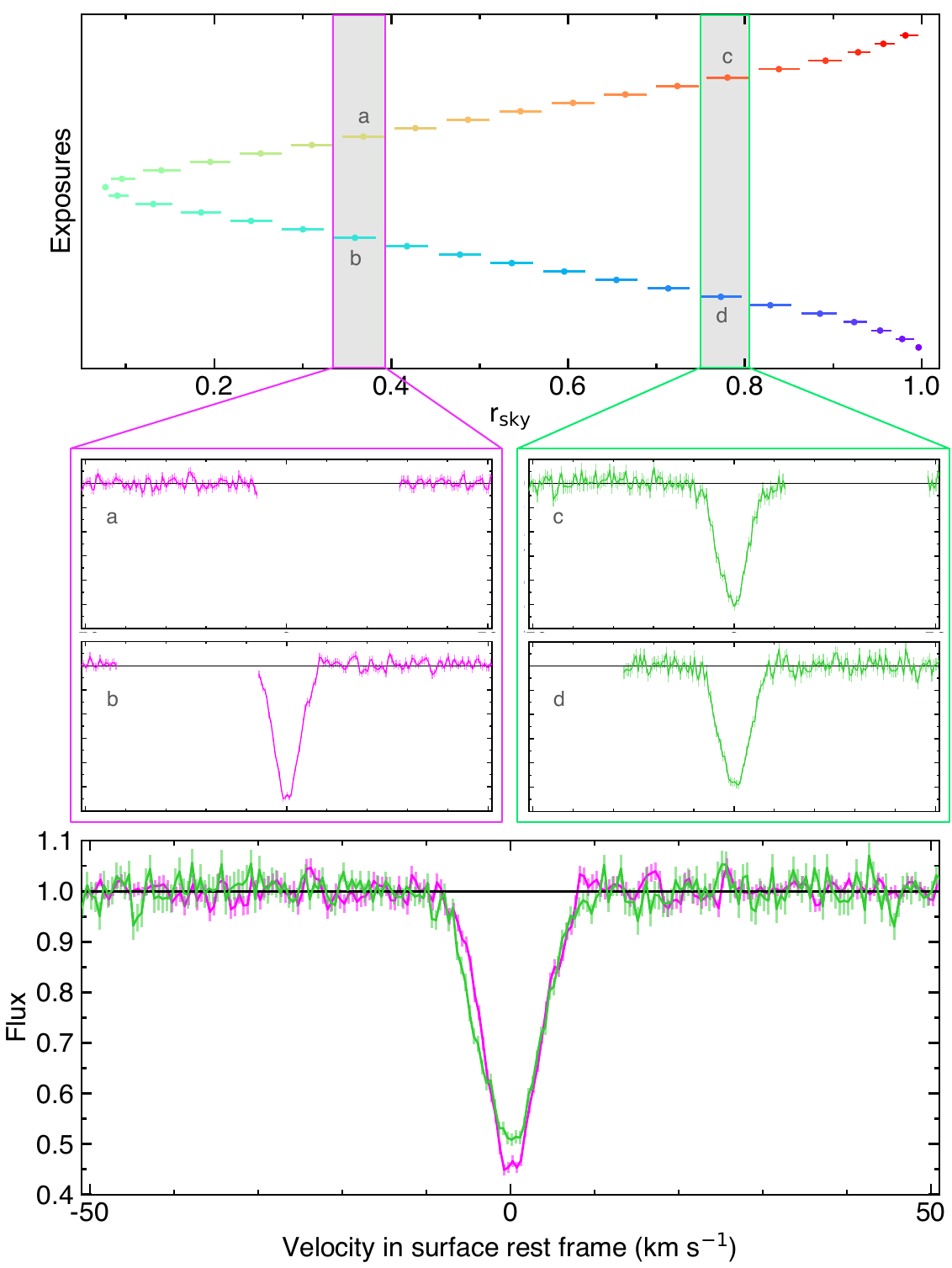}
\centering
\caption[]{Binning of WASP-76 intrinsic CCFs in epoch 1. The top panel shows the $r_\mathrm{sky}$ windows covered by each exposure (offset vertically for visibility, and colored over the rainbow scale with increasing orbital phase). Middle panels show the intrinsic CCFs of exposures overlapping with the binning windows chosen as example. The bottom panel shows the resulting binned profiles, corrected for planet contamination and tracing variations in shape across the stellar surface. 
}
\label{fig:Resampling}
\end{figure}

Disk-integrated profiles are binned over orbital phase (temporal resampling), typically to define a reference spectrum for the unocculted star. Intrinsic profiles can be binned over phase to build a similar series at higher S/N, which may allow analyzing stellar lines in individual new exposures (Sect.~\ref{sec:ana_Intr}) at the expense of blurring. Alternatively, intrinsic profiles can be binned over $r_\mathrm{sky}$ (spatial resampling) to build a series that maps the emission of the photosphere more directly. Indeed, with the assumption that stellar lines only vary with distance to the star center, intrinsic profiles at opposite phases around the time of inferior conjunction are symmetrical and can be combined into a higher-S/N profile in which planet-contaminated ranges are filled in. This approach is illustrated in Fig.~\ref{fig:Resampling}. Intrinsic profiles resampled as a function of $r_\mathrm{sky}$, or into a single master, can be used to tile \textsc{antaress} stellar grid (Sect.~\ref{sec:st_grid}) or to correct residual profiles for stellar contamination. 

%%%%%%%%%%%%%%%%%%%%%%%%%%%%%%%%%%%%%%%%%%%%%%%%%%%%%%%%%%%%%%%%%%%%%%%%%%%%%

\subsection{Conversion from 2D to 1D spectra}
\label{sec:2D_1D}

This method is applied to disk-integrated or intrinsic stellar spectra once they have gone through their latest processing stage (Fig.~\ref{fig:flowchart}), to delay as much as possible resolution loss and the introduction of correlated errors. A converted disk-integrated spectrum can be seen in Fig.~\ref{fig:CCF_mask_HD209_strict}, and the procedure is detailed in Sect.~\ref{apn:2D_1D}. The \textsc{antaress} workflow resumes with the 1D spectral time
series  after conversion. One of the interests of this method is also to generate order-merged 1D spectra, with associated uncertainties, which can be used beyond \textsc{antaress} (comparison with stellar models, template-matching to derive $rv$, analysis of interstellar medium absorption, etc). \\

%%%%%%%%%%%%%%%%%%%%%%%%%%%%%%%%%%%%%%%%%%%%%%%%%%%%%%%%%%%%%%%%%%%%%%%%%%%%%%%%%%%%%%%%%%%%%

\subsection{Cross-correlation}
\label{sec:CCFs}

This method generates CCF profiles by cross-correlating spectra with a binary mask that contains the central wavelengths and weights of a set of lines. Only mask lines that contribute to the cross-correlation in all exposures are kept, so that CCFs remain comparable. The method can be applied to disk-integrated or intrinsic stellar spectra, after which the \textsc{antaress} workflow resumes with the CCF time series. In that case, we recommend generating masks specific to the processed datasets (Sect.~\ref{sec:CCF_masks}). The method is also used to compute telluric CCFs from measured and theoretical spectra (Sect.~\ref{sec:tell_corr}). The procedure is detailed in Sect.~\ref{apn:CCFs}, with examples of CCFs shown in Fig.~\ref{fig:Resampling}.

It is advised to compute CCFs at the $rv$ resolution corresponding to the pixel width of the original spectra, to limit correlations introduced between CCF pixels. Nonetheless, the \textsc{antaress} workflow propagates these correlations into the covariance matrices associated with CCFs, and accounts for them in their analysis, so that smaller $rv$ steps can be used to gain in spectral resolution (Sect.~\ref{apn:test_resamp}).

%%%%%%%%%%%%%%%%%%%%%%%%%%%%%%%%%%%%%%%%%%%%%%%%%%%%%%%%%%%%%%%%%%%%%%%%%%%%%

\section{Analysis methods}
\label{sec:ana_mod}

We describe here the use and scope of the \textsc{Analysis} methods, which exploit the data resulting from a given sequence of \textsc{Processing} methods in spectral or CCF format, over a broad spectral range or a specific spectral line, and in individual, resampled, or joined exposures.

\subsection{Stellar mask generator}
\label{sec:CCF_masks}

Custom masks tailored to a given star have been shown to yield CCFs of improved quality (e.g., \citealt{Bourrier2023}). Thus, we propose a method, detailed in Sect.~\ref{apn:CCF_masks}, to generate cross-correlation masks directly from the processed data. It was adapted to the \textsc{antaress} workflow from a version of the \textsc{KitCat} method, and we refer to \citet{Cretignier2020a} for the underlying concepts. The linelist is built by identifying and selecting absorption lines in a 1D, normalized master spectrum, calculated over all epochs of a given instrument to keep their CCFs comparable and avoid biases. Masks are however specific to each instrument because their different spectral coverage and line spread functions change the shape and distribution of stellar lines. Each mask line is associated with a weight representative of the photonic error on its $rv$ shift. 

While masks are better-defined from high-S/N disk-integrated spectra, it may be interesting to derive them from intrinsic spectra in the case of fast-rotating stars, when disk-integrated lines are blended and difficult to isolate. Masks built from an entire spectrum allow generating white CCFs to analyze global properties of the photosphere (e.g., its rotational velocity, Sect.~\ref{sec:ana_indiv_star_spec}). By cross-matching the white linelist with a VALD linelist specific to the target star (\citealt{Piskunov1995,Kupka2000,Ryabchikova2015}), it is further possible to define masks specific to a given species, which trace different layers in the stellar atmosphere\footnote{Lines of species with differing atomic properties, or different excitation and ionization states, may form at different temperatures in the stellar atmosphere, see e.g., \citealt{AlMoulla2022}.}. The interest of this approach is enhanced by \textsc{antaress}'s method to define intrinsic CCFs as a function of stellar surface coordinates (Sect.~\ref{sec:temp_resamp}). Series of such intrinsic CCFs, built for a suitable set of species would, in effect, map the stellar atmosphere vertically and horizontally.

\begin{figure}
\includegraphics[trim=0cm 0cm 0cm 0cm,clip=true,width=\columnwidth]{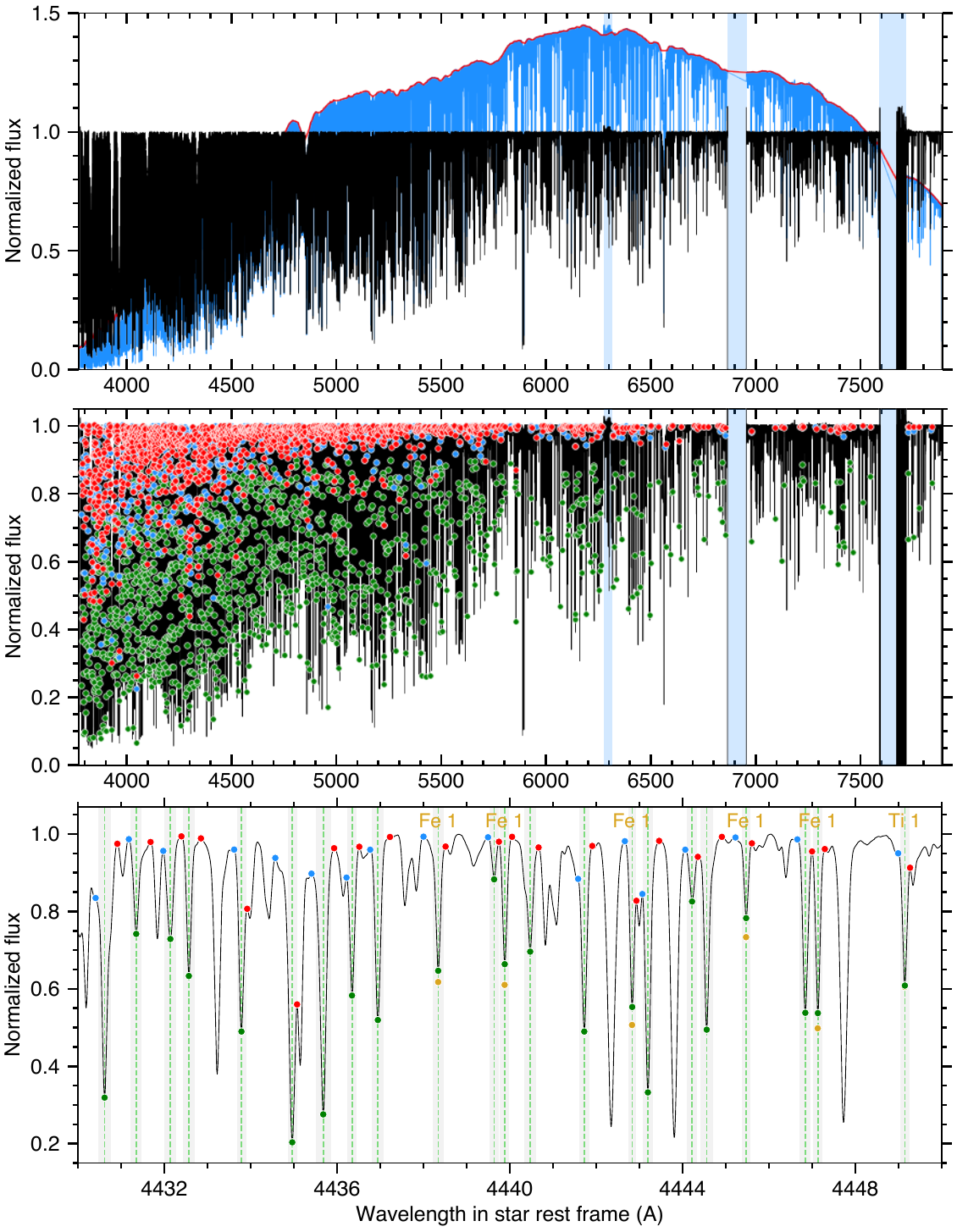}
\centering
\caption[]{Generation of a CCF mask for HD\,209458. \textit{Top panel}: Master disk-integrated spectrum over all out-of-transit 1D spectra, before (blue) and after (black) normalization with the stellar continuum (red). Shaded blue areas are excluded due to strong telluric contamination. \textit{Middle panel:} Final mask lines, identified in the master spectrum by their minimum (green disks) and surrounding maxima (red disks). \textit{Bottom panel}: Zoom highlighting the range of each mask line (shaded bands) and, when cross-matched with VALD, their origin species.}
\label{fig:CCF_mask_HD209_strict}
\end{figure}

\subsection{Analysis of individual stellar profiles}
\label{sec:ana_indiv_CCF}

We propose methods to derive properties from disk-integrated (Sect.~\ref{sec:ana_indiv_star_spec}) or intrinsic (Sect.~\ref{sec:ana_Intr}) stellar profiles in individual exposures. Analyzing original time series allows assessing their quality and identifying variations in the derived stellar properties (Sect.~\ref{sec:lprof_var}). Analyzing new series, resampled within an epoch or over several epochs (Sect.~\ref{sec:temp_resamp}), allows performing these analysis with higher precision at the expense of temporal/spatial resolution. 

Properties are directly measured on a stellar line (e.g., bissector, equivalent width) or derived by fitting the stellar profile with one of the models described in Sect.~\ref{apn:mod_prof}. Models can be defined analytically, to estimate first-order properties controlling the shape and position (e.g., FWHM, contrast, $rv$) of a single stellar line, or numerically to reproduce complex line shapes or to fit a portion of the stellar spectrum. When the analysis aims at deriving correct physical properties or comparing line properties between different instruments, the model stellar profile is convolved with the relevant instrumental responses.

%%%%%%%%%%%%%%%%%%%%%%%%%%%%%%%%%%%%%%%%%%%%%%%%%%%%%%%%%%%%%%%%%%%%%%%%%%%%%%%%%%%%%%

\subsubsection{Disk-integrated stellar lines}
\label{sec:ana_indiv_star_spec}

\begin{figure}
\includegraphics[trim=0cm 0cm 0cm 0cm,clip=true,width=\columnwidth]{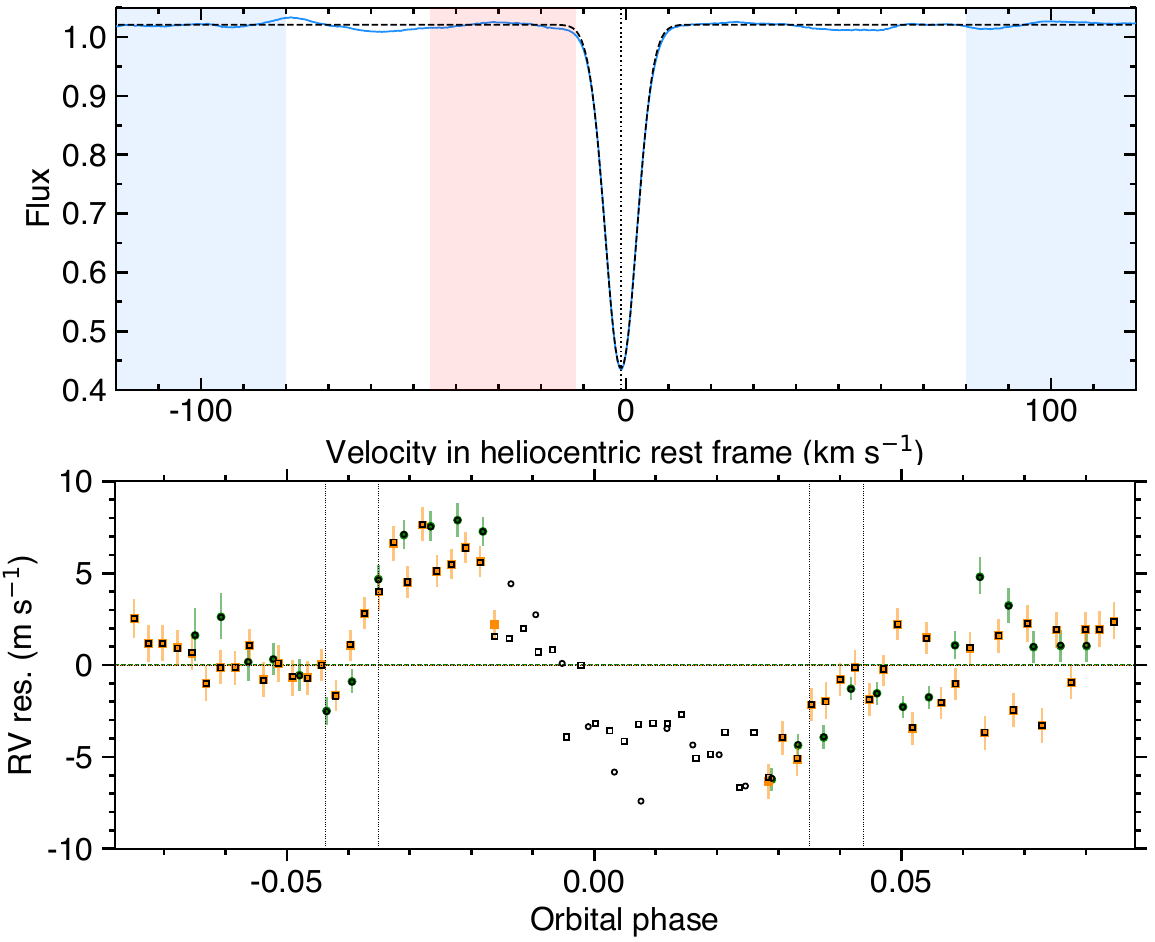}
\centering
\caption[]{Fit to WASP-76 disk-integrated CCFs, computed with the ESPRESSO DRS F9 mask. \textit{Top panel}: The dashed black profile is the Gaussian best-fit to the CCF in the last exposure (blue profile, phase -0.0162) before the stellar line is too contamined by the planetary signal (red area) to be accurately fitted. Blue areas are never contaminated by the planet, and used to measured the continuum. \textit{Bottom panel}: $rv$ residuals from the Keplerian model. Orange squares and green disks were derived from CCFs with no strong planetary contamination in epochs 1 and 2. Matching black symbols (shown without errors for clarity) were derived without excluding planetary ranges from the fits.}
\label{fig:DI_WASP76_fit}
\end{figure}

Most disk-integrated lines of G and K-type stars are well fitted with Gaussian profiles, especially when cross-correlated over many transitions (Fig.~\ref{fig:DI_WASP76_fit}), while lines with damping wings or asymmetrical shapes can be reproduced with Voigt or skewed Gaussian profiles. Most M dwarf lines are blended and difficult to fit individually, but result in sidelobed CCFs that are well reproduced with Double-Gaussian profiles (\citealt{Bourrier_2018_Nat}). The disk-integrated lines of early-type stars may be shaped by rotational broadening or even gravity darkening, and cannot be reproduced with these simple analytical profiles. In that case, the lines can still be fitted numerically to constrain local line properties, the photosphere geometry, and its intensity variations. Fitting the disk-integrated CCF of fast-rotating stars may constrain $v$\,sin\,$i_\mathrm{\star}$ independently of the value derived from the RM analysis (e.g., \citealt{Bourrier2020_HEARTSIII}). 

\begin{figure}
\includegraphics[trim=0cm 0cm 0cm 0cm,clip=true,width=\columnwidth]{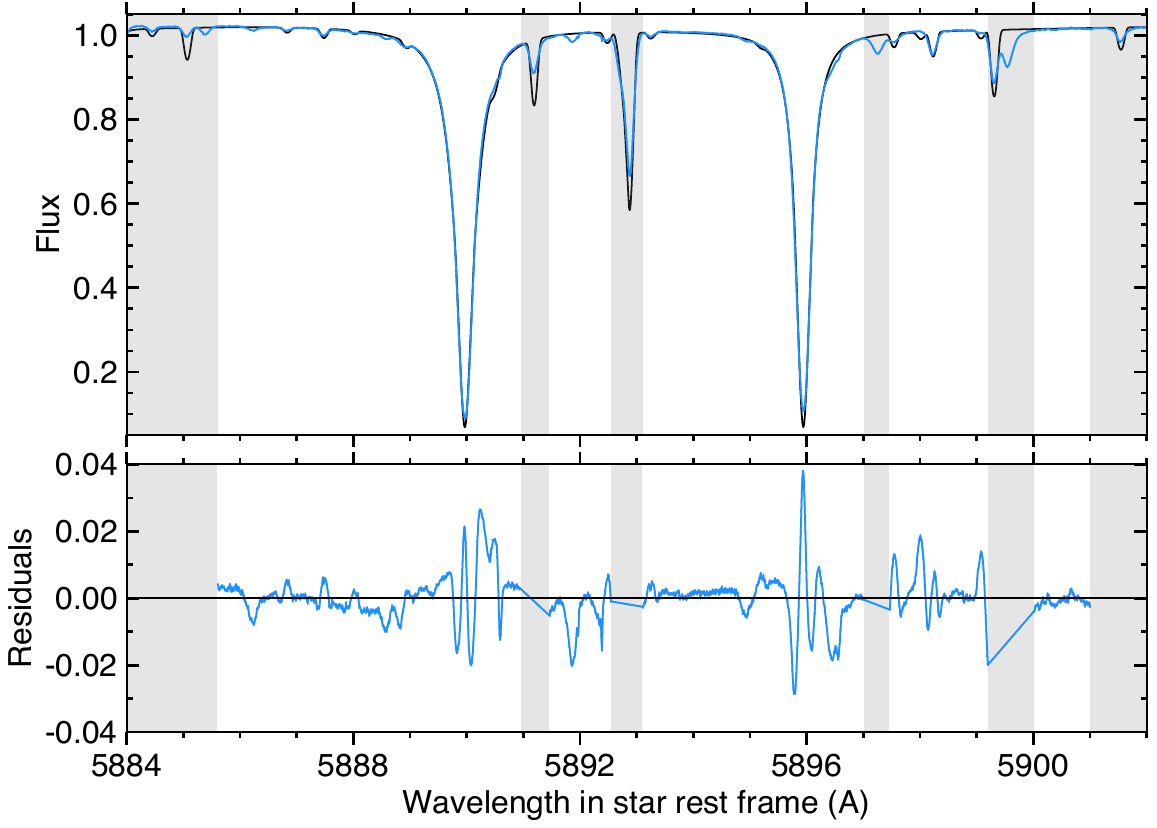}
\centering
\caption[]{Fit to the disk-integrated sodium doublet of HD\,209458. \textit{Top panel}: The black profile shows the best-fit \textsc{pySME} theoretical spectrum to the measured spectrum (blue profile) over the unshaded regions. Only the sodium abundance was varied in this example, so that other deep stellar lines were excluded from the fit. \textit{Bottom panel}: residuals from the fit.  }
\label{fig:DI_Na_fit}
\end{figure}

Numerical profiles can further be used to constrain the abundance of a line species. To illustrate this we fitted the sodium doublet of HD\,209458 in the out-of-transit 1D spectrum (Sect.~\ref{sec:2D_1D}) averaged over both epochs (Sect.~\ref{sec:temp_resamp}). Profiles were tiled over \textsc{antaress} stellar grid (Sect.~\ref{sec:st_grid}) using the rotational velocity and limb-darkening properties from Table~\ref{tab:sys_prop}, and series of theoretical intrinsic profiles calculated with a plane-parallel MARCS2012 stellar atmosphere model ($\xi$ = 1\,km\,s$^{-1}$), a VALD linelist for HD\,209458, an overall metallicity [Fe/H] = 0.02, and the NLTE departure coefficients for sodium from \citet{Anish2022_dataset}. The best fit (Fig.~\ref{fig:DI_Na_fit}) is obtained for a model sodium abundance of 6.05. While the NLTE grid reproduces better the overall shape of the sodium doublet (see also \citealt{CasasayasBarris2021,Dethier2023}), the core and wings of the lines cannot be well fitted at the same time, leaving residuals on the order of 2\%. This discrepancy could be due to the contribution of the chromosphere in the core of the sodium lines (\citealt{Bruls1992}), highlighting the need for refinements in stellar atmosphere models that can be driven by \textsc{antaress} measurements of spatially resolved line profiles (Sect.~\ref{sec:RMR_spec}).  

Stellar properties are typically derived from out-of-transit exposures to prevent biases induced by the transiting planet. Series of line shape properties (Sect.~\ref{sec:lprof_var}) allow detrending the disk-integrated profiles from systematics over the full epoch (Sect.~\ref{sec:st_line_detrend}), and $rv$ time
series  allow us to check the quality of the Keplerian model and evaluating the presence of an RM anomaly. In-transit disk-integrated profiles are then used to extract planet-occulted profiles, rather than to analyze their anomalous properties. However, useful information can still be derived from those profiles if they can be masked for the spectral ranges contaminated by the planetary atmosphere (Sect.~\ref{sec:atm_mask}) and planet-occulted lines (Sect.~\ref{apn:line_fit}). The masked disk-integrated stellar lines are then not distorted by the planet anymore, and can be fitted with the same model as for the unocculted star. This allows deriving stellar properties homogeneously over the entire epoch, which can be of particular interest to study stellar variability at short timescales (e.g., pulsations, \citealt{Wyttenbach2020}). This approach is only applicable with disk-integrated lines broad enough to remain defined despite the masking. We illustrate this with the $rv$ derived from WASP-76 CCFs, masked for planetary absorption lines (Fig.~\ref{fig:DI_WASP76_fit}). The orbital motion of the planet combined with the width of its absorption signature result in many in-transit profiles being too contaminated to be accurately fitted (Fig.~\ref{fig:Intr_CCF_map_WASP76}).

%%%%%%%%%%%%%%%%%%%%%%%%%%%%%%%%%%%%%%%%%%%%%%%%%%%%%%%%%%%%%%%%%%%%%%%%%%%%%

\subsubsection{Intrinsic stellar lines}
\label{sec:ana_Intr}

\begin{figure}
\includegraphics[trim=0cm 0cm 0cm 0cm,clip=true,width=\columnwidth]{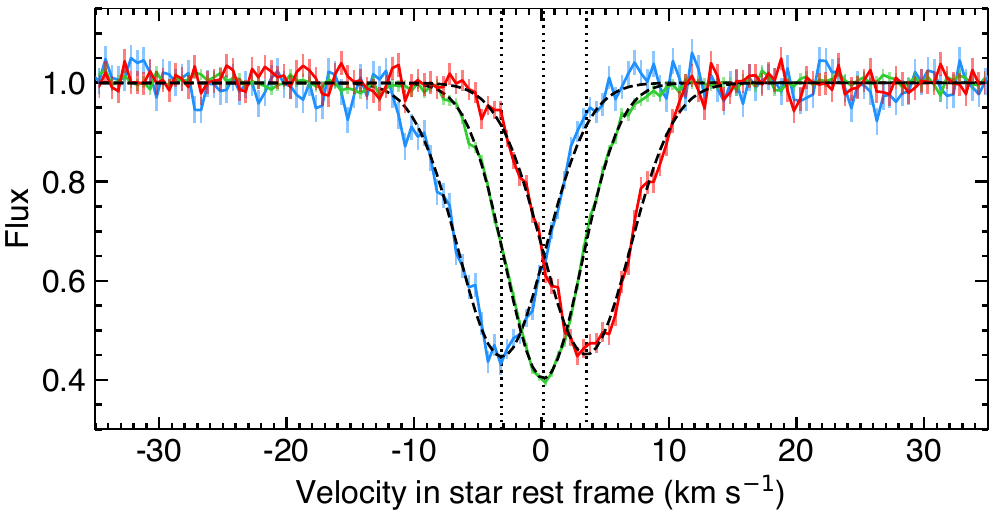}
\centering
\caption[]{Average line profile of HD\,209458 in the white CCFs calculated with ESPRESSO F9 mask (epoch 2). Exposures at orbital phases -0.014 (ingress, blue profile), 0 (conjunction, green profile), and 0.014 (egress, red profile) illustrate the change in line shape between stellar limb and center. Because of the aligned orbit, profiles occulted at ingress and egress have opposite $rv$ and similar shapes. Dashed black lines show the Gaussian profiles used to derive the line properties (Fig.~\ref{fig:Intr_prop_HD209_W76}).}
\label{fig:Intr_prof_HD209}
\end{figure}

Intrinsic stellar lines are measured at lower S/N than disk-integrated lines and have simpler shapes, especially when averaged over several electronic transitions, such that a Gaussian profile typically reproduces them well (Fig.~\ref{fig:Intr_prof_HD209}). For planets that are large or occult broad areas during an exposure, compared to the spatial scale of line profile and $rv$ field variations over the photosphere, intrinsic profiles should however be fitted with numerical models (Sect.~\ref{sec:st_grid}). 

By construction (Sect.~\ref{sec:intr_prof}), ranges contaminated by the planetary atmosphere are excluded from intrinsic profiles (Fig.~\ref{fig:Intr_CCF_map_WASP76}). Our fitting method, based on a Bayesian approach (Sect.~\ref{apn:line_fit}), allows retrieving information on the local stellar line even when it is partially masked (Fig.~\ref{fig:Intr_prop_HD209_W76}) or hidden in the noise (typically at the stellar limbs), and assessing which exposures can be exploited in time
series  analysis (Sect.~\ref{sec:fit_series}). For both WASP-76 and HD\,209458 only the very first and last exposures at the limbs had to be excluded, plus those exposures of WASP-76 in which the intrinsic line is fully contamined by the planet (Fig.~\ref{fig:Intr_CCF_map_WASP76}).

Series of properties derived from individual intrinsic profiles can be analyzed with a dedicated method (Sect.~\ref{sec:lprof_var}) to map the local stellar photosphere. The line centroids trace the surface $rv$ field, controlled by the stellar rotational velocity, convective blueshift, and differential rotation. Morphological properties, such as width and contrast, trace center-to-limb variations in their spectral profile. Average properties of local stellar lines, or a species abundance, can further be derived by fitting a master of all intrinsic profiles. Comparing the width of master intrinsic and disk-integrated CCFs, or fitting the latter using the stellar grid tiled with an intrinsic profile series, provides complementary constraints on stellar rotation. Since intrinsic lines may vary in shape along the transit chord, it is nonetheless preferred to fit them jointly with a global model (Sect.~\ref{sec:fit_series}).

%%%%%%%%%%%%%%%%%%%%%%%%%%%%%%%%%%%%%%%%%%%%%%%%%%%%%%%%%%%%%%%%%%%%%%%%%%%%%

\subsubsection{Line profile variations}
\label{sec:lprof_var}

Once properties have been derived from stellar lines in individual exposures, they can be studied as a series to characterize their variations. The method proposed here allows quantifying anomalous trends in disk-integrated properties to inform line detrending (Sect.~\ref{sec:st_line_detrend}), defining the models best describing intrinsic properties along the transit chord (Sect.~\ref{sec:fit_series}), and identifying deviations to the series from noise-related outliers or activity-related features (spots, pulsations, etc). We detail in Sect.~\ref{apn:lprof_var} the various models that can be fitted to line property series, as a function of ambient, data-related, or stellar variables for disk-integrated properties, or as a function of stellar coordinates for intrinsic properties.

\begin{figure*}
\begin{minipage}[tbh!]{\textwidth}
\includegraphics[trim=0cm 0cm 0cm 0cm,clip=true,width=\columnwidth]{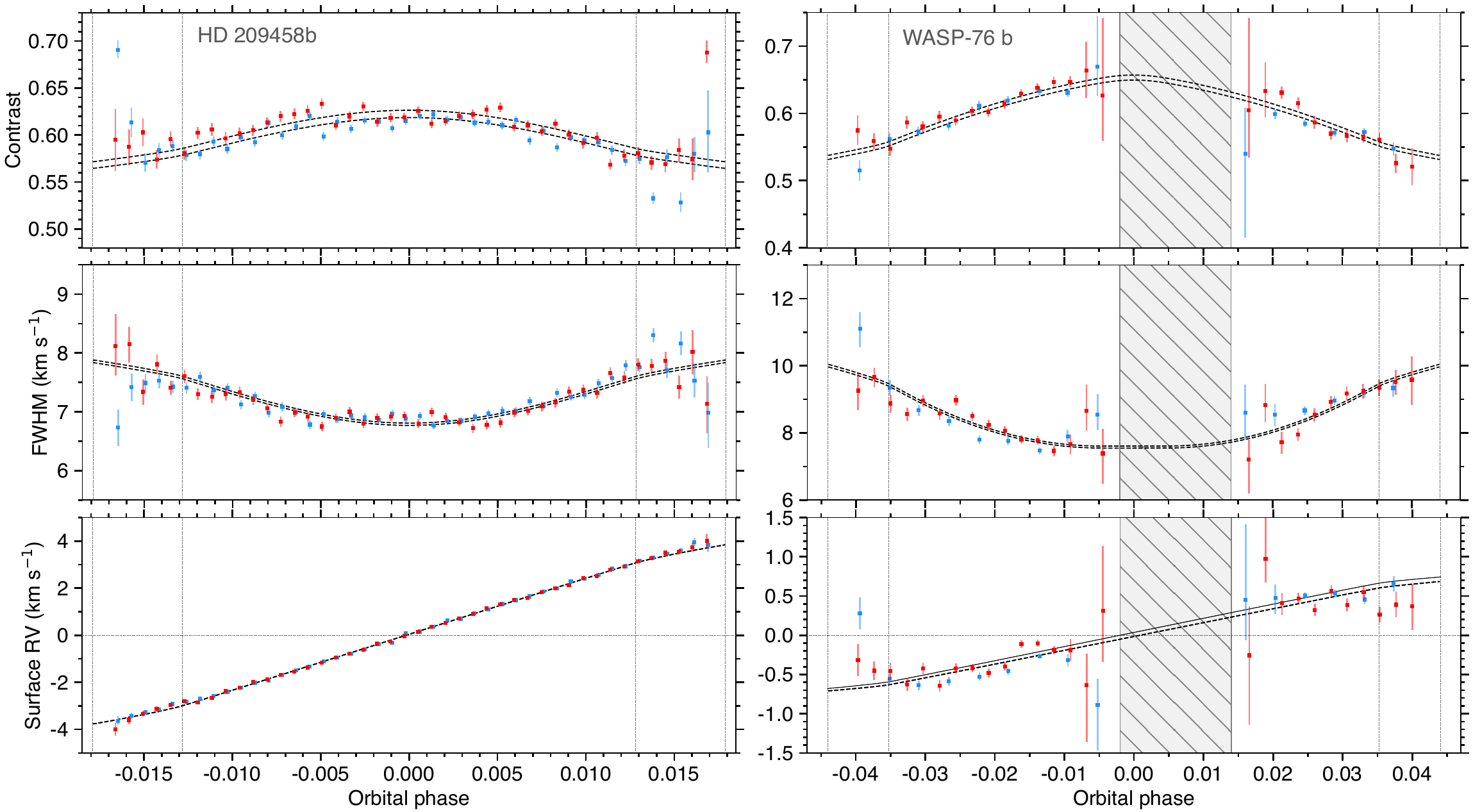}
\centering
\end{minipage}
\caption[]{Properties of stellar regions occulted by HD\,209458b and WASP-76b. Dotted vertical lines show transit contacts. Measured values (first epochs in blue, second epochs in red) are derived from best-fit profiles to each intrinsic CCF, before convolution with ESPRESSO LSF (Sect.~\ref{sec:ana_Intr}). They are thus comparable to the models (dashed black curves) derived from fits to intrinsic CCF series (Sect.~\ref{sec:RMR}). The black curve in WASP-76 surface $rv$ is the model from \citealt{Ehrenreich2020}. The two outliers in contrast at the limbs of HD\,209458 are consistent with the FWHM and $rv$ series, and thus kept in the analysis. WASP-76 local stellar lines are fully contaminated by the planet over the shaded range. }
\label{fig:Intr_prop_HD209_W76}
\end{figure*}

\paragraph{White-light analysis:} Fig.~\ref{fig:Intr_prop_HD209_W76} shows the intrinsic line contrast, FWHM, and $rv$ in white CCFs from the regions occulted by the two planets. The surface $rv$ of HD\,209458 are consistent with those from \citet{CasasayasBarris2021}, but more similar between the two epochs at the limbs. The surface $rv$ of WASP-76 are overall consistent with those from \citet{Ehrenreich2020}, but differ sufficiently to impact the global fits to intrinsic profiles (see the next section). For both systems we found no evidence for convective blueshift. While there is a hint of differential rotation for HD\,209458, it is driven by measurements at the limb that are uncertain due to imprecise transit timing and impact parameter (\citealt{CasasayasBarris2021}). Hereafter, we thus assume that HD\,209458 and WASP-76 rotate as solid bodies. Their line shape properties show clear variations along the transit chords, well described by polynomials of the distance to star center (linear for HD\,209458, quadratic for WASP-76). The contrast of the line decreases toward the limbs as we probe cooler stellar layers while the presence of convective cells, which scan different velocities across the different depths of the layers, is likely the reason for the increase in line FWHM toward the limbs. We note that WASP-76 surface properties can be retrieved in more exposures than disk-integrated properties (Fig.~\ref{fig:DI_WASP76_fit}), because intrinsic lines are not broadened by stellar rotation and overlap less with planet-contaminated ranges. Nonetheless, some intrinsic profiles that are partially masked cannot be accurately fitted in individual exposures, highlighting the interest for a joint fit to the profile series. One of the aims of analyzing line profile variations is to initialize this joint fit, from which are derived final properties for the stellar photosphere and planet architecture.

\begin{figure}
\includegraphics[trim=0cm 0cm 0cm 0cm,clip=true,width=\columnwidth]{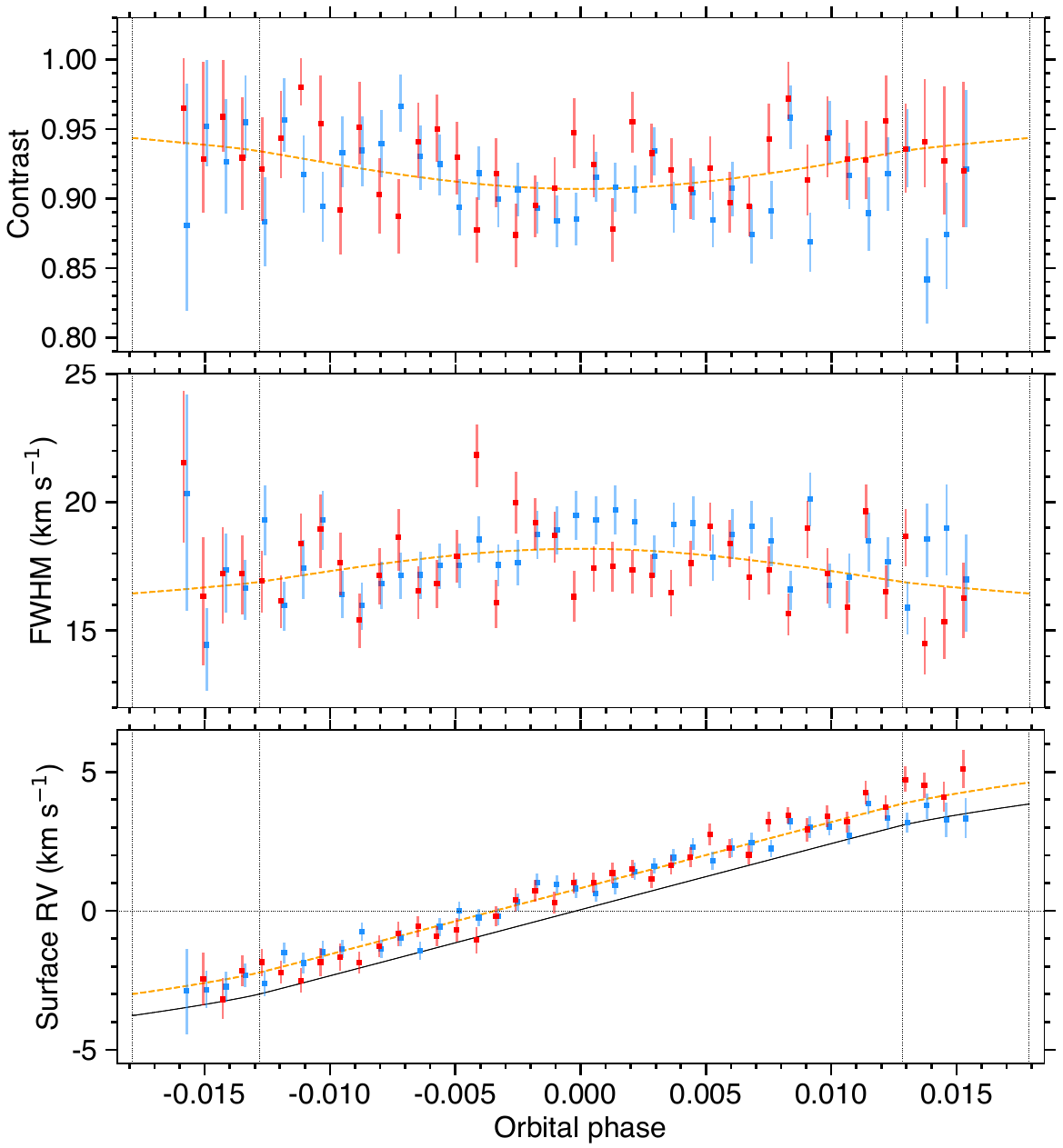}
\centering
\caption[]{Sodium doublet properties of stellar regions occulted by HD\,209458b (same format as Fig.~\ref{fig:Intr_prop_HD209_W76}). A common model to both epochs (dashed orange curves) is preferred for all properties. The photospheric $rv$ model is shown as a solid black curve
}
\label{fig:Intr_prop_HD209_Na}
\end{figure}

\paragraph{Spectral analysis:} Fig.~\ref{fig:Intr_prop_HD209_Na} shows the contrast, FWHM, and $rv$ of the intrinsic sodium doublet from HD\,209458. Gaussian profiles were fitted to CCFs of the doublet, using a mask with equal weights to match the similar line depths (Fig.~\ref{fig:Intr_prof_HD209_Na_RMR}). The sodium line width and contrast are larger than for iron lines in white CCFs (Fig.~\ref{fig:Intr_prop_HD209_W76}), and vary oppositely with distance to the star center (they get narrower and deeper toward the stellar limb). Furthermore, while the sodium line $rv$ follow the same trend as the average photosphere they display a significant redshift common to both epochs.  
The same analysis performed on deep and narrow stellar lines in the vicinity of the sodium lines shows $rv$ series consistent with the photospheric model, demonstrating that this redshift traces processes behind the formation of the doublet. 

%%%%%%%%%%%%%%%%%%%%%%%%%%%%%%%%%%%%%%%%%%%%%%%%%%%%%%%%%%%%%%%%%%%%%%%%%%%%%%%%%%%

\subsection{Analysis of joint intrinsic stellar profiles}
\label{sec:fit_series}

This method builds on the RMR technique (\citealt{Bourrier2021}), whose philosophy is to fit jointly series of intrinsic profiles (Sect.~\ref{apn:glob_mod}) rather than fitting individual profiles and their derived properties. The \textsc{antaress} workflow offers two major updates: model intrinsic profiles are calculated with higher precision (Sect.~\ref{sec:pl_occ_grid}) and can be fitted to both CCF and spectral profiles. The latter point allows constraining first the overall properties of the stellar surface (Sect.~\ref{sec:RMR}), and then investigating how a given spectral line compares with them (Sect.~\ref{sec:RMR_spec}).

\subsubsection{White-light analysis (RM revolutions)}
\label{sec:RMR}

Rather than applying the RMR to profiles derived from input disk-integrated CCFs (e.g., \citealt{Bourrier2021,Bourrier2022}), the \textsc{antaress} workflow exploits CCFs calculated from extracted intrinsic spectra (Sect.~\ref{sec:CCF_intr}), which trace directly the properties of planet-occulted regions. White intrinsic CCFs, calculated at high S/N with a mask of quiet and well-defined stellar lines (Sect.~\ref{apn:CCF_masks}), can constrain the architecture of the system (sky-projected $\lambda$ or 3D $\psi$ spin-orbit angle, stellar inclination $i_\star$, scaled semi-major axis $a_p/R_{\star}$, orbital inclination $i_p$), the global velocity field of the photosphere (stellar rotational velocity, $v_{\rm eq} \sin i_\star$, differential rotation, convective blueshift), and the average shape of its absorption lines (contrast, FWHM, etc). We detail in Sect.~\ref{apn:RMR} how and when these various properties can be constrained. 

\begin{figure*}
\begin{minipage}[tbh!]{\textwidth}
\includegraphics[trim=0cm 0cm 0cm 0cm,clip=true,width=\columnwidth]{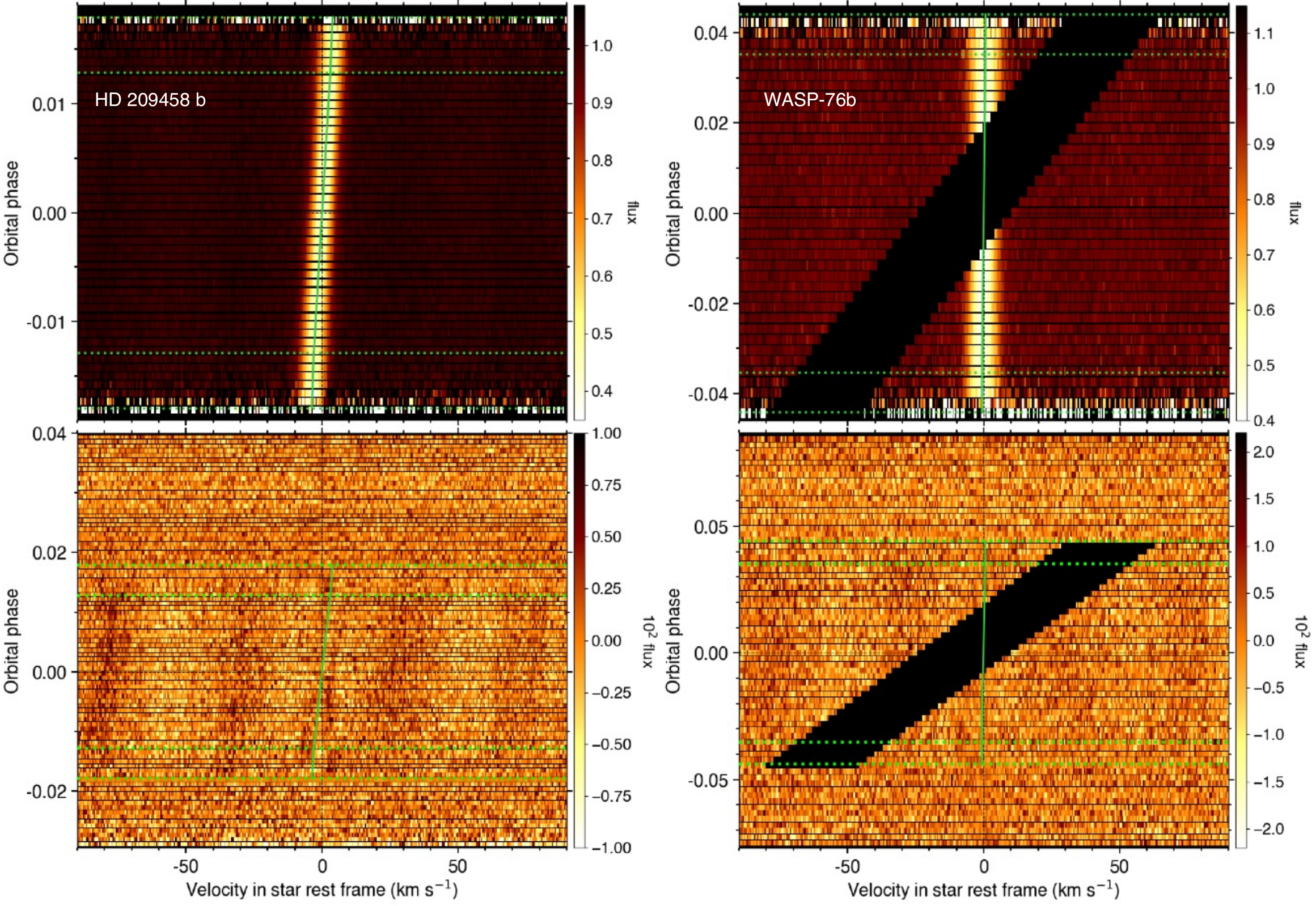}
\centering
\end{minipage}
\caption[]{RMR fits of HD\,209458b (shown for epoch 1) and WASP-76b (shown for epoch 2). Maps are plotted as a function of $rv$ in the star rest frame (in abscissa) and orbital phase (in ordinate). Green dotted lines show transit contacts. The masked streak in WASP-76 maps is contaminated by the planetary atmosphere. \textit{Top panels:} CCFs of intrinsic spectra, colored with flux. Green solid lines shows the best stellar surface $rv$ models. \textit{Bottom panels:} Residuals over the full epochs. Out-of-transit values show the CCFs of residual spectra, and in-transit values the difference between the CCFs of intrinsic spectra and their best-fit models (scaled to the out-of-transit level). The slanted in-transit features aligned with the surface $rv$ model of HD\,209458 may arise from stellar lines close to the CCF mask lines.}
\label{fig:RMR_white}
\end{figure*}

Here, we apply a RMR fit to the joined epochs of HD\,209458 and WASP-76, using intrinsic CCFs calculated with ESPRESSO F9 mask. Based on the analysis in Sect.\ref{sec:lprof_var} we assume solid-body rotation for the stars, so that $\lambda$ and  $v_{\rm eq} \sin i_\star$ are the only free parameters of the surface $rv$ model. We further define the contrast and FWHM of the theoretical line profiles as polynomials of distance to star center, modulated by coefficients specific to each epoch. Spatial oversampling is applied to the model exposures (Sect.~\ref{sec:pl_occ_grid}), although the blurring induced by the planets' motion and the photospheric areas they occult is negligible. Results from the best-fits, which yield reduced $\chi^2$ = 1.1, are highlighted in Fig.~\ref{fig:RMR_white}. The stability of the stars, the quality of the processed data, and the good agreement between CCFs and their best-fit is further visible through the homogeneity and low dispersion of their residuals.

\paragraph{HD\,209458 :} The best fit yields $\lambda$ = $1.055\stackrel{+0.073}{_{-0.077}}^{\circ}$ and $v_\mathrm{eq} \sin i_\star$ = 4.272$\pm$0.007\,km\,s$^{-1}$ with a similar precision as \citet{CasasayasBarris2021} ($\lambda$ = $1.58\pm0.08^{\circ}$ and $v_\mathrm{eq} \sin i_\star$ = 4.228$\pm0.007$\,km\,s$^{-1}$), who used the same datasets and system properties but applied a reloaded RM analysis (\citealt{Cegla2016}) to input disk-integrated CCFs. Our results are likely consistent, considering that more realistic error bars on $\lambda$ and $v_\mathrm{eq} \sin i_\star$ (accounting for the uncertainty on the transit timing and impact parameter) are on the order of 1$^{\circ}$ and 100\,m\,s$^{-1}$, respectively (\citealt{CasasayasBarris2021}).

%----------------------------

\begin{figure}
\includegraphics[trim=0cm 2.5cm 0cm 0.5cm,clip=true,width=\columnwidth]{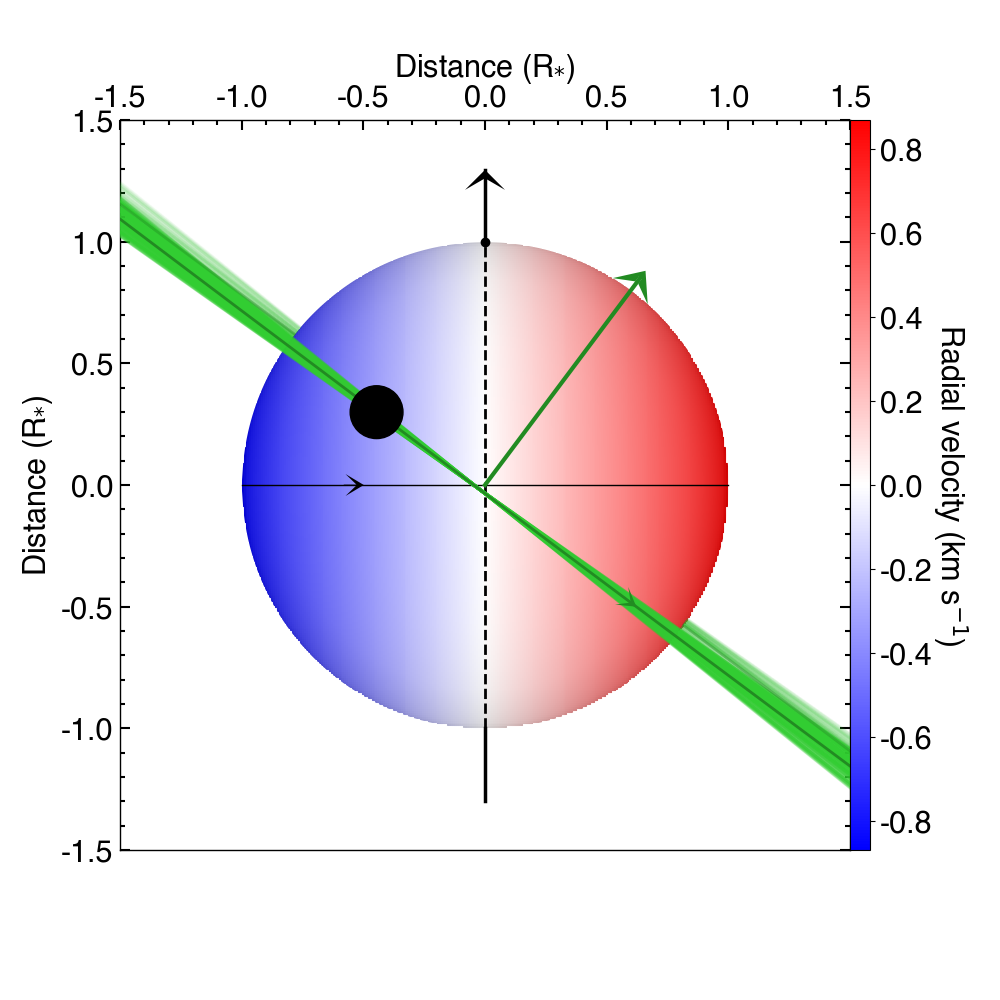}
\centering
\caption[]{Sky-projected view of WASP-76 for the best-fit orbital architecture (see plot details in Fig.~\ref{fig:System_view}). Thin lines surrounding the orbital trajectory (thick curve) show orbits obtained for orbital inclination, semi-major axis, and sky-projected obliquity values drawn randomly within 1$\sigma$ from their probability distributions.}
\label{fig:WASP76_system}
\end{figure}

\paragraph{WASP-76 :} The configuration of the system, with an impact parameter close and consistent with 0 (\citealt{Ehrenreich2020}), is such that the planet-occulted surface $rv$ and their derivative approximate during transit (when orbital phase $\phi\sim$0, see Eq.~\ref{eq:rv_surf}):
\begin{equation}
\begin{split}
rv_\mathrm{surf/\star}(\phi) &= v_{\rm eq}\sin i_{\star} \, a_{p}/R_{\star} \, \cos(\lambda)\,2\,\pi\,\phi \\ 
\dot{rv}_\mathrm{surf/\star}(\phi) &=v_{\rm eq}\sin i_{\star} \, a_{p}/R_{\star} \, \cos(\lambda)\,2\,\pi
\end{split}
\end{equation}
Thus, $v_{\rm eq}\sin i_{\star} \cos(\lambda)$ is constrained in the same way by the slope and level of the $rv$ series and there is a strong degeneracy between $\lambda$ and $v_\mathrm{eq} \sin i_\star$ (Fig.~\ref{fig:Corr_diag_WASP76}). This is in part why we derive ($\lambda$ = -37.1$\stackrel{+12.6}{_{-20.6}}^{\circ}$ ; $v_\mathrm{eq} \sin i_\star $ = 0.86$\stackrel{+0.13}{_{-0.19}}$\,km\,s$^{-1}$) while \citet{Ehrenreich2020} derive ($\lambda$ = 61$\pm$7$^{\circ}$ ; $v_\mathrm{eq} \sin i_\star$ = 1.5$\pm$0.3\,km\,s$^{-1}$), since both yield $v_{\rm eq}\sin i_{\star} \cos(\lambda) \sim$ 0.71\,km\,s$^{-1}$. Furthermore our derived $rv$ series is slightly offset compared to \citet{Ehrenreich2020} (Fig.~\ref{fig:Intr_prop_HD209_W76}), crossing 0\,km\,s$^{-1}$ after mid-transit rather than before, which requires the planet to transit the blueshifted stellar hemisphere longer ($\lambda$ in between -180 and 0$^{\circ}$, Fig.~\ref{fig:WASP76_system}) than the redshifted one ($\lambda$ in between 0 and 180$^{\circ}$). These differences in $rv$ series likely arise from our use of CCFs calculated from intrinsic spectra, rather \citet{Ehrenreich2020}'s use of intrinsic CCFs derived from disk-integrated CCFs (Sect.~\ref{apn:perf_intr_DI}). Finally, in an attempt to break the degeneracy, \citet{Ehrenreich2020} set a strong prior on $v_\mathrm{eq} \sin i_\star$ (1.61$\pm$0.28\,km\,s$^{-1}$) derived as the quadratic difference between the FWHM of the stellar disk-integrated and planet-occulted master CCFs. Combining the average FWHM of our planet-occulted lines with the width of ESPRESSO line spread function (LSF), we derive a similar $v_\mathrm{eq} \sin i_\star$ = 1.75\,km\,s$^{-1}$. However, our in-depth analysis show that the FWHM of WASP-76 intrinsic lines vary by at least 2\,km\,s$^{-1}$ between the center and limbs of the star (Fig.~\ref{fig:Intr_prop_HD209_W76}). As a result, the rotational velocity estimated by \citet{Ehrenreich2020} from the average planet-occulted line, and in turn their spin-orbit angle, was biased. 

To try and mitigate the degeneracy between $\lambda$ and $v_\mathrm{eq} \sin i_\star$ we applied a RMR fit with $a_p/R_{\star}$ and $i_p$ free to vary, using photometry priors from \citet{Ehrenreich2020} ($a_p/R_{\star}$ = 4.08$\stackrel{+0.02}{_{-0.06}}$, $i_p$ = 89.623$\stackrel{+0.005}{_{-0.034}}^{\circ}$). We derived values consistent with the priors ($a_p/R_{\star}$ = 4.059$\stackrel{+0.044}{_{-0.033}}$ ; $i_p$ = 89.604$\stackrel{+0.024}{_{-0.015}}^{\circ}$), showing that the intrinsic CCFs alone cannot constrain these orbital properties. Refining the orbital architecture of the WASP-76 system thus requires both additional transit spectroscopy and photometry. We note that our results change neither the conclusion from \citet{Ehrenreich2020} that WASP-76b is on a misaligned orbit, nor their analysis of the planet atmospheric signal since their surface $rv$ model, used to mask the planet-occulted stellar lines, remain within 100\,m\,s$^{-1}$ from our best-fit.

%%%%%%%%%%%%%%%%%%%%%%%%%%%%%%%%%%%%%%%%%%%%%%%%%%%%%%%%%%%%%%%%%%%%%%%%%%%%%%%%%%%

\subsubsection{Spectral analysis}
\label{sec:RMR_spec}

Analyzing intrinsic line series in specific transitions allows measuring properties of the corresponding species and its formation layer. We illustrate this approach with 1D intrinsic spectra of HD\,209458 in the sodium doublet (Sect.~\ref{sec:2D_1D}). Numerical profiles were calculated like the white RMR fit (Sect.~\ref{sec:RMR}) but using theoretical intrinsic spectra, calculated with the disk-integrated properties and a free sodium abundance (Sect.~\ref{sec:ana_indiv_star_spec}). These theoretical spectra were Doppler-shifted according to the derived photospheric $rv$ model (Sect.~\ref{sec:RMR}), allowing for a free offset as hinted in Sect.~\ref{sec:lprof_var}.   

\begin{figure}[tbh!]
\includegraphics[trim=0cm 0cm 0cm 0cm,clip=true,width=\columnwidth]{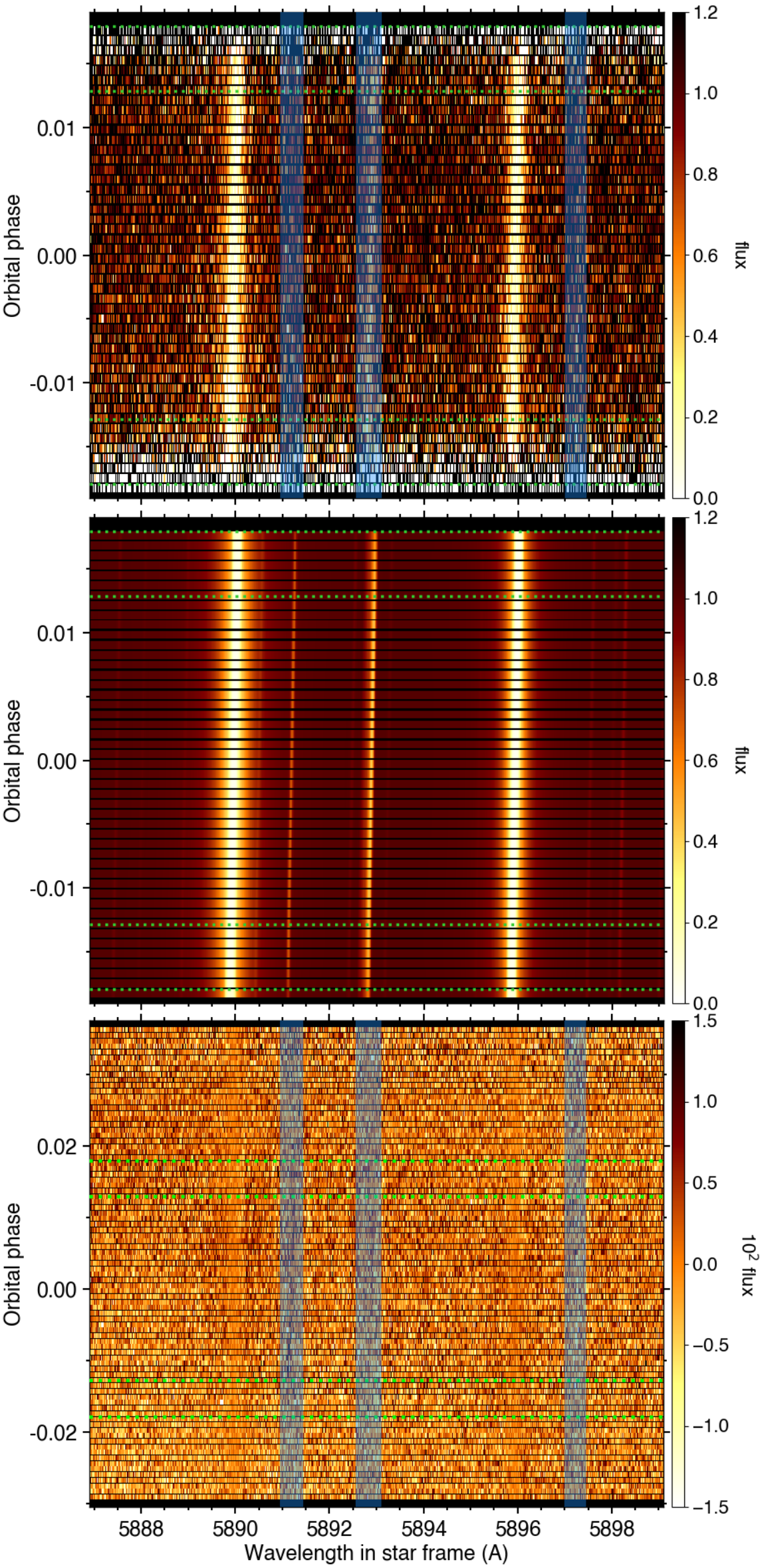}
\centering
\caption[]{Joint fit to HD\,209458 intrinsic sodium doublet, shown for epoch 2 as a function of wavelength in the star rest frame (in abscissa) and orbital phase (in ordinate). Green dotted lines are transit contacts. \textit{Top panel:} Intrinsic spectra, colored with flux. Stellar lines kept to solar abundance were excluded from the fit (shaded regions). \textit{Middle panel:} Best-fit theoretical model. \textit{Bottom panels:} Residuals over the full epoch, calculated as in Fig.~\ref{fig:RMR_white}.}
\label{fig:Intr_map_HD209_Na_RMR}
\end{figure}

\begin{figure}[tbh!]
\includegraphics[trim=0cm 0cm 0cm 0cm,clip=true,width=\columnwidth]{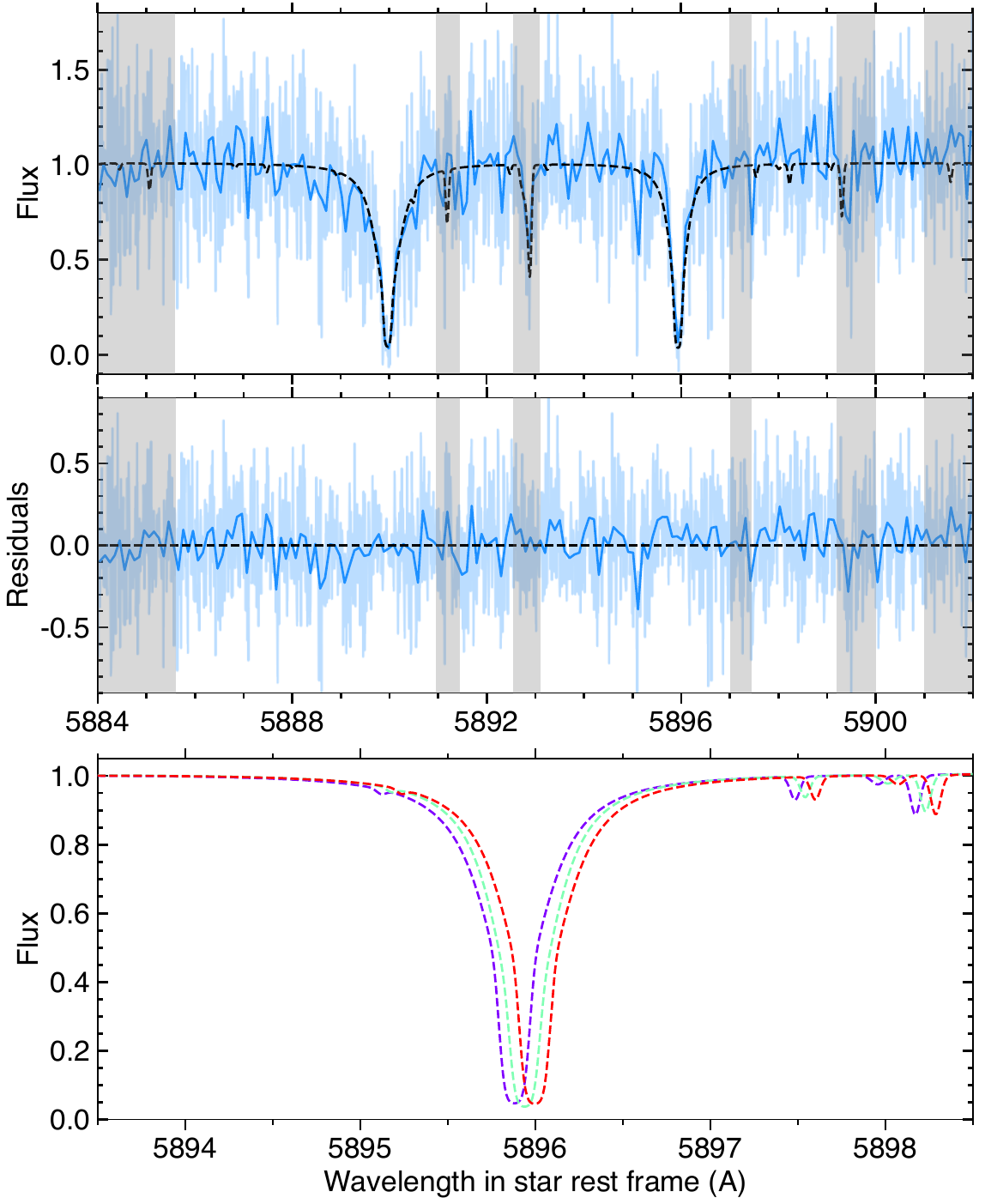}
\centering
\caption[]{Intrinsic sodium doublet of HD\,209458, for the joint fit to both epochs. \textit{Top panel:} Measured (blue) and model (black) profiles at mid-transit in epoch 2. Original data (semi-transparent spectrum) was downsampled (solid spectrum) for clarity. Shaded regions were excluded from the fit. \textit{Middle panel:} Residual spectrum. \textit{Bottom panel:} Zoom on the D1 line, illustrating center-to-limb variations from the NLTE pySME/MARCS model (profiles are colored in purple, green, and red at orbital phases -0.0126, -0.0002, and 0.0123).}
\label{fig:Intr_prof_HD209_Na_RMR}
\end{figure}

The best fit (Fig.~\ref{fig:Intr_map_HD209_Na_RMR}) is obtained for an abundance of 6.056$\pm$0.007 and a Doppler offset of 849$\pm$27\,m\,s$^{-1}$, consistent with the values derived from disk-integrated lines. This suggests that the sodium line profiles retrieved along the transit chord of HD\,209458b are representative of the photosphere in the observed epochs, with the star displaying few active/spotted regions in this spectral band. While the NLTE pySME/MARCS stellar model matches well the sodium line shape it does not account for a Doppler offset like the one we measured. Furthermore, the theoretical model reproduces the decrease in line width observed from center to limb (Sect.~\ref{sec:lprof_var}) but simulates a decrease in contrast opposite to the measured variation (Fig.~\ref{fig:Intr_prof_HD209_Na_RMR}). These discrepancies emphasize the interest of confronting stellar models (e.g., \citealt{Plez2012,Wehrhahn2023}) to measurements of local stellar lines, as provided by the \textsc{antaress} workflow. We also note that our sodium abundance for HD\,209458 is lower than the solar value ($\sim$6.17) used in theoretical transmission spectra accounting for stellar contamination (\citealt{CasasayasBarris2021,Dethier2023}). The extraction and characterization of planet-occulted stellar lines is thus also of strong interest to the interpretation of transits.

%%%%%%%%%%%%%%%%%%%%%%%%%%%%%%%%%%%%%%%%%%%%%%%%%%%%%%%%%%%%%%%%%%%%%%%%%%%%%

\section{Performance analysis}
\label{sec:perf}

We investigated the impact of various methods in the \textsc{antaress} workflow on the quality of the outputs, in particular intrinsic CCFs and their RMR fits. We summarize here the results of this investigation, detailed in Sect.~\ref{apn:perf}:
\begin{itemize}%[wide=0pt] %,label={},  itemindent = 0pt]
\item Resampling: we propose the first workflow that resamples spectra on a common grid as late as possible, reducing blurring and correlated noise to a minimum. Resampling has negligible impact on methods that modify the low-frequency stellar spectral profile, such as the color effect or wiggles, but it smears sharp spectral features at various steps of the workflow. While variations in stellar line properties and RMR results remain marginal for the studied datasets, they would be stronger and more significant for stars with larger Keplerian motion, narrower or asymmetrical stellar lines, and higher S/N spectra. This makes our approach particularly relevant for next-generation spectrographs on the ELT. 
\item Covariance: the optimal processing starts introducing correlations between pixels when extracting residual profiles, in which case the maximum covariance in cross-correlated intrinsic stellar lines remains small enough that it does not impact their analysis. Even if profiles are resampled earlier in the workflow, covariances are propagated efficiently as banded matrices and accounted for in fits, which is a major novelty of our approach. This allows oversampling CCFs and better resolving the stellar line profiles, which may be of interest with lower instrumental resolution than ESPRESSO and when lines are asymmetrical.
\item Weights: neglecting the contributions of instrumental flux calibration and telluric lines (for disk-integrated spectra), or disk-integrated stellar lines (for intrinsic spectra), to the true error estimated in a given pixel can bias significantly its weighted spatial/temporal mean.
\item Flux balance: a spectrum-wide correction of the color effect is critical for broadband spectral analysis and the computation of unbiased stellar CCFs. Capturing smaller-scale flux balance variations, while it may further improve the shape of CCFs, is however not critical for $rv$ derivation and the interpretation of the RM effect.
\item Wiggles: a global, homogeneous correction is necessary to remove biases when analyzing ESPRESSO spectra. However, wiggles appear to smooth out when computing stellar CCFs. While their correction may still improve the shape of the CCFs, it is not critical for $rv$ derivation and the interpretation of the RM effect.
\item Intrinsic CCFs: our workflow offers the possibility to keep data in spectral format as long as desired, so that CCFs can be directly calculated from intrinsic spectra rather than derived from disk-integrated CCFs. We recommend using the former, as the latter distorts the shape of intrinsic stellar lines and may bias the interpretation of the RM effect.
\item Chromatic scaling: accounting for broadband variations in planetary absorption and stellar emission when rescaling stellar spectra may remove biases from their CCF profiles and the interpretation of the RM effect. We recommend using chromatic scaling with high-S/N data and planets with strong broadband atmospheric features.
\item CCF masks: generating custom masks from the processed data allows building CCFs of comparable or higher quality than standard masks, and performing self-consistent RM analysis.  
\end{itemize}

%%%%%%%%%%%%%%%%%%%%%%%%%%%%%%%%%%%%%%%%%%%%%%%%%%%%%%%%%%%%%%%%%%%%%%%%%%%%%

\section{Summary and conclusions}
\label{sec:conclu}

The \textsc{antaress} workflow consists of a set of rigorous methods designed to process high-resolution transit datasets homogeneously, ensuring a reproducible extraction and analysis of stellar and planetary spectra. Particular care has been taken to remain as close as possible to the original data, by reverting to raw flux units when relevant, propagating covariances, and avoiding spectral resampling unless necessary. 

We introduce a set of methods meant to correct echelle spectra obtained from the ground for environmental and instrumental effects, which are mainly related to telluric absorption and flux balance variations induced by Earth's atmosphere, cosmic rays, and optical interferences. A first-order correction for these effects is sufficient to analyze stellar line $rv$ and perform RM analysis, but a finer correction reduces biases in spectral profiles and increases CCF quality. We then describe our methods for correcting trends in disk-integrated stellar lines, so that they are comparable and can be processed similarly by aligning disk-integrated spectra in the star rest frame, rescaling them to their relative chromatic flux level, and extracting planet-occulted spectra.\\

The workflow can be used to analyze stellar lines, in which case the planet is considered as a contaminant. Planet-occulted spectra are corrected for flux and $rv$ modulations induced by the planetary continuum and then they are masked for planetary atmospheric lines. The resulting series of intrinsic spectra directly trace the properties of the photosphere along the transit chord, without any residual contamination by the planet. We highlight that the flux contrast between planet-occulted regions and the full stellar disk is better at the bottom of disk-integrated stellar lines, making slow-rotating stars with deep lines optimal cases to measure intrinsic stellar lines at high precision. We propose several methods to analyze disk-integrated and intrinsic stellar lines, using simple analytical profiles or a numerical model of the photosphere. This model, which is used as part of the data processing and to generate mock datasets, is generic enough that the workflow can be applied to any type of stars, from cool and slow-rotating dwarfs to oblate and gravity-darkened giants. 

The analysis of intrinsic stellar profiles is first performed on individual exposures, to assess their variations across the stellar surface. Then, the profile series is fitted with a joint model informed by the first step, to derive properties at higher precision. This two-step RMR analysis can be applied to ``white'' CCFs computed over the full spectra, to constrain the $rv$ field of the photosphere and the system orbital architecture. Once these global properties have been derived, they can be fixed and the analysis can then be applied to individual spectral lines, or even to CCFs over a subset of lines, to derive properties of a specific species or stellar atmospheric layer. We emphasize that the workflow allows us to cross-correlate intrinsic spectra directly, avoiding biases associated with the processing of disk-integrated CCFs. 

To illustrate the use of the workflow, we analyzed archival ESPRESSO datasets from the iconic stars HD\,209458 and WASP-76. A white-line analysis of their intrinsic spectra revealed average stellar lines that get shallower and broader toward the limbs. We confirmed the published values for the sky-projected spin-orbit angle and stellar rotational velocity of HD\,209458 (\citealt{CasasayasBarris2021}) and unveiled biases in the derivation of these properties for WASP-76 by \citet{Ehrenreich2020}. While our revision does not change the planet atmospheric signature derived by the latter authors, it highlights the need for a robust processing of transit spectroscopy datasets. We then carried out a spectral analysis of HD\,209458's sodium doublet, detecting center-to-limb variations of the intrinsic lines opposite to those of the average photospheric lines. The intrinsic sodium abundance is consistent with the disk-integrated one, but lower than the solar abundance used in previous studies of the planetary atmosphere. Finally, we found divergences in shape and position compared to theoretical spectra, highlighting the interest of the workflow to confront stellar atmospheric models with spatially resolved stellar spectra.\\

The \textsc{antaress} methods are generic and can be used outside of the workflow. For example, in this work we show how a proper calculation of weighted means between different spectra is critical to avoid biases. Combined with our method to convert 2D spectra into 1D spectra and CCFs, it allows for the calculation of accurate master profiles of the disk-integrated and local photosphere. Such profiles can be used for comparison with stellar models or to generate binary CCF masks specific to a given star. The workflow is modular, so that new methods can be added in a straightforward way. Because the concepts behind the processing and analysis methods do not depend on a given instrument, their use can be extended to other ground-based or space-borne, high-resolution spectrographs as long as the relevant correction and formatting methods are added. In this work, we present the application of  the \textsc{antaress} workflow to planet-occulted stellar spectra. In a companion paper, we will describe the subsequent methods aimed at extracting and analyzing series of atmospheric spectra uncontaminated by the star to sample  the planetary limbs spatially and
dynamically.

%%%%%%%%%%%%%%%%%%%%%%%%%%%%%%%%%%%%%%%%%%%%%%%%%%%%%%%%%%%%%%%%%%%%%%%%%%%%%%%%%%%%%%%%%%%%%%%%%%%%%%%%%%%%%%%%%%%%%%%%%%

\begin{acknowledgements}
We highlight our appreciation of the referee's work, considering the depth in which they read this comprehensive manuscript and the attention they gave to understanding its contents.

We thank G. Fu and N. Nikolov for providing the low-resolution transmission spectrum of WASP-76b, K. Lind and A. Amarsi for their help with stellar NLTE grids for sodium, J. Hoeijmakers for discussions about transmission spectroscopy, A. Deline for inspiring discussions about wiggles, D. Luc for comments about coding, Y. Zhao for his help with GSL.

This work has been carried out within the framework of the NCCR PlanetS supported by the Swiss National Science Foundation under grants 51NF40$\_$182901 and 51NF40$\_$205606. This project has received funding from the European Research Council (ERC) under the European Union's Horizon 2020 research and innovation programme (project {\sc Spice Dune}, grant agreement No 947634; project {\sc SCORE}, grant agreement No 851555). X.D acknowledges the support from the Swiss National Science Foundation under the grant SPECTRE (No 200021$\_$215200). R.A. is a Trottier Postdoctoral Fellow and acknowledges support from the Trottier Family Foundation. This work was supported in part through a grant from the Fonds de Recherche du Québec - Nature et Technologies (FRQNT). This work was funded by the Institut Trottier de Recherche sur les Exoplanètes (iREx). H.M.C acknowledges funding from a UKRI Future Leader Fellowship, grant number MR/S035214/1.

This work has made use of the VALD database, operated at Uppsala University, the Institute of Astronomy RAS in Moscow, and the University of Vienna.

\textsc{antaress} makes use of the following packages: \textsc{Astropy}, \citet{Astropy2022}; \textsc{batman}, \citet{Kreidberg2015}; \textsc{emcee}, \citet{Foreman2013};\textsc{lmfit}, \citet{Newville2016};\textsc{Matplotlib}, \citet{Hunter2007};
\textsc{numpy}, \citet{Harris2020}; \textsc{pandas}, \citet{reback2020};
\textsc{pathos}, \citet{McKerns2012}; \textsc{PyAstronomy}, \citet{Czesla2019}; \textsc{pySME}, \citet{Wehrhahn2023}; \textsc{SciPy}, \citet{Virtanen2020}.

\end{acknowledgements}

\bibliographystyle{aa} % style aa.bst
\bibliography{biblio} % your references Yourfile.bib

\begin{appendix}

\section{Resampling, error propagation, and fitting}
\label{apn:resamp}

\subsection{Resampling and error propagation}
\label{apn:resamp_error}

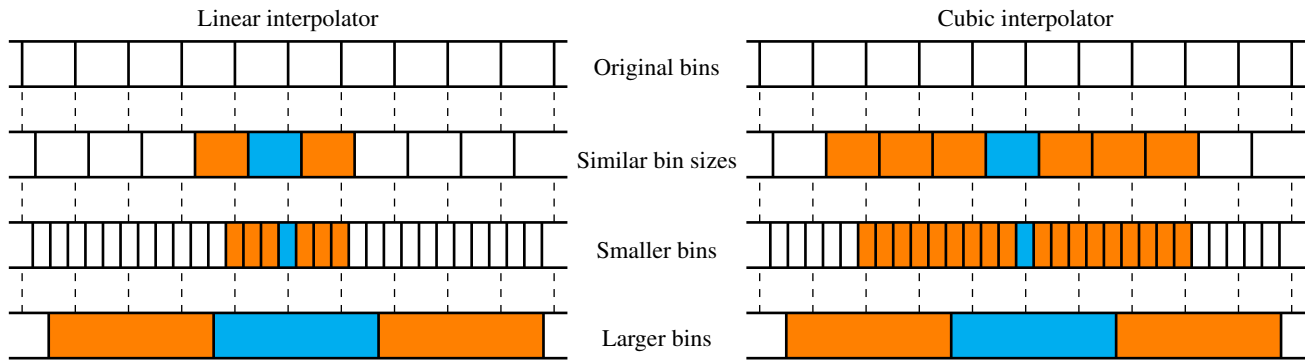
\begin{figure*}
  \centering
  \def\lwa{0.5pt}
  \def\lwb{1pt}
  \def\npix{10}
  \def\w{7mm}
  \def\h{6mm}
  \def\nspec{4}
  \def\edges{%
    {{0.0, 1.0}},%
    {{0.25, 1.0, 5, 4, 6}},%
    {{0.20, 0.33, 15, 12, 18}},%
    {{0.5, 3.1, 2, 1, 3}}%
  }
  \begin{tikzpicture}
    \foreach \edgek [count=\kspec] in \edges {
      \pgfmathsetmacro{\ea}{\edgek[0]}
      \pgfmathsetmacro{\step}{\edgek[1]}
      \pgfmathsetmacro{\eb}{\edgek[0]+\edgek[1]}
      \ifnum\kspec>1
        \pgfmathsetmacro{\ref}{\edgek[2]}
        \pgfmathsetmacro{\cora}{\edgek[3]-1}
        \pgfmathsetmacro{\corb}{\edgek[4]+1}
        \foreach \x in {0,1,...,\npix} {
            \draw[line width=\lwa, dashed] (\x*\w, 2*\nspec*\h-2*\kspec*\h+\h) -- (\x*\w, 2*\nspec*\h-2*\kspec*\h+2*\h);
          }
      \fi
      \foreach \x [count=\i] in {\ea,\eb,...,\npix} {
          \ifnum\kspec>1
            \ifnum\i=\ref
              \draw[fill=cyan] (\x*\w, 2*\nspec*\h-2*\kspec*\h) rectangle (\x*\w+\step*\w, 2*\nspec*\h-2*\kspec*\h+\h);
            \else
              \ifnum\i>\cora
                \ifnum\i<\corb
                  \draw[fill=orange] (\x*\w, 2*\nspec*\h-2*\kspec*\h) rectangle (\x*\w+\step*\w, 2*\nspec*\h-2*\kspec*\h+\h);
                \fi
              \fi
            \fi
          \fi
          \draw[line width=\lwb] (\x*\w, 2*\nspec*\h-2*\kspec*\h) -- (\x*\w, 2*\nspec*\h-2*\kspec*\h+\h);
        }
      \draw[line width=\lwb] (-0.25*\w,2*\nspec*\h-2*\kspec*\h) -- (\npix*\w+0.25*\w,2*\nspec*\h-2*\kspec*\h);
      \draw[line width=\lwb] (-0.25*\w,2*\nspec*\h-2*\kspec*\h+\h) -- (\npix*\w+0.25*\w,2*\nspec*\h-2*\kspec*\h+\h);
    }
    \node[above] at (current bounding box.north) {Linear interpolator};
  \end{tikzpicture}%
  \def\labels{%
    {Original bins},%
    {Similar bin sizes},%
    {Smaller bins},%
    {Larger bins}%
  }%
  \begin{tikzpicture}
    \foreach \labelk [count=\kspec] in \labels {
      \node at (0,2*\nspec*\h-2*\kspec*\h+0.5*\h) {\labelk};
    }
  \end{tikzpicture}%
  \def\edges{%
    {{0.0, 1.0}},%
    {{0.25, 1.0, 5, 2, 8}},%
    {{0.20, 0.33, 15, 6, 24}},%
    {{0.5, 3.1, 2, 1, 3}}%
  }%
  %\hspace{3mm}%
  \begin{tikzpicture}
    \foreach \edgek [count=\kspec] in \edges {
      \pgfmathsetmacro{\ea}{\edgek[0]}
      \pgfmathsetmacro{\step}{\edgek[1]}
      \pgfmathsetmacro{\eb}{\edgek[0]+\edgek[1]}
      \ifnum\kspec>1
        \pgfmathsetmacro{\ref}{\edgek[2]}
        \pgfmathsetmacro{\cora}{\edgek[3]-1}
        \pgfmathsetmacro{\corb}{\edgek[4]+1}
        \foreach \x in {0,1,...,\npix} {
            \draw[line width=\lwa, dashed] (\x*\w, 2*\nspec*\h-2*\kspec*\h+\h) -- (\x*\w, 2*\nspec*\h-2*\kspec*\h+2*\h);
          }
      \fi
      \foreach \x [count=\i] in {\ea,\eb,...,\npix} {
          \ifnum\kspec>1
            \ifnum\i=\ref
              \draw[fill=cyan] (\x*\w, 2*\nspec*\h-2*\kspec*\h) rectangle (\x*\w+\step*\w, 2*\nspec*\h-2*\kspec*\h+\h);
            \else
              \ifnum\i>\cora
                \ifnum\i<\corb
                  \draw[fill=orange] (\x*\w, 2*\nspec*\h-2*\kspec*\h) rectangle (\x*\w+\step*\w, 2*\nspec*\h-2*\kspec*\h+\h);
                \fi
              \fi
            \fi
          \fi
          \draw[line width=\lwb] (\x*\w, 2*\nspec*\h-2*\kspec*\h) -- (\x*\w, 2*\nspec*\h-2*\kspec*\h+\h);
        }
      \draw[line width=\lwb] (-0.25*\w,2*\nspec*\h-2*\kspec*\h) -- (\npix*\w+0.25*\w,2*\nspec*\h-2*\kspec*\h);
      \draw[line width=\lwb] (-0.25*\w,2*\nspec*\h-2*\kspec*\h+\h) -- (\npix*\w+0.25*\w,2*\nspec*\h-2*\kspec*\h+\h);
    }
    \node[above] at (current bounding box.north) {Cubic interpolator};
  \end{tikzpicture}
  \caption{Sketch of correlated pixels after resampling a spectrum with \bindensity{},
  depending on the chosen interpolator and the new bins size.
  We highlight in orange all the pixels which are correlated with a reference pixel (marked in blue).
  The original spectrum is assumed to be uncorrelated.}
  \label{fig:bindensity}
\end{figure*}

We developed for \textsc{antaress} an open-source C library with python bindings, called \bindensity{}%
\footnote{The \bindensity{} package is installable with pip and conda and the sources are publicly available at \url{https://gitlab.unige.ch/jean-baptiste.delisle/bindensity}}. It implements two resampling algorithms as well as basic operations on spectra, with error propagation using banded covariance matrices. We defined a linear and a cubic interpolator of the cumulative flux, which corresponds to assuming a uniform and a quadratic density over each pixel, respectively. In both cases, the cumulative flux is assumed to be continuous, and the integral of the flux on each original pixel is preserved. The linear interpolator can be used for fast, preliminary analyses, but the cubic interpolator should be preferred to limit blurring sharp spectral features. We do not propose higher order interpolators to prevent the risk of overfitting and creating spurious signals in the resampled spectra.

In the linear case, this rule is sufficient to determine the two coefficients of the interpolator, since it defines the value of the cumulative at each edge of the original pixels. With this linear interpolator, only new pixels that overlap the same original pixel will be correlated (see Fig.~\ref{fig:bindensity}). If the size of the new pixels is similar to the size of the original pixels, it typically generates correlation only between two consecutive pixels, and the covariance matrix will be tridiagonal (the main diagonal and the first sub/super-diagonal are non-zero). If several successive linear interpolations are performed, the covariance matrix's bandwidth (number of non-zero sub/super diagonals) keeps increasing at each step. This bandwidth growth should be avoided because of the increased computational cost, but also because it is associated to a loss of resolution. This is why we avoid as much as possible to perform successive changes of rest frames, but instead apply spectral shifts directly to the spectra.

In the cubic case, one needs to specify two additional constraints to determine the four coefficients of the interpolating polynomial on each bin. A widely spread method is to use cubic splines, which require the interpolator to be $C^{2}$ (twice continuously differentiable). The polynomial coefficients can then be obtained by solving a tridiagonal system of equations, which can be implemented with dedicated efficient algorithms, but which introduces correlations between all pixels in the spectrum. Moreover, assuming the cumulative flux to be $C^{2}$ means that the flux density is continuously differentiable. Such an hypothesis might not be physically justified if spectra present breaks, in particular in saturated absorption lines. We thus decided to relax the $C^{2}$ condition and to require the derivative at each edge between two pixels to be equal to the average density of the two pixels. Our cubic interpolator is thus only $C^{1}$ in cumulative flux, which means that the flux density is continuous but not differentiable. Using these modified constraints, solving for the polynomial coefficients is straightforward, and the interpolation introduces correlations between new pixels that overlap the same original pixel or two consecutive original pixels. If the sizes of the new and original pixels are similar, the resulting covariance matrix would typically have two sub/super-diagonals (see Fig.~\ref{fig:bindensity}).

\subsection{Fitting}
\label{apn:fit_cov}

Model fitting in the \textsc{antaress} methods is performed using the log-probability function of a set of parameters $\theta$:
\begin{equation}
\ln p(\theta | d , M) = \ln \varpi(\theta) + \ln \mathcal{L}(\theta) 
\end{equation}
Where $m$ represents the family of models fitted to the data $d$, $\varpi(\theta) = p(\theta | m)$ the prior probability of the model parameters, and $\mathcal{L}(\theta) = p(d| m , \theta) $ the likelihood. Under the assumption of Gaussian noise we have:
 \begin{equation}
\ln \mathcal{L}(\theta) = - \ln( \sqrt{ \det(2\pi\Sigma)})  -\frac{1}{2} r^t \Sigma^{-1} r
,\end{equation}
Where $r$ = $d-m(\theta)$ is the vector of residuals and $\Sigma$ the noise covariance matrix.

We use the Cholesky decomposition of the banded covariance matrix $C$ implemented in the workflow, so that:
\begin{equation}
ln \mathcal{L}(\theta) = -\sum_{n} \ln( \sqrt{2 \pi} L[n,n])   - \frac{1}{2} u^{t} u
.\end{equation}
Here, $L$ is the lower triangular matrix satisfying $C = L L^{t}$, and $u = L^{-1} r$. It is banded with the same width as $C$, so that the cost of all calculations scales with the number of spectral pixels. 

If we assume that the data is dominated by white noise and correlations can be neglected, then $\Sigma$ reduces to the variance $\sigma^2$ along its diagonal and
\begin{equation}
ln \mathcal{L}(\theta) = -\sum_{n} \ln( \sqrt{2 \pi} \sigma(n))   - \frac{1}{2} \sum_{n}  \left( \frac{r(n)}{\sigma(n)} \right)^2 
.\end{equation}

The MCMC approach implemented in the \textsc{antaress} pipeline samples the posterior probability distributions (PDF) of the free parameters describing $m$ using \textsc{emcee} (\citealt{Foreman2013}). We propose standard priors to set on the parameters (Uniform and Gaussian) as well as original priors : ``Gaussian Halves'' (defined as two half-Gaussian profiles with the same center and different widths, or ``Complex'' (defined through a custom function that uses combinations of the model parameters). The number of walkers and the burn-in phase is adjusted based on the degrees of freedom of the considered problem and the convergence of the chains. Best-fit values for the parameters are set to the median of their PDFs, and their 1$\sigma$ uncertainty ranges are defined using highest density intervals to better account for multimodal and asymmetric PDFs.

\section{Description of planetary systems}

\subsection{Coordinates}
\label{apn:pl_coord}

We describe here the orthonormal reference frames centered on the star that are used to define the planet coordinates and introduce various orbital properties of interest. The first frame has its $x$ axis as the major axis of the orbit oriented toward the periapsis and its $y$ axis obtained by rotating $x$ within the plane along the orbital motion: 
\begin{equation}
\begin{split}
x_{\rm orb}(t) =& \aRs \, (\cos(E(t))-e), \\
y_{\rm orb}(t) =& \aRs \, \sqrt{1-e^2} \, \sin(E(t)). \\
\end{split} 
\end{equation}
Here, $E$ is the eccentric anomaly, derived numerically at a given orbital phase (or time $t$). The corresponding orbital velocity coordinates are:
\begin{equation}
\begin{split}
x'_{\rm orb}(t) =& -\frac{2 \pi}{P} \, \left( \aRs \right)^2 \, \frac{1}{D_{\rm p}(t)} \, \sin(E(t)), \\
y'_{\rm orb}(t) =&  \frac{2 \pi}{P} \, \left( \aRs \right)^2 \, \frac{1}{D_{\rm p}(t)} \sqrt{1-e^2} \, \cos(E(t)), \\
\mathrm{with} \, D_{\rm p}(t) =& \aRs \, (1 - e \, \cos(E(t))). \\
\end{split} 
\label{eq:der_orb}
\end{equation}
Rotating the orbital coordinate system around the orbital plane normal by the argument of periastron $\omega$, so that $x$ becomes the node line, yields:
\begin{equation}
\begin{split}
x_{\rm LOS}(t) =& -x_{\rm orb}(t) \, \cos(\omega) +  y_{\rm orb}(t) \, \sin(\omega),  \\       
y_{\rm LOS}(t) =&  x_{\rm orb}(t) \, \sin(\omega) +  y_{\rm orb}(t) \, \cos(\omega).  \\     
\end{split} 
\label{eq:orb2los}
\end{equation}
Rotating the coordinate system around the node line by the orbital inclination $i_p$ then yields the sky-projected ``orbital'' reference frame in which $y$ is along the sky-projected orbital normal and $z$ is along the LOS:
\begin{equation}
\begin{split}
x_{\rm sky}(t) =& x_{\rm LOS}(t), \\       
y_{\rm sky}(t) =& -y_{\rm LOS}(t) \, \cos(i_p),  \\
z_{\rm sky}(t) =&  y_{\rm LOS}(t) \, \sin(i_p).  \\      
\end{split} 
\label{eq:los2sky}
\end{equation}
Rotating the latter reference frame around the LOS by the sky-projected obliquity $\lambda$ of the planet yields the sky-projected ``stellar'' frame, where $x$ is the node line of the stellar equator and $y$ the sky-projected stellar spin:   
\begin{equation}
\begin{split}
x_{\star \rm sky}(t) =& x_{\rm sky}(t) \, \cos(\lambda) - y_{\rm sky}(t) \, \sin(\lambda),   \\ 
y_{\star \rm sky}(t) =& x_{\rm sky}(t) \, \sin(\lambda) + y_{\rm sky}(t) \, \cos(\lambda),   \\ 
z_{\star \rm sky}(t) =& z_{\rm sky}(t).    \\ 
\end{split} 
\end{equation}   
Here, we caution that some equations in the workflow (e.g, that of the $rv$ of planet-occulted regions) require the coordinates of the sky-projection of the planet onto the stellar surface. In that case, $z_{\star \rm sky}$ is calculated as $\sqrt{1-x_{\star \rm sky}^2-x_{\star \rm sky}^2}$ for a spherical star and by solving the quadratic describing the photosphere for an oblate star. A final rotation around the star node line by $i_{\star}$, inclination of the stellar spin with respect to the LOS, yields the position of the planet or occulted region in the star rest frame, in which $x$ now follows the stellar equator, and $y$ is the actual stellar spin:
\begin{equation}
\begin{split}
x_{\star}(t) =& x_{\star \rm sky}(t),   \\ 
y_{\star}(t) =& y_{\star \rm sky}(t) \, \sin(i_{\star}) + z_{\star \rm sky}(t) \, \cos(i_{\star}),   \\ 
z_{\star}(t) =& -y_{\star \rm sky}(t) \, \cos(i_{\star}) + z_{\star \rm sky}(t) \, \sin(i_{\star}).    \\   
\end{split} 
\end{equation}

Another useful coordinate is the Keplerian $rv$ of the star, relative to the barycenter of the planetary system:
\begin{equation}
\rm rv_{\star/B_{\star}}(t) = \sum_{body \, p} K_{\rm p} \, [\cos(\theta_{\rm p}(t) \, + \, \omega_{\rm p})+e \, \cos(\omega_{\rm p})],
\end{equation}
\label{eq:kep}
where $K_{\rm p}$ is the Keplerian $rv$ semi-amplitude induced by a given body orbiting the star and $\theta_{\rm p}$ the true anomaly of its orbit.

\begin{figure*}
\begin{minipage}[tbh!]{\textwidth}
\includegraphics[trim=0cm 0cm 0cm 0cm,clip=true,width=0.9\columnwidth]{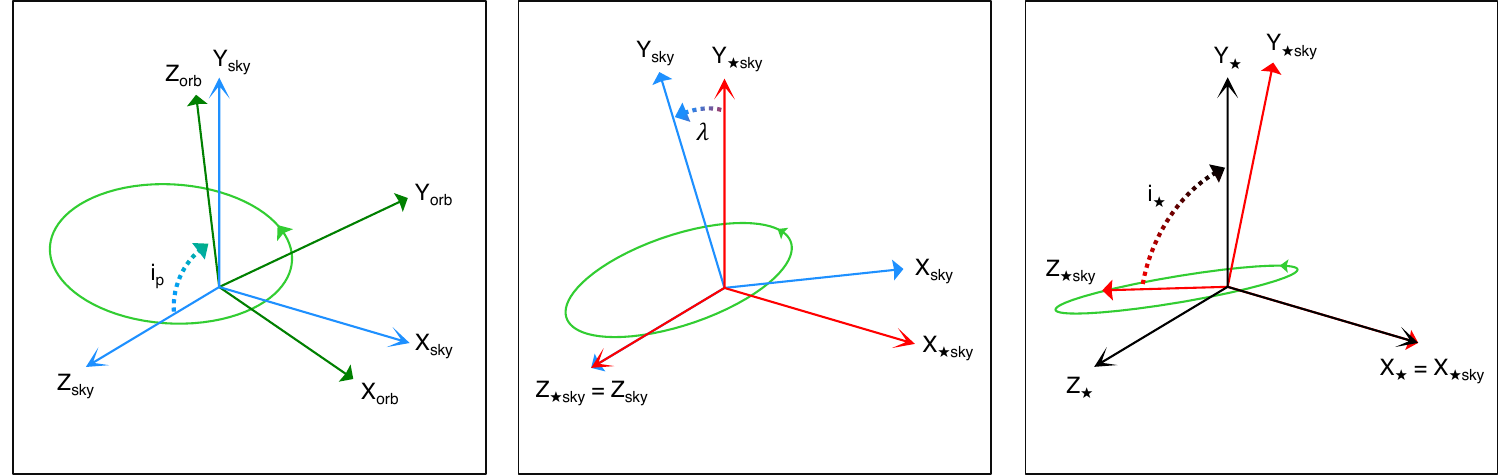}
\centering
\end{minipage}
\caption[]{Main \textsc{antaress} reference frames (green axis, orbital; blue axis, sky-projected orbital; red axis, sky-projected stellar; black, stellar). The star is at the origin. An eccentric planetary orbit is shown as a green curve, oriented in the direction of the planet motion.  }
\label{fig:Frames}
\end{figure*}

\subsection{Planet-occulted profiles}
\label{apn:pl_occ_grid}

We describe here the detailed calculation of line profiles. Without oversampling of the transit chord, properties of the planet-occulted region are calculated for a single position of the planetary grid corresponding to the center of the chord. With oversampling, the planetary grid is positioned at regular intervals $k$ along the exposure chord. The disk of planet $p$ is discretized at each position $k$, calculating the relevant stellar surface properties $x_{i}$ in each cell $i_{k,\wb}$ of the grid. Only cells that fall within the sky-projected photosphere, and outside of the planetary disks already processed, are considered. We note the chromatic dependance of the planet-occulted cells and properties, linked to broadband variations of the stellar intensity and planet radius. 

Planet-occulted line profiles can then be calculated at two precision levels (Fig.~\ref{fig:Precisions}):
\begin{itemize}
\item Low: properties of the region occulted during the time window $\overline{t}$ are approximated as the brigthness-weighted average of all $x_{i}$: 
\begin{equation}
x(\wb,\overline{t}) = \frac{\sum_{k,p} \sum_{i_{k,\wb}} x_{i} I_{i}(\wb) S_{i} }{\sum_{k,p}  \sum_{i_{k,\wb}} I_{i}(\wb) S_{i} }
\label{eq:bav}
,\end{equation}
where $I_{i}(\wb)$ and $S_{i}$ designate a cell intensity level and area. The intrinsic profile associated with the occulted region is then defined as $I(x(\wb,\overline{t}),\lambda)$, with $I$ one of the models described in Sect.~\ref{sec:st_grid}. Because the photospheric $rv$ field may vary substantially over the occulted region, $I$ is broadened with a rotational kernel defined by the range of velocities covered by the exposure chord. The intrinsic profile is then continuum-scaled using the chromatic planet-occulted flux calculated with Eq.~\ref{eq:bav} for $x_{i}=1$, and Doppler-shifted using the chromatic planet-occulted $rv(\wb,\overline{t})$.

\item High: the line profile of each oversampled region is calculated as $F_{k}(\lambda) = \sum_{i_{k}} F_{i}(\lambda)$, where $F_{i}(\lambda) = I(x_{i},\lambda) S_{i}$ is the local profile from a given cell. The exposure profile is then calculated as the average of the $F_{k}$ over all regions.

\end{itemize}

Chromatic planet-occulted profiles can only be defined in low precision, as the local profiles summed in high precision already depend on wavelength.

\section{Spectral correction models}
\label{apn:init_corr}

Here, we describe the models used by the \textsc{Corrections} methods of the \textsc{antaress} workflow.

\subsection{Spectral flux calibration model}
\label{apn:count_sc}

Because DRS conversion and correction tables are not necessarily available, we estimate the detector flux calibration profile directly from the input spectra. Assuming that $g_{\rm cal}$ varies slowly with wavelength, Eq.~\ref{eq:gdet} yields:
\begin{equation}
g_{\rm cal}(\wb,E,t) = \frac{  \sum_{\wb} \sigma^{\rm meas}(\lambda,E,t)^2 }{ \sum_{\wb} F^{\rm meas}(\lambda,E,t) } 
.\end{equation}
Values of $g_{\rm cal}(\wb,E,t)$ measured in each spectral order are fitted with a piece-wise model made of three polynomials continuous in value and derivative. They cover the central part of the order, as well as its blue and red edges where the count level decreases sharply, ensuring that the shape of $g_{\rm cal}$ is properly captured. Outliers, typically in regions of deep telluric or stellar lines, are automatically excluded from the fit. The derived profiles show slight variations in time, especially in regions of low S/N. Our final estimate of the flux calibration profile is thus $\infsymb g \supsymb_{\rm cal}$, the median of the calibration models fitted to all exposures of a processed instrument.

\subsection{Telluric spectra model}
\label{apn:tell_corr}

Properties of the Earth atmospheric model are derived for each molecule by fitting its output telluric spectrum to the spectrum measured in each exposure, using Levenberg-Marquard minimization. Direct measurements of the ambient temperature, airmass, and integrated water vapor at zenith are available at the facilities of ESPRESSO, HARPS, HARPS-N, CARMENES, and NIRPS. They are used to fix the model temperature, and to initialize the fitted pressure and column density of the species along the LOS, for spectra obtained with these instruments. Uniform priors can further be set on these fitted properties, which can be useful in low-S/N datasets where some exposures are poorly adjusted.

First, the spectral grid of the observed spectrum is temporarily shifted back to the Earth rest frame, and if necessary converted into vacuum wavelengths. Then, a high-resolution telluric model is calculated in wavenumber space at each step of the minimization using a subset of deep telluric absorption lines. The wavelength, width, and intensity of the lines are adjusted depending on the fitted atmospheric properties, and used to compute the line profiles (Lorenzian for H$_2$O, Voigt for other species; see \citealt{Allart2022}). The optical depth spectrum of the species is obtained through multiplication of its column density by the co-added absorption line profiles. The resulting telluric absorption spectrum is convolved with the instrumental response and sampled to the instrumental wavelength grid. 2D maps providing the width of a Gaussian resolution kernel as a function of detector pixel and spectral order were already available for the different modes and epochs of ESPRESSO and NIRPS. We derived maps for HARPS and HARPS-N following the same approach as in \citet{Allart2022}, applying 2D polynomial fits over unresolved thorium lines measured across the instrument detector. For other instruments, a constant resolution is used. The comparison between model and data is performed over a CCF calculated with a binary mask composed of the selected telluric lines, with weights unity. 

An interest of this approach is the possibility to determine whether the correction of a specific molecule is warranted, by evaluating the amplitude of its telluric CCF in the raw and corrected data (Fig.~\ref{fig:tell_corr}). We improve upon the original ATC by using the \textsc{antaress} method to compute CCFs (Sect.~\ref{sec:CCFs}), applying accurate spectral resampling and error propagation (Sect.~\ref{sec:resamp}), and deriving more realistic uncertainties through the use of the covariance matrix (Sec.~\ref{sec:resamp}, Fig.~\ref{fig:tell_corr}). 

Once best-fit atmospheric properties are found for all molecules, we co-add their high-resolution optical depth spectra using the full HITRAN linelist (which include micro-tellurics), calculate the resulting telluric absorption spectrum, convolve it with the instrumental response, and resample it over the spectral grid of the processed exposure. Corrected spectra are obtained through division with this final telluric absorption spectrum. We set a threshold on telluric line depth above which the telluric-absorbed flux is considered too low and noisy to be retrieved. The threshold is typically set to 90\% (\textit{i.e.} 10\% of the flux being transmitted but is adjustable depending on the dataset. Flagged pixels are masked.

\subsection{Flux balance model}
\label{apn:col_bal}

We describe here the approach used to model flux balance variations (Sect.~\ref{sec:col_bal}). 

We first define the spectra $F_{\rm ref}$ that are used as reference for the flux balance of the star. A selection of spectra in each epoch are aligned in the star rest frame (shifting them by the stellar Keplerian and systemic $rv$ motion), resampled on a common spectral grid, and normalized in each order to their average flux level over the selection. Reference spectra in each epoch are then calculated as the median of the processed spectra, to avoid distortions induced by spurious features (e.g., cosmics, Sect.~\ref{sec:cosm_corr}). Reference spectra for each instrument are calculated as the mean of their epoch-specific median spectra, naturally giving more weight to epochs with higher flux levels. If a single instrument is processed, using this instrument reference to set the final flux balance for all its spectra limits the amplitude of the correction and keeps corrected spectra closer to their original balance. If several instruments are processed, their common flux balance can be set to one of the instrument-specific references, or to measured/theoretical stellar spectra independent of the processed datasets. 

We then determine the color balance variations for each exposure. First, the median reference spectrum is shifted to the rest frame of the exposure spectrum and resampled over its spectral grid. Then, the reference and exposure spectra are converted back from extracted $F$ to raw $N$ photoelectron counts (Sect.~\ref{sec:count_sc}) and binned at a lower resolution in each order. The ratio $R^{\rm meas}$ between the binned reference and exposure spectra is fitted with a smoothing spline or polynomial $\rm R^{\rm theo}$. A preliminary fit is performed to automatically identify and exclude outliers, and the final fitted model (Fig.~\ref{fig:ColorBalanceModel}) is an estimate of:

\begin{equation}
\begin{split}
R^{\rm meas}(\nubsun,E,t) &= \frac{N^{\rm meas}(\nubsun,E,t)}{N_{\rm ref}(\nubsun,E)} \\
    &=  \frac{ \delta_{\rm p}(\nubsun,E,t) \, c(\nubsun,E,t) }{C_{\rm ref}(\nubsun,E)}.\\
\end{split}
\end{equation}
Here, we assume that the reference spectrum and the true stellar spectrum deviate at low resolution, so that $C_{\rm ref}$ and $N_{\star}$ remain separated when binning $\rm N_{\rm ref} = C_{\rm ref} N_{\star}$. Light frequency $\nu$ is used as fit coordinate to stretch the data and its variations over a more regular grid. Finally, we extrapolate $R^{\rm theo}(\nubsun,E,t)$ over the native spectral grid of the exposure to define its correction. The same approach is used when determining the color balance variations between the median reference and the final reference for each epoch (Sect.~\ref{sec:col_bal}). We note that this final reference is not used directly to correct flux balance variations within each epoch because we found it introduces high-frequency variations in the correction of individual spectra. 

When correcting for flux balance variations we recommend that orders and ranges that are too contaminated by telluric absorption (typically toward longer wavelengths in optical spectra and between the J and H bands in NIR spectra), or where noise is too large to fit properly the color variations (typically the bluest orders in low-S/R optical data), are excluded. The bin size, as well as the smoothing factor or polynomial degree, must be adapted for each dataset to minimize the broadband dispersion of $F^{\rm corr}/F_{\rm ref}$ while preventing the overcorrection of spectra and the introduction of artificial variations. Bin size must be smaller than the typical range of variations of the color balance $c$ and planetary contribution $\delta_{\rm p}$ but large enough to smooth out high-frequency features, especially those caused by variations in narrow stellar and planetary lines. Variations at the level of each order linked to the physical size of the detector in wavelength space (Sect.~\ref{sec:col_bal}) must be sampled with a smaller binwidth but using a low-order polynomial to only capture order-wide variations.

\subsection{Cosmics}
\label{apn:cosm_corr}

We detail here the processing of the datasets for the cosmics correction method. Each epoch is treated independently. The number of adjacent exposures used for comparison can be adjusted depending on the dataset, with one half selected before the processed exposure and the other half after. In cases where the planet induces a strong RM effect, we advise choosing a small number of adjacent exposures so that their spectra remain similar (ie, with similar RM distortions). Spectra used for comparison are temporarily aligned in the stellar rest frame using the Keplerian $rv$ model for the system (Sect.~\ref{sec:align_star}), $rv$ provided by the instrument DRS, or delta-$rv$ measured via cross-correlation of the adjacent exposure spectra with the processed one. In general, we recommend using the Keplerian model since it is not sensitive to the photonic $rv$ dispersion, but the other approaches can be useful in high-S/R datasets where the $rv$ dispersion around the Keplerian motion is caused by stellar activity (e.g., granulation). The aligned spectra of adjacent exposures are resampled on the spectral grid of the processed exposure at time $t$, and set to the same global flux level. We can then calculate the mean $F_{\rm comp}(\wstar,E)$ and standard deviation $\sigma_{\rm F_{\rm comp}}(\wstar,E)$ for each bin, over all adjacent exposures, and search for cosmics as described in Sect.~\ref{sec:cosm_corr}.

\subsection{Persistent peaks}
\label{apn:perspeak_corr}

To identify persistent peaks, spectra are first temporarily shifted back to the Earth rest frame so that detector features and telluric lines are aligned. We calculate a master stellar spectrum as the median of all spectra in an epoch to estimate the stellar continuum (Sect.~\ref{sec:st_cont}) and calculate the residuals between the spectrum of each exposure and this continuum. Then we identify bins where residuals deviate by more than a threshold times their flux error, and flag all surrounding bins falling within a chosen window (with typical width that of the peaks to be excluded). Finally, a bin is masked if it is flagged in all exposures or in a chosen number of consecutive exposures.

\subsection{Wiggle model}
\label{apn:wig_module}

\subsubsection{Description}
\label{apn:wig_mod_descr}

From the analysis of several datasets besides those in the present study, we determined that the wiggles are best described as the sum of multiple sinusoidal components $k$ (typically two): 
\begin{equation}
W(\nu,t) = 1 + \sum_{k} A_{k}(\nu,t) \sin( 2 \pi \int\left(F_{k}(\nu,t) d\nu \right) - \Phi_{k}(t)).
\end{equation}
The amplitude $A_{k}$ and frequency\footnote{We highlight that the phase of the sine term is calculated as the integral of the frequency $F_{k}(\nu)$. This is because the rate at which a sinusoid phase changes is equal to the frequency at which this sinusoid oscillates, in other words a sinusoid frequency is the derivative of its phase.} $F_{k}$ of each component show fewer variations as a function of light frequency $\nu = c/\lambda$ than wavelength, and are described as: 

\begin{equation}
\begin{split}
A_{k}(\nu,t) &= \sum_{i=0}^{da,k} a_\mathrm{chrom,k,i}(t)(\nu-\nu_\mathrm{ref})^i, \\
F_{k}(\nu,t) &= \sum_{i=0}^{df,k} f_\mathrm{chrom,k,i}(t)(\nu-\nu_\mathrm{ref})^i, \\
\end{split}
\end{equation}
where the polynomials are defined relative to $\nu_\mathrm{ref}$, corresponding to 6000\,\AA\,, to decrease leverage in the fits. The chromatic amplitude and frequency coefficients $a_\mathrm{chrom,k,i}(t)$ and $f_\mathrm{chrom,k,i}(t)$, as well as the phase reference $\Phi_{k}(t)$, are expressed as a linear combination of the cartesian pointing coordinates $g(\left\{\alpha_\mathrm{point,k}\right\},t) =$
\begin{align}
\alpha_\mathrm{u/west} &+ \alpha_\mathrm{x} x(t) + \alpha_\mathrm{y/west} y(t) + \alpha_\mathrm{z/west} z(t) \, &\mathrm{if \, \theta(t) > \theta_\mathrm{mer}}\nonumber,\\
\alpha_\mathrm{u/east} &+ \alpha_\mathrm{x} x(t) + \alpha_\mathrm{y/east} y(t) + \alpha_\mathrm{z/east} z(t) \, &\mathrm{if \, \theta(t) < \theta_\mathrm{mer}}, \nonumber\\
\mathrm{with\,}&x(t) = \sin(\theta(t)),\nonumber \\
                          &y(t) = \cos(\theta(t)), \nonumber\\
                          &z(t) = \sin(\phi(t)), \nonumber\\
\end{align}
\label{eq:point_func}
where $\theta$ and $\phi$ are the azimuth and altitude of the telescope. We found that the model coefficients change when the LOS crosses the meridian ($\theta_\mathrm{mer}=180^{\circ}$), but that the model remains continuous in value ($\alpha_\mathrm{y/west}  = \alpha_\mathrm{y/east}  + \sin(\phi_\mathrm{mer})(\alpha_\mathrm{z/west}  - \alpha_\mathrm{z/east}) - (\alpha_\mathrm{u/west}- \alpha_\mathrm{u/east} )  
$) and derivative ($\alpha_\mathrm{x/west} = \alpha_\mathrm{x/east} = \alpha_\mathrm{x}$). The full model is thus controlled by a set of pointing parameters $\left\{a_\mathrm{point,k,i}\right\}$, $\left\{f_\mathrm{point,k,i}\right\}$, and $\left\{\Phi_\mathrm{point,k,i}\right\}$ for each component $k$. 

\subsubsection{Fitting procedure}
\label{apn:wig_car}

\begin{figure*}
\begin{minipage}[tbh!]{\textwidth}
\includegraphics[trim=0cm 0cm 0cm 0cm,clip=true,width=0.8\columnwidth]{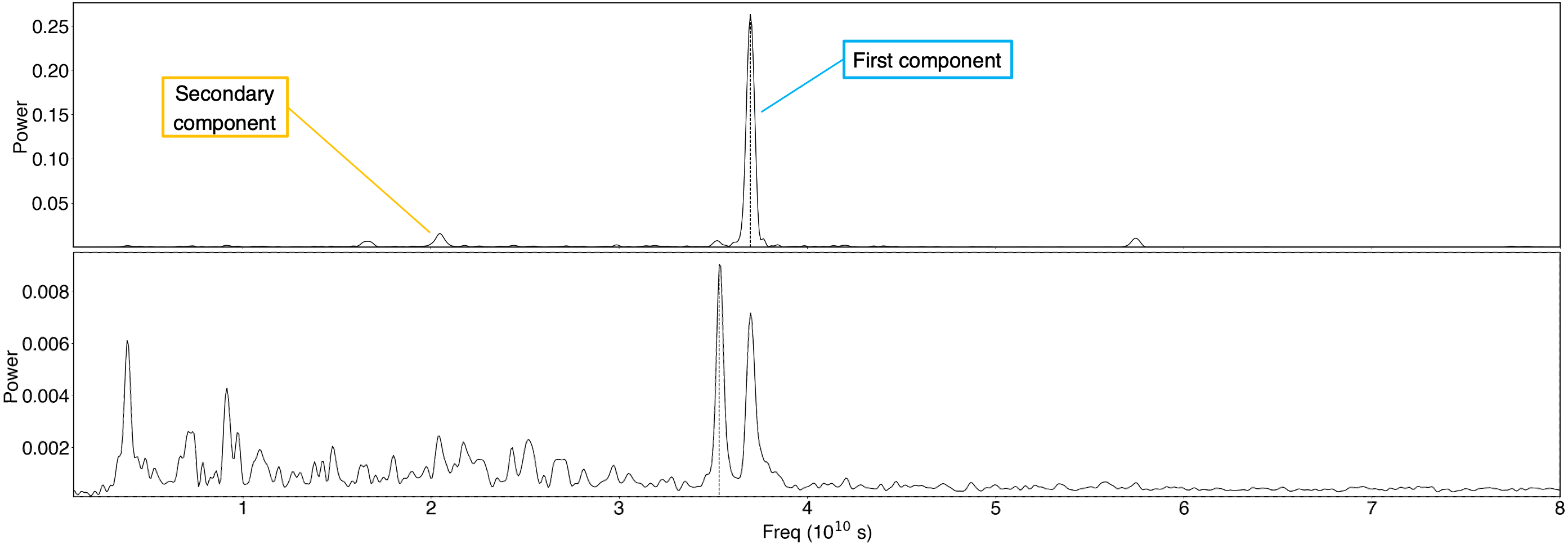}
\centering
\end{minipage}
\caption[]{Periodogram cumulated over all transmission spectra in HD\,209458b epoch 2, before (top panel) and after (bottom panel) wiggle correction. The two components included in the model are highlighted. Although a peak to the left of the first component may be removed by including a third component to the model, we emphasize the change in power scale between the two panels.}
\label{fig:apn_Global_perio_wig}
\end{figure*}

\begin{figure*}
\begin{minipage}[tbh!]{\textwidth}
\includegraphics[trim=0cm 0cm 0cm 0cm,clip=true,width=\columnwidth]{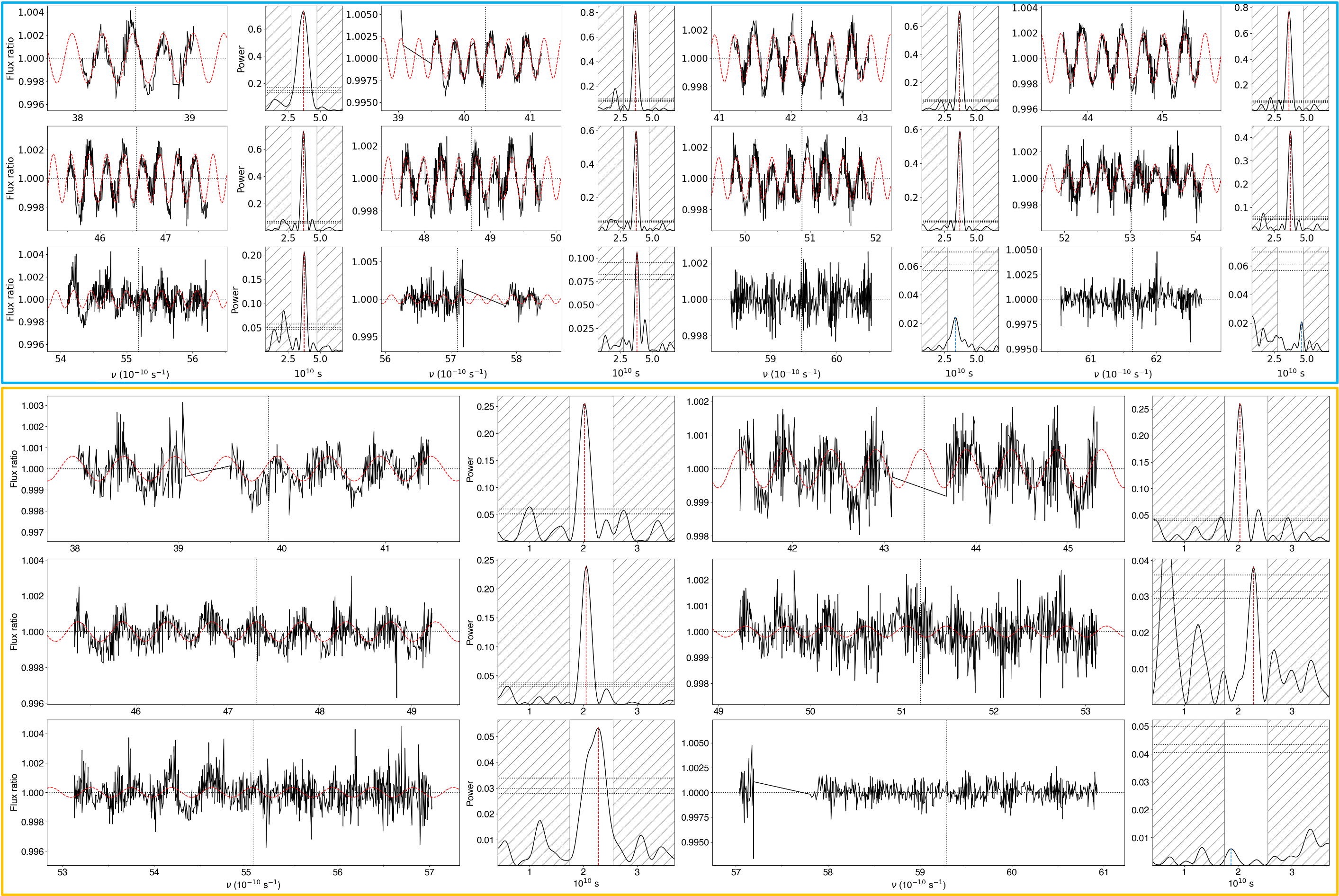}
\centering
\end{minipage}
\caption[]{Chromatic sampling, illustrated with exposure at index 14 in HD\,209458b epoch 2. Subpanels show the processed transmission spectrum for consecutive positions of the sampling window, with the periodogram used to determine the local frequency $A_\mathrm{k}(\nu_\mathrm{window})$ of the considered wiggle component (the unhatched range is set by the user to limit the search around the frequency expected from the screening step). The red model shows the sine model fitted at this frequency. The blue and orange boxes respectively show the sampling of the first and second wiggle components, whose frequencies are distinct enough that they can be processed successively. The width of the sliding window is adjusted to cover several cycles of the considered wiggle component while remaining small enough that its frequency is constant over the band.}
\label{fig:apn_Sampling_wig}
\end{figure*}

\begin{figure*}
\begin{minipage}[tbh!]{\textwidth}
\includegraphics[trim=0cm 0cm 0cm 0cm,clip=true,width=0.8\columnwidth]{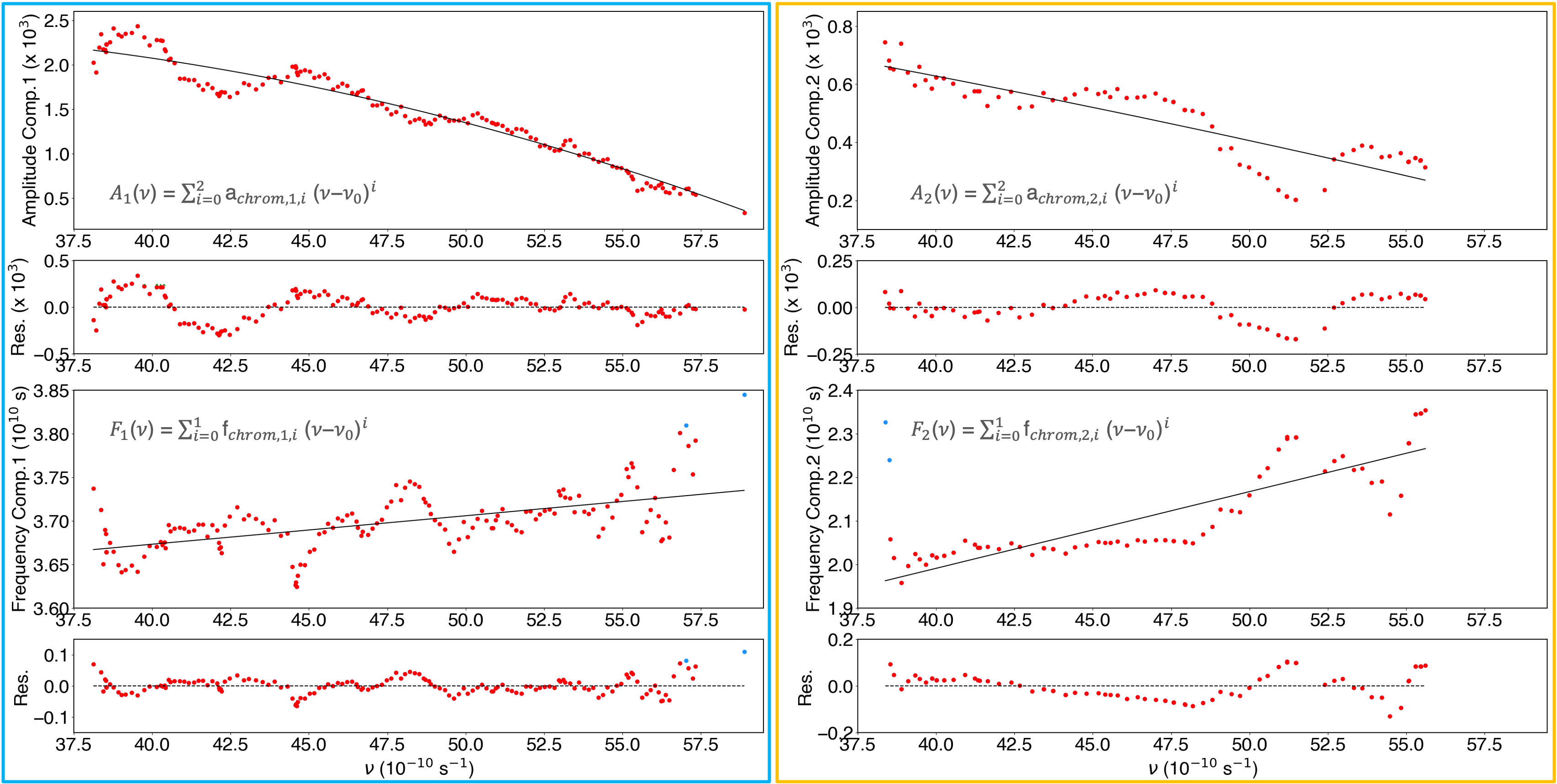}
\centering
\end{minipage}
\caption[]{Chromatic analysis. Red points in the top subpanels show measurements from the sampling step in Fig.~\ref{fig:apn_Sampling_wig}, for the amplitude (top panels) and frequency (bottom panels) of the first (blue box) and second (orange box) wiggle components. Bottom subpanels show residuals between the measurements and their best-fit model (solid black curves). Blue points are outliers automatically excluded from the fit through sigma-clipping. The wave-like variations in the measurements are an artifact of the sampling method.}
\label{fig:apn_Chrom_ana_wig}
\end{figure*}

\begin{figure*}
\begin{minipage}[tbh!]{\textwidth}
\includegraphics[trim=0cm 0cm 0cm 0cm,clip=true,width=0.8\columnwidth]{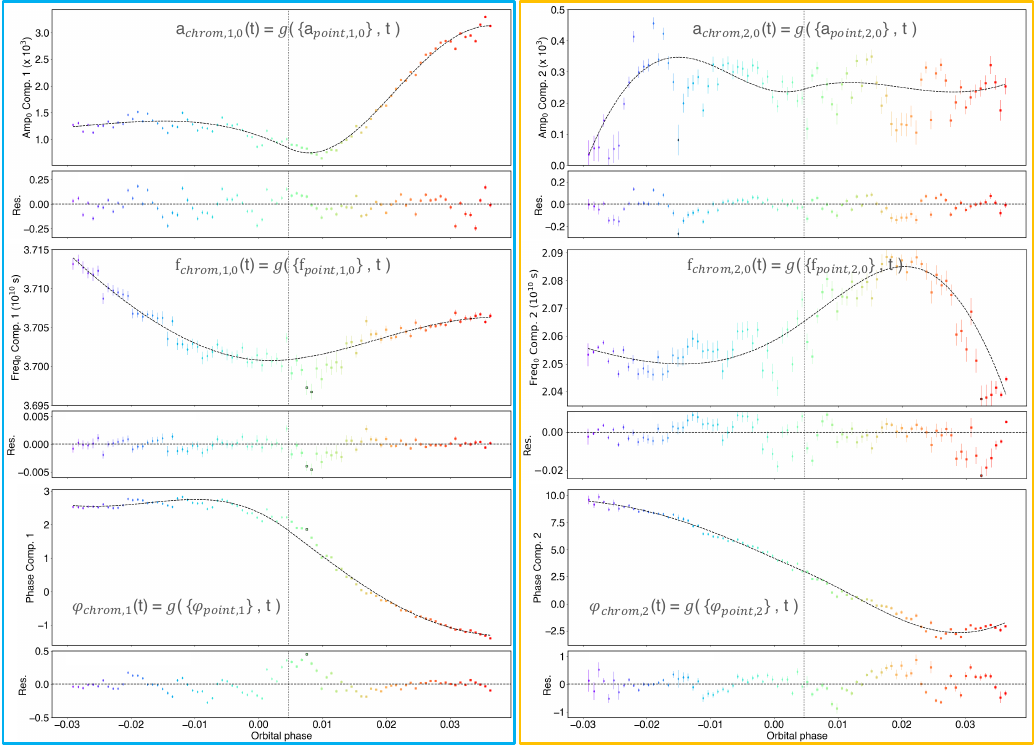}
\centering
\end{minipage}
\caption[]{Pointing analysis. Diamonds (resp. squares) show the chromatic amplitude (top), frequency (middle) and phase (bottom) coefficients derived east (resp. west) of the meridian from the fit to individual exposures, plotted as a function of orbital phase. The meridian crossing is indicated as a vertical dotted line. Bottom subpanels show residuals between the measurements and their best-fit pointing coordinate model (dashed black curves). The blue and orange boxes correspond to the first and second wiggle components.}
\label{fig:apn_Pointing_ana_wig}
\end{figure*}

Each epoch is processed independently. First, 2D spectra are aligned in the star rest frame using a Keplerian RV model (Sect.~\ref{sec:align_star}), so that stellar lines are removed and the wiggles isolated when calculating transmission spectra. A master spectrum is computed for the epoch, using all exposures to smooth out as much as possible the wiggles. The master and spectra are converted back from extracted to raw photoelectron counts (Sect.~\ref{sec:st_cont}), and downsampled over a common spectral grid. The downsampling binwidth is large enough to remove possible correlations between bins but small enough to sample all wiggle components. In-transit planetary lines are ignored in our analysis, considering that they are narrower than the period of the wiggles and typically smaller in amplitude. A 1D transmission spectrum is computed for each exposure by dividing their spectrum with the master in each spectral order, and grouping them together. Transmission spectra for all exposures are shifted into the Earth rest frame, which we expect traces more directly the wiggles, and run through the following steps: 
\begin{enumerate}
\item Screening: required to exclude spectral ranges of poor quality (e.g., from telluric contamination) or unaffected by the wiggles (bluest orders) from the analysis. A periodogram cumulated over all exposures allows identifying which wiggle components are detectable (Fig.~\ref{fig:apn_Global_perio_wig}). The dominant and weak components always contain most of the wiggle power, but additional components may be present and included in the model if relevant (although its power does not justify a correction we identified a third component with frequency $\sim156\times10^{10}\,s$ in the HD\,209458 datasets, which matches the high-frequency signal noted by \citet{CasasayasBarris2021}). 
\item Chromatic sampling: we run a sliding window over each transmission spectrum, identify the strongest peak in a periodogram calculated at each window position, and fit a sine to the window spectrum using the peak frequency (Fig.~\ref{fig:apn_Sampling_wig}). Because the wiggles can be approximated by sines with constant frequency over narrow bands, this step allows sampling the frequency $F_{k}(\nu)$ and amplitude $A_{k}(\nu)$ of a given component in each exposure. The sampling windows must be broad to contain several cycles of a wiggle component, so that we make the windows overlap between successive positions to oversample the measurements. Once a component is processed, its piecewise model built from all windows is used to correct temporarily the transmission spectrum so that the next component can be sampled. 
\item Chromatic analysis: the frequency and amplitude of each component sampled in the previous step are fitted with polynomial models of $\nu$ (Fig.~\ref{fig:apn_Chrom_ana_wig}). This allows determining their degree and deriving guess values for the chromatic coefficients $a_\mathrm{chrom,k,i}$ and $f_\mathrm{chrom,k,i}$ in each sampled exposure. In most datasets that we analyzed, the amplitude and frequency could be described as a quadratic and linear function of $\nu$, respectively decreasing and increasing toward shorter wavelengths. For high-S/R wiggles, which can be characterized over the full ESPRESSO range, higher degrees may be necessary to describe the wiggle variations in the bluest orders.
\item Exposure fit: the spectral wiggle model $W(\nu)$ is initialized using the results from the previous step and fitted to all exposures individually (Fig.~\ref{fig:wig_trans}). This step provides accurate estimates of the $a_\mathrm{chrom,k,i}(t)$, $f_\mathrm{chrom,k,i}(t)$, and $\Phi_{k}(t)$ coefficients. 
\item Pointing analysis: the time series of chromatic amplitude and frequency coefficients, as well as the phase, derived independently from the fit to each exposure are fitted with the pointing coordinate function (Eq.~\ref{eq:point_func}), to evaluate the pointing parameters $\left\{a_\mathrm{point,k,i}\right\}$, $\left\{f_\mathrm{point,k,i}\right\}$, and $\left\{\Phi_\mathrm{point,k,i}\right\}$ (Fig.~\ref{fig:apn_Pointing_ana_wig}).
\item Global fit: the full spectro-temporal wiggle model $W(\nu,t)$ is initialized using the results from the previous step and fitted to all exposures together. Given the large number of free parameters and the complexity of a model made of cumulated sines, guess values close to the best-fit are essential to the convergence of the fit.
\end{enumerate}

\begin{figure}
\includegraphics[trim=0cm 0cm 0cm 0cm,clip=true,width=0.9\columnwidth]{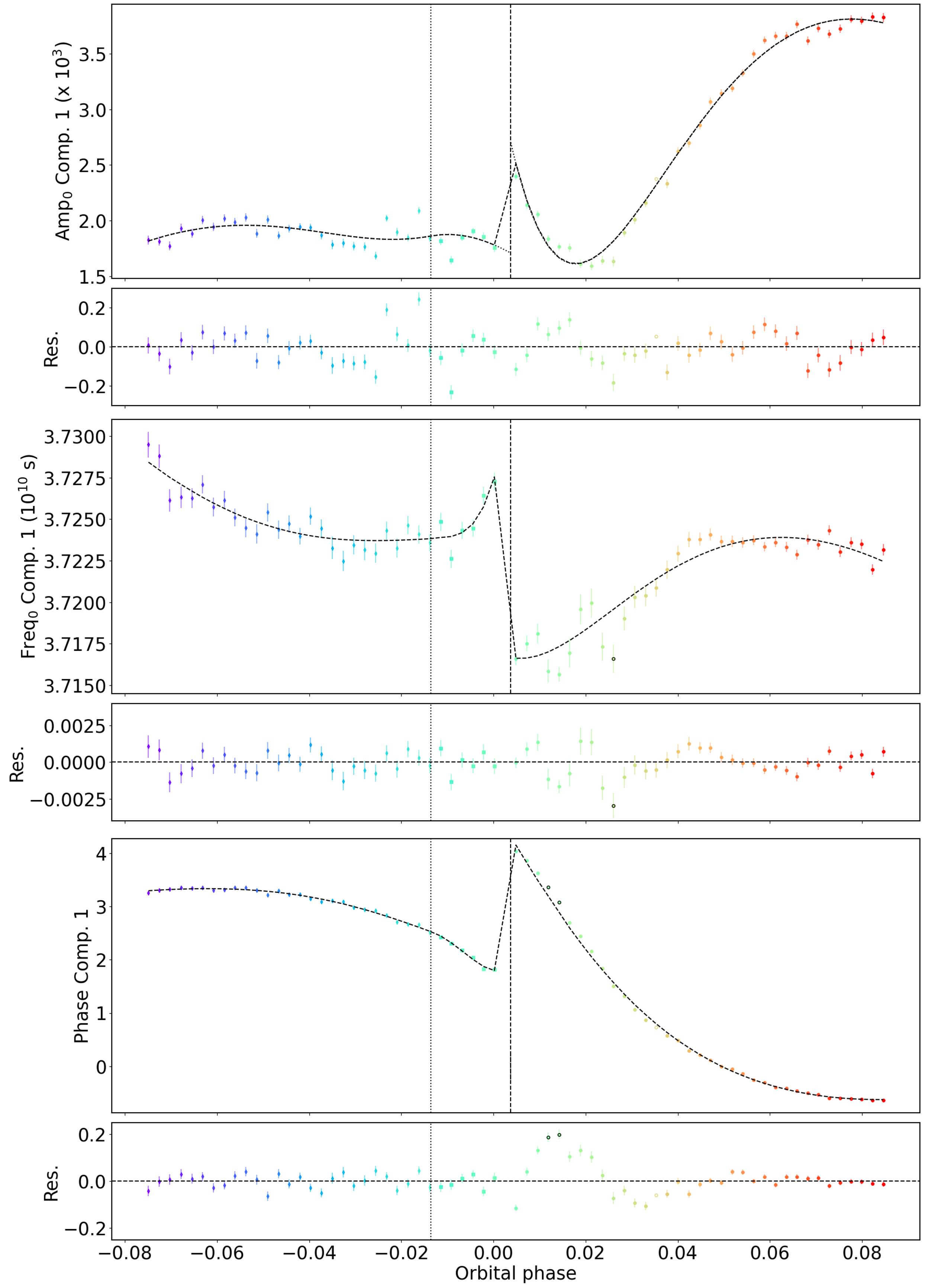}
\centering
\caption[]{Same as Fig.~\ref{fig:apn_Pointing_ana_wig}, for the dominant wiggle component in WASP\,76 epoch 2. These plots illustrate the reset of pointing coefficients following a change in guide star, indicated as a vertical dashed line. }
\label{fig:apn_Pointing_ana_wig_guidshift}
\end{figure}

\section{Detailed processing methods}
\label{apn:proc_mod}

\subsection{Stellar line detrending}
\label{apn:st_line_detrend}

We describe here our approach to correct individual stellar lines (\textit{i.e.}, an isolated spectral line or a cross-correlated line) or a full spectrum for systematic trends characterized using the out-of-transit cross-correlated line properties:
\begin{itemize}
\item $rv$ correction: profiles are shifted using the $rv$ detrending model, following the same approach as in Sect.~\ref{sec:align_star}.
\item Contrast correction: profiles are temporarily normalized and offset to a null continuum. A ``vertical stretching'' is then applied by dividing the flux with the contrast detrending model, before returning profiles to their original continuum. Normalization is performed using the average continuum flux for single line profiles, or the stellar spectral continuum (Sect.~\ref{sec:st_cont}) for full spectra.
\item FWHM correction: profiles are temporarily aligned on a null velocity (spectral profiles are first passed into RV space relative to the line wavelength). A ``horizontal stretching'' is then applied by dividing the aligned $rv$ grid with the FWHM detrending model, before profiles are shifted back to their original frame. This correction can only be performed on single line profiles. 
\end{itemize} 
FWHM and contrast models are normalized over each epoch to maintain the properties of the corrected profiles to their original mean value.

\subsection{Broadband flux scaling}
\label{apn:flux_scaling}

The $F^{\rm meas}$ spectra posterior to the \textsc{Flux balance} correction have the same low-frequency profile as the chosen reference $F_{\rm ref}$:
\begin{equation}
\begin{split}
F^{\rm meas}(\wbstar,E,t) &= F_{\rm ref}(\wbstar,E) R^{\rm norm}(E,t) \\
                                                 &= C_{\rm ref}(\wbstar,E) F_{\star}(\wbstar,E) R^{\rm norm}(E,t). \\ 
\end{split}
\end{equation}
First, we divide all spectra by their corresponding exposure duration, converting them in the flux density units to which $F$ will correspond to hereafter. Then we correct spectra for $R^{\rm norm}$ (ratio between the total flux of the spectrum and that of the reference, Eq.~
\ref{eq:fbal}) through multiplication by 
\begin{equation}
C_{\rm F}(E,t) =\frac{\mathrm{TF}_{\rm ref}}{\sum_{\wstar} F^{\rm meas}(\wstar,E,t) d\wstar} 
.\end{equation}
The integrated flux of the resulting profiles is thus normalized to TF$_{\rm ref}$, which can either be set to the median flux over the epoch $<t_k,\sum_{\wstar} F^{\rm meas}(\wstar,E,t) d\wstar>$, or to $F_{\rm ref} \sum_{\wstar} d\wstar$ with $F_{\rm ref}$ an arbitrary flux density. For consistency, the integration is performed over the same spectral range used to calculate $R^{\rm norm}$. We note that regions affected by the planet should not be excluded from this spectral range, since measured light curves (see below) are integrated over bands that include those regions. 

Then we re-inject the broadband contributions of the star and planet via a correction $c(\wstar,E,t)$, so that:
\begin{equation}
c(\wstar,E,t) \, F^{\rm meas}(\wstar,E,t) = F(\wstar,E,t).
\end{equation}
Here, the time
series  of $F$ spectra have the correct relative flux level. To define the final flux scaling we can write the above equation at low frequency:
\begin{equation}
c(\wbstar,E,t) C_{\rm ref}(\wbstar,E) \sum_{\wstar} F_{\star}(\wstar,E) d\wstar =  
\sum_{\wstar} F(\wstar,E,t) d\wstar 
.\end{equation}
We remind that the factor $C_{\rm ref}$ represents the unknown deviation from the true stellar spectrum. Because this deviation profile has low-frequency variations and is affecting all exposures in an epoch in the same way, it only matters when extracting planetary emission using absolute fluxes or if the stellar and reference spectra deviate differently between epochs (see Sect.~\ref{sec:col_bal}). In general, we can thus ignore $C_{\rm ref}$ in the final correction $lc$, which is defined as:  
\begin{equation}
lc(\wbstar,E,t) = \frac{\sum_{\wstar} F(\wstar,E,t) d\wstar}{\sum_{\wstar} F_{\star}(\wstar,E) d\wstar}
\label{eq:sc_lc}
.\end{equation}
The scaling light curve $lc(\wbstar,E,t)$ can be defined chromatically, namely, over a set of bands at $\wbstar$, to account for broadband variations in the planetary absorption/emission and stellar emission (Fig.~\ref{fig:LC_scaling_module_chrom}). We propose three methods to calculate $lc$:
\begin{itemize}
\item \textit{import}: we import a measured or theoretical light curve into the \textsc{antaress} workflow. This approach allows accounting for time-variable features, possibly specific to a given epoch, such as spot crossings, planetary thermal emission, or reflection of starlight. 
\item \textsc{batman}: we use this package from \citet{Kreidberg2015} to compute a transit light curve with simple shape, controlled by the transit depth of a single planet, one of the available limb-darkening laws (uniform, linear, quadratic, square-root, logarithmic, exponential, power2, nonlinear) and its associated coefficients, and time resolution.
\item \textit{numerical}: we use \textsc{antaress} stellar surface model (Sect.~\ref{sec:st_surf}). The light curve is calculated by subtracting from the disk-integrated emission the flux from regions occulted by the planet(s) in each exposure, calculated following the approach described in Sect.~\ref{sec:pl_occ_grid}. This method allows accounting for planets transiting simultaneously and for complex stellar photospheres (e.g., oblate gravity-darkened stars, or spotted stars). 
\end{itemize}
Those three methods allow defining light curves specific to a given epoch, to account for changes in the planetary or photospheric properties (in particular variations in the stellar emission due to unocculted spots). In all cases, it is important to define the light curve at high temporal resolution so that it can be averaged within the time window of each exposure and avoid blurring.

We note that algorithms like \textsc{batman} or \textsc{antaress}'s numerical approach actually return $\delta_{\rm p}(\wbstar,E,t)$ (Sect.~\ref{sec:workflow}), which is calculated assuming a constant stellar emission and planetary absorption within a band. This quantity is not strictly equivalent to the integrated flux ratio between each exposure and the star in Eq.~\ref{eq:sc_lc}, since $F$ differs from $F^{\star}$ at high resolution (due to the planet-occulted and atmospheric-filtered stellar lines, Eq.~\ref{eq:Fp}) and $F^{\star}$ thus does not simplify in this ratio. This does not matter, however, as long as the theoretical light curves match the ratio.

Finally, we interpolate the chromatic $lc(\wbstar,E,t)$ over the high-resolution spectral grid of each exposure to calculate the rescaled spectra as:
\begin{equation}
\begin{split}
F^{\rm sc}(\wstar,E,t) &= F^{\rm meas}(\wstar,E,t) C_{\rm F}(E,t) lc(\wstar,E,t)   \\
                                          &= F(\wstar,E,t) C^{\rm norm}_{\rm ref}(\wbstar,E)   \\
\rm where \, C^{\rm norm}_{\rm ref}(\wbstar,E) &=\frac{TF_{\rm ref}}{\sum_{\wstar} F_{\rm ref}(\wstar,E) d\wstar}, \\
\end{split}
\end{equation}

\subsection{Residual profile extraction}
\label{apn:res_prof}

We detail here the procedure to retrieve residual profiles. For each exposure at time $t$ we build a master disk-integrated profile of the unocculted star, $F_{\star}^{t}$, as the weighted average (Sect.~\ref{sec:temp_resamp}) of a selection of out-of-transit $F^{\rm sc}$ profiles representative of the star during transit. Instead of resampling the out-of-transit profiles on a common grid, averaging them into a single master, and resampling this master on the individual spectral grid of each exposure, we directly resample the out-of-transit profiles on the spectral grid of the processed exposure before they are averaged. The master is thus built from profiles that are only resampled once in our approach, preventing blurring and the introduction of correlated noise (Sect.~\ref{sec:resamp}). 

Spectral ranges contaminated by the planetary atmosphere (\textit{i.e.}, where $F_p$ and $S_i^{p[thin]}$ are not null, Eq.~\ref{eq:Fsc_Fres}) are identified as described in Sect.~\ref{sec:atm_mask} and masked so that the master represents the pure stellar emission (see Sect.~\ref{sec:temp_resamp}). Future versions of the \textsc{antaress} workflow could account for possible variations in the disk-integrated profile of the unocculted star, for example in the presence of spots moving over the duration of an epoch, so that $F_{\star}^{t}$ remains equivalent to the disk-integrated profile occulted at time $t$.

\subsection{Intrinsic profile extraction}
\label{apn:intr_prof}

We detail here the procedure to retrieve intrinsic profiles, which are derived from residual profiles as:
\begin{equation}
\begin{split}
F^{\rm intr}(\wstar,E,t) &= \frac{F^{\rm res}(\wstar,E,t)}{1-lc(\wstar,E,t)}            \\
                                                &= C^{\rm norm}_{\rm ref}(\wbstar,E) \,  \frac{\sum_{i \in \iocc} f_i(\wstar,E) \, S_i^{\rm p[thick]}(\wbstar,E)  }{1-lc(\wstar,E,t)}.                                                    
\end{split}
\end{equation}
Here, the spectral ranges absorbed by the optically thin planetary atmosphere have been masked, unless intrinsic profiles are spectral and converted into CCFs later on in the workflow. Indeed, if the planetary atmosphere absorbs at transitions present in the CCF mask used to cross-correlate intrinsic spectra then its lines shift over time in the star rest frame because of the planet orbital motion (Sect.~\ref{sec:atm_mask}). In that case, masking those contaminating lines when defining intrinsic spectra (Sect.~\ref{sec:intr_prof}) would exclude mask lines from the cross-correlation in some exposures but not others, yielding CCFs that are not equivalent over the time series. To circumvent this issue, the masking of the planetary signal is performed in velocity space, after intrinsic spectra have been converted into CCFs (Fig.~\ref{fig:Intr_CCF_map_WASP76}). 
 
The spatial accuracy at which stellar line profiles can be sampled depends on the blurring induced by the planet size and orbital motion. If we can assume that the stellar emission does not vary substantially over the planet-occulted region during an exposure, then the above equation simplifies as\begin{equation}
F^{\rm intr}(\wstar,E,t) \sim C^{\rm norm}_{\rm ref}(\wbstar,E) \, \frac{f_\mathrm{occ}(\wstar,E,t) \, S_\mathrm{occ}^{\rm p[thick]}(\wbstar,E,t) }{1-lc(\wstar,E,t)}                                                       
.\end{equation}
The surface density flux from the region occulted during the exposure at time $t$ writes as:
\begin{equation}
f_\mathrm{occ}(\wstar,E,t) = I_\mathrm{occ}(\wstar,E,t) \, \alpha_\mathrm{occ}(\wbstar,E,t)                     \label{eq:Iocc}                 
,\end{equation}
where $I$ is the specific intensity emitted by the occulted region and $\alpha$ its broadband intensity variations (such as limb- and gravity-darkening). During transit, the spectral flux scaling $lc$ over a given band at $\wbstar$ corresponds to (Sect.~\ref{sec:flux_scaling}):
\begin{equation}
1-lc(\wbstar,E,t) = \frac{\overline{I}(v) \, \alpha_\mathrm{occ}(\wbstar,E,t) \, \overline{S}_\mathrm{occ}(E,t) }{\sum_{\wstar} F_{\star}(\wstar,E)  d\wstar}                                         
,\end{equation}
where $\overline{S}_\mathrm{occ}$ and $\overline{I}$ are the achromatic planetary surface and stellar intensity values that allow reproducing the integrated flux ratio $\sum_{\wstar} F(\wstar,E,t) d\wstar / \sum_{\wstar} F_{\star}(\wstar,E)  d\wstar$ measured in this band, as noted in Sect.~\ref{apn:flux_scaling}. While $\overline{S}_\mathrm{occ}$ does not necessarily match the true $S_\mathrm{occ}^{\rm p[thick]}$, because of the spectral variations of $I_\mathrm{occ}(\wstar,E,t)$ over the band, we assume that the ratio $\overline{S}_\mathrm{occ}/S_\mathrm{occ}^{\rm p[thick]}$ remains constant during transit, so that:
\begin{equation}
\begin{split}
F^{\rm intr}(\wstar,E,t) &= C^{\rm norm}_{\rm ref}(\wbstar,E) \, \sum_{\wstar} F_{\star}(\wstar,E)  d\wstar \times \\
                                                & \frac{I_\mathrm{occ}(\wstar,E,t) \, \alpha_\mathrm{occ}(\wbstar,E,t) \, S_\mathrm{occ}^{\rm p[thick]}(\wbstar,E,t) }{\overline{I}(v) \, \alpha_\mathrm{occ}(\wbstar,E,t) \, \overline{S}_\mathrm{occ}(E,t) },               
\end{split}                                     
\end{equation}
Can be rewritten as:
\begin{equation}
\begin{split}
F^{\rm intr}(\wstar,E,t) &= F^{\rm norm}_{\rm ref}(\wbstar,E) I_\mathrm{occ}(\wstar,E,t).
\end{split}                                     
\end{equation}
Here, all the components that are constant over time throughout the transit and with low-frequency spectral variations have been combined into $F^{\rm norm}_{\rm ref}$. 

Planet-occulted stellar spectra are distorted by broadband spectral variations in the planet size. Indeed, the planet occults a larger area of the stellar photosphere - with a broader $rv$ distribution and a different average $rv$ - at wavelengths where it appears larger, for example due to Rayleigh scattering. We use \textsc{antaress} stellar surface model to calculate the surface $rv$ associated with each planet-occulted region as a function of wavelength (Sect.~\ref{sec:st_surf}), using the same chromatic planetary transit depths and stellar intensity variations used for broadband flux scaling (Sect.~\ref{sec:flux_scaling}). Intrinsic spectra are then corrected for chromatic deviations around the ``white'' $rv$ of planet-occulted regions, so that differences in the $rv$ of intrinsic stellar lines across the spectrum only trace variations due to the stellar atmosphere. WASP-76b yields maximum chromatic deviations on the order of 5\,m\,s$^{-1}$ at ingress/egress, which is much smaller than the RV precision that can be reached on stellar lines in a local spectral band with ESPRESSO. This effect is thus likely negligible with present instrumentation, but may need to be taken into account with high-resolution spectrographs on the ELT.

\section{Generic methods}
\label{apn:gen_mod}

\subsection{Stellar continuum}
\label{apn:st_cont}

We detail here our method, adapted from \citet{Cretignier2020b}, to determine the spectral continuum of the target star using the processed data. A master spectrum of the unocculted star is first built from each epoch or a combination of epochs, resampling spectrally (Sect.~\ref{sec:resamp}) and then temporally (Sect.~\ref{sec:temp_resamp}) a selection of out-of-transit exposures. The master spectrum is smoothed with a Savitzky-Golay filter to remove noise with higher frequency than a chosen scale. Local maxima are identified as bins whose smoothed flux value is the highest in their closest neighbourhood, defined as a window of chosen width. The resulting spectrum of local maxima is stretched so that variations in the flux direction become smaller than those in the wavelength direction. An alpha-shape algorithm is then applied over the stretched spectrum to remove maxima formed by blended lines, setting the radius of the rolling pin to a larger size than the typical spectral line width. Finally, outlying maxima are removed and the remainder used to define the continuum of the master spectrum via linear interpolation. Smoothing and local maxima windows, stretching parameter, and rolling pin radius can be adjusted depending on the instrumental resolution and stellar line properties to better control the continuum determination.

\subsection{Spectral weight profiles}
\label{apn:spec_weights}

\subsubsection{True error estimate}
\label{apn:est_true}

The flux in a given pixel is estimated as $F^{\rm meas}(\lambda)$, which follows a Poisson distribution that can be approximated by a normal distribution with mean the true flux $F^{\rm true}(\lambda)$ and standard deviation $\sigma^{\rm true}(\lambda)$, if enough photoelectrons are measured. Because the uncertainty $\sigma^{\rm meas}(\lambda)$ associated with a measured pixel flux is proportional to $\sqrt{F^{\rm meas}(\lambda)}$, it is a biased estimate of the true error and should thus not be used to weigh the flux. Indeed, pixels where $F^{\rm meas}(\lambda)$ is lower (resp. larger) than $F^{\rm true}(\lambda)$ due to statistical variations would be associated with a lower  (resp. larger) uncertainty than their true error, resulting in artifically larger (resp. lower) weights and a weighted mean that underestimates (resp. overestimates) the true flux. 

We thus weigh the flux in a given pixel using an estimate of its true error. Our underlying assumption is that the measured flux averaged over a large spectral band, or over a large number of exposures, is close enough to its true value that its measured error can be considered as an estimate of its true error. The definition of the weights then depends on the type of profile, disk-integrated or intrinsic, it is applied to. Since time
series  of disk-integrated and intrinsic spectra are aligned in the star or surface rest frame before being averaged, all medium- and high-frequency components involved in the weight profiles are shifted in the same way as their associated spectrum. For example telluric spectra originally defined in the barycentric rest frame are shifted to the same frame as the averaged exposures. Due to these shifts, the common flux calibration profile defined in the detector rest frame becomes specific to each exposure. 

\subsubsection{Definition of weight profiles}
\label{apn:w_DI}

At their latest processing stage disk-integrated spectra write as (see Eq.~\ref{eq:Fsc}):
\begin{equation}
F^{\rm sc}(\wstar,E,t) = F^{\rm meas}(\wstar,E,t) C_{\rm corr}(\wstar,E,t) lc(\wstar,E,t)  
,\end{equation}
where $C_{\rm corr}$ groups all spectral corrections that were applied to the measured spectra (Sect.~\ref{sec:init_corr}), except for cosmics and wiggles. For the sake of clarity we only report the dependency on $\wstar$ in the following expressions. Uncertainties on the measured flux density can be expressed as a function of the raw photoelectron count (Eq.~\ref{eq:gdet}), yielding the following uncertainty on the true flux $F^{\rm sc,true}$ measured during $\Delta t$: 
\begin{equation}
\sigma^{\rm sc,true}(\wstar) = \tilde{g}_{\rm cal}(\wstar) \sqrt{N^{\rm true}(\wstar)} C_{\rm corr}(\wstar) lc(\wstar) / \Delta t
.\end{equation}
\label{eq:sc_true}
The spectra $F^{\rm sc}/lc$ have the same low-resolution profile as the spectrum of the unocculted star $F_{\star}$, allowing us to estimate the raw number of photoelectrons as:
\begin{equation}
N^{\rm true}(\wstar) \sim \frac{ F^{\rm meas}_{\star}(\wstar) \Delta t}{\tilde{g}_{\rm cal}(\wstar) C_{\rm corr}(\wstar) }    
,\end{equation}
where the stellar spectrum $F^{\rm meas}_{\star}$, averaged over many out-of-transit exposures, is assumed to be a good estimate of $F^{\rm true}_{\star}$. Replacing $N^{\rm true}$ with the above expression in Eq.~\ref{eq:sc_true} then yields the weight profile on a disk-integrated spectrum defined in Eq.~\ref{eq:weights_prof}.

Then, based on Eq.~\ref{eq:fres} and \ref{eq:intr}, the error on the true intrinsic flux is:
\begin{equation}
\begin{split}
\sigma^{\rm intr,true}(\lambda) &= \frac{ \sigma^{\rm res,true}(\lambda) }{ 1 - lc(\lambda)   }  \\
\rm where \, \sigma^{\rm res,true}(\lambda) &= \sqrt{ \sigma_{\rm \star}^{\rm true}(\lambda)^2 + \sigma^{\rm sc,true}(\lambda)^2 }.       \\
\end{split}
\end{equation}
Since the measured unocculted stellar spectrum is assumed to approximate the true spectrum (Sect.~\ref{apn:w_DI}), we only use its variance with $\sigma_{\rm \star}^{\rm true} \sim \sigma_{\rm \star}^{\rm meas}$, yielding the weight profile on an intrinsic spectrum defined in Eq.~\ref{eq:weights_prof}.

\subsection{Temporal and spatial resampling}
\label{apn:temp_resamp}

Temporal/spatial resampling is performed over a selection of exposures, typically within the out-of-transit window for disk-integrated profiles and within the transit window for intrinsic profiles. All profiles are resampled on a common spectral grid prior to being binned. The start and end positions of input exposures along the binning dimension, $x_\mathrm{start}(t)$ and $x_\mathrm{end}(t)$, are defined using \textsc{antaress} numerical grid (Sect.~\ref{sec:st_surf}). The start and end positions of new exposures, $x_\mathrm{start}^\mathrm{in}$ and $x_\mathrm{end}^\mathrm{in}$, are defined for each exposure or by specifying a $x$ range and number of exposures. The effective start and end positions of new exposures are then calculated as $\min(\max_t(x_\mathrm{start}(t),x_\mathrm{start}^\mathrm{in}))$ and $\max(\min_t(x_\mathrm{end}(t),x_\mathrm{end}^\mathrm{in}))$, considering all input exposures that overlap with the new exposure. The central positions of new exposures are calculated as the mean of their start/end positions. Input exposures are weighted along the binning dimension by $\Delta x$, the overlap between this exposure's window and the window of the new exposure. We prefer to set $x$ = $r_\mathrm{sky}$ rather than $\mu$ (corresponding to $\sqrt{1-r_\mathrm{sky}^2}$ for a spherical star) as a binning coordinate for intrinsic profiles because it samples more regularly the stellar surface (Sect.~\ref{sec:lprof_var}). For a transiting planet on a circular orbit, with $i_p \sim$90$^{\circ}$ and orbital phases $\phi$ close to 0, $r_\mathrm{sky}$ approximates as 2\,$\pi$\,$\phi$\,$a_p$/$R_{\star}$. For the same exposure time, the $r_\mathrm{sky}$ ranges covered by the planet at two different positions are thus more similar than in $\mu$. 

The temporal/spatial resampling method does not interpolate input profiles along $x$. Instead, Eq.~\ref{eq:av_prof} assumes a uniform flux density over the window of each exposure. This allows resampling the flux in a given spectral bin even if it is undefined in several input exposures. Undefined spectral bins in input profiles, for example masked because of contamination by the planetary atmosphere (Sect.~\ref{sec:atm_mask}), have their flux and weights set to zero so that they do not contribute to the mean. The covariance matrix of the mean profile is calculated as described in Sect.~\ref{sec:resamp}, based on Eq.~\ref{eq:av_prof}. We do not account for covariance over the temporal and spatial dimension, as successive exposures are assumed to be independent.

\subsection{Conversion from 2D to 1D spectra}
\label{apn:2D_1D}

Spectra are converted over a 1D spectral grid common to all exposures, specific to a given instrument, and with uniform resolution in $ln\,\lambda$. Spectra from overlapping orders have to be equivalent in flux to prevent the introduction of biases, highlighting the importance of a proper flatfielding, deblazing, and flux balance correction. We first resample spectra in each spectral order over the common 1D grid (Sect.~\ref{sec:resamp}). Spectral weight profiles, mainly shaped here by the flux calibration profiles (Sect.~\ref{sec:spec_weights}), are then calculated in each order and normalized by the total weight in each pixel common to overlapping orders. Flux spectra and covariance matrixes from all spectral orders are finally multiplied by their corresponding weight profiles, and coadded.

\subsection{Cross-correlation}
\label{apn:CCFs}

We detail here \textsc{antaress} cross-correlation method. It can be applied over all spectral orders or a selection, for example to generate a chromatic set of CCFs (\citealt{Cristo2022}). CCFs are calculated over a $rv$ grid of constant step and chosen range.

In a first step we identify lines of the binary mask that can contribute to the cross-correlation in all processed exposures. Only lines for which the requested $rv$ range (once converted into a wavelength range around this line) is fully within the spectral range of the processed spectrum (which is not the case, for example, if the line is at the edge of a spectral order) and contains no undefined pixels are kept. The spectrum of each exposure is then cross-correlated with the filtered mask as follows:
\begin{equation}
CCF(rv,E,t) = \sum_{l} F( \lambda_l , E,t) \frac{w_l }{\infsymb g_l \supsymb_{\rm cal} }                                  
\label{eq:CCF}
.\end{equation} 
For each mask line $l$ the flux spectrum and covariance matrix are resampled on the CCF grid, converted from $rv$ to wavelengths $\lambda_l = \lambda^0_l \sqrt{ (1 + \mathrm{rv/c}) / (1 - \mathrm{rv/c}) } $ relative to the line transition $\lambda^0_l$. They are then scaled back approximately to raw count units using the mean of the flux calibration profile over the line grid $\infsymb g_l \supsymb_{\rm cal}$ (Sect.~\ref{sec:count_sc}), so as to naturally give more weight to lines measured with a larger number of raw photoelectons without modifying the local flux balance of the spectrum. The local flux spectrum is then multiplied by the line weight $w_l$ and added to the CCF flux grid, while the covariance matrix is multiplied by the squared line weight and stored. Once all lines have been processed the module co-adds their weighted covariance matrixes, using the largest one to set the minimum number of diagonals in the CCF matrix (Sect.~\ref{sec:resamp}). When calculating CCFs of stellar spectra, the associated stellar spectral weight (Sect.~\ref{sec:spec_weights}) and broadband flux scaling grid (Sect.~\ref{sec:flux_scaling}) are cross-correlated in the same way.

\section{Analysis methods}

\subsection{Stellar mask generator}
\label{apn:CCF_masks}

Prior to using this method the \textsc{antaress} workflow must have been set to align spectra in a common rest frame, using either the systemic\footnote{Since a precise systemic $rv$ is later derived from disk-integrated CCFs, we only use an approximate value at this stage and keep in mind that masks are defined with respect to this reference.} plus Keplerian RV for disk-integrated spectra (Sect.~\ref{sec:align_star}), or the surface $rv$ along the transit chord for intrinsic spectra (Sect.~\ref{sec:align_intr_prof}). Aligned spectra must then have been converted from 2D to 1D (Sect.\ref{sec:2D_1D}) and resampled into a master spectrum (Sect.~\ref{sec:temp_resamp}). No in-transit exposures should be used when computing a disk-integrated master, to avoid biases due to planet-occulted stellar lines.

The method starts by normalizing the master spectrum with the estimated stellar continuum (Sect.~\ref{sec:st_cont}), smoothing it to remove high-frequency noise, and oversampling it on a regular spectral grid. Local maxima and minima are identified as bins whose flux value is the highest (resp. lowest) in their closest neighbourhood, defined as a fraction of the lines FWHM. Lines can then be defined as the unique successions of distinct maximum-minimum-maximum, with the line central wavelength set to the minimum position. Lines that fall within spectral ranges known to be contaminated by deep telluric lines, or ranges selected by the user, are excluded. Successive selections can then be applied to refine the linelist, with thresholds that can be adjusted depending on the quality of the data and the type of the star (the process is illustrated in Fig.~\ref{fig:CCF_mask_HD209_strict}):
\begin{enumerate}
\item Lines must have a depth (relative to the global continuum, and to their local continuum) within a given range, and a minimum depth and width (relative to the closest maximum) larger than a threshold.
\item An independent estimate of the line positions, from a polynomial fit to their core, must not deviate from their minimum position by more than a threshold.   
\item Lines must be deeper than contaminating telluric lines by a threshold. A telluric line is a contaminant if, as it shifts between its minimum and maximum spectral positions, it crosses the range covered by a stellar line. The use of transit time series grants a finer control over this step than with long-term $rv$ series. We retrieve the telluric spectra (Sect.~\ref{sec:tell_corr}) associated with the exposures used to build the master stellar spectrum, align them in the Earth rest frame, and bin them into a master telluric spectrum. Telluric lines are then identified following the same approach as for the stellar spectrum, and checked for overlap with stellar lines by considering the stellar Keplerian and Earth barycentric motion over the processed exposures.
\item Lines must not be too asymmetric, with criteria based on the width and depth differences between their maxima lower than a threshold.
\item Lines must have a low $rv$ dispersion. A line-by-line analysis (\citealt{Dumusque2018,Bouchy2001}) is applied to each exposure used in the master spectrum, cross-correlating their spectra to derive $rv$ time series. Each line must have a $rv$ dispersion over the series lower than the mean error by a threshold (this criteria should not be used when the number of exposures is too small), and an average $rv$ calculated as the weighted mean over the series lower than a threshold.
\item A final selection is applied to exclude more finely stellar lines contaminated by tellurics.
\end{enumerate}
Weights representative of the photonic error on the mask line positions (\citealt{Bourrier2021}) are defined as the squared inverse of the error on the line $rv$s (\citealt{Bouchy2001}). 

Once a general list of stellar lines is defined, a cross-matching can be applied with a VALD linelist generated with the host star properties and for the instrument spectral range. We adjust the depth of the VALD lines so that they better match the stellar lines, then identify the unblended VALD lines most consistent with each mask line based on their relative difference in depth and position (Fig.~\ref{fig:CCF_mask_HD209_strict}). Once species have been attributed to the mask lines, a refined selection can be made.

\begin{figure*}
\begin{minipage}[tbh!]{\textwidth}
\includegraphics[trim=0cm 0cm 0cm 0cm,clip=true,width=0.8\columnwidth]{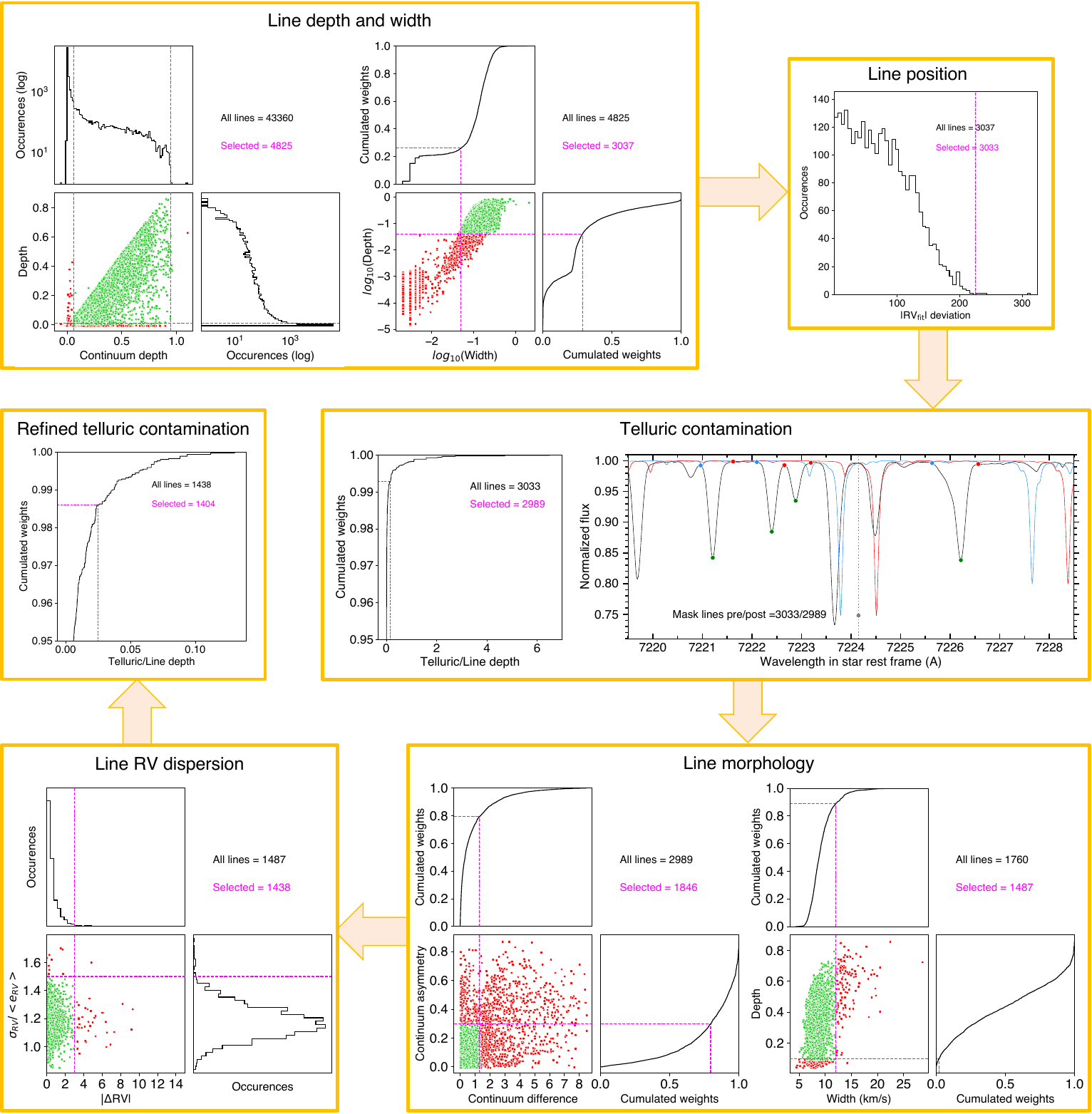}
\centering
\end{minipage}
\caption[]{Flowchart of CCF mask generation, illustrated with HD\,209458. Thresholds applied in each step (magenta lines) can be adjusted based on the resulting line selection displayed as 2D distributions (green and red disks show lines kept and excluded, respectively), occurence histograms, and cumulative functions of the line weights. We illustrate how telluric contamination is flagged by showing the master telluric spectrum at its minimum (blue) and maximum (red) doppler shifts during the two epochs, relative to the master stellar spectrum (black profile). The stellar line at $\sim$7223.7\,\AA\, is excluded due to contamination by the telluric line highlighted at its mean position by the dashed grey line. }
\label{fig:CCF_mask_HD209_strict_flowchart}
\end{figure*}

\subsection{Analysis of individual stellar profiles}
\label{apn:ana_indiv_CCF}

\subsubsection{Model line profiles}
\label{apn:mod_prof}

All stellar profiles are defined with a polynomial continuum, multiplied by spectral absorption lines calculated analytically or numerically. Analytical profiles provide a direct way to fit a single stellar line with simple shape in disk-integrated or intrinsic spectra. Numerical profiles are calculated using \textsc{antaress} stellar surface model (Sect.~\ref{sec:st_surf}), summing elementary profiles over the full star to fit disk-integrated profiles or over planet-occulted regions to fit intrinsic profiles. As described in Sect.~\ref{sec:st_grid} the elementary profiles can be analytical, theoretical, or measured. The last two options allow fitting several lines over a portion of the stellar spectrum (Fig.~\ref{fig:DI_Na_fit}). 

Analytical (either used directly or in the numerical model) and theoretical profiles are defined over regular pixel grids so that they can be convolved with the instrumental response function (assumed to be a Gaussian with FWHM preset for each instrument). When working with low-resolution data, profiles are calculated over an oversampled grid before convolution, and then resampled on the pixel grid of the fitted profile, to avoid blurring. Measured profiles are already broadened by instrumental convolution and kept on their grid of origin (Sect.~\ref{sec:intr_prof}).  

\subsubsection{Line profile fitting}
\label{apn:line_fit}

Disk-integrated profiles from original time series are analyzed in the rest frame of the input data, where their derived centroids can be used to check the Keplerian model and measure the systemic $rv$ of the system. After being aligned, profiles can be analyzed in the star rest frame. We highlight that masking narrow planetary atmospheric features does not remove the effect of planetary occultation on disk-integrated profiles. Indeed, the opaque planet continuum generates spectral features through its absorption of the local stellar line profile (e.g., \citealt{Dethier2023}). These features yield the so-called ``Doppler shadow'' (\citealt{cameron2010a}) that distorts in-transit line profiles and induces the RM anomaly when deriving their centroid with an undistorted model of the unocculted star. We propose an original method to circumvent this bias, which consists in excluding the range absorbed by the planet-occulted stellar line. This range can be defined in a given exposure as a $rv$ window covering the breadth of the line, centered around the line transition Doppler-shifted by the surface $rv$ of the planet-occulted region (Eq.~\ref{eq:rv_surf}). Masked disk-integrated stellar line profiles are not distorted anymore by the planetary continuum, and can be fitted with the unocculted stellar line model. 

Intrinsic profiles from original time series are analyzed in the star rest frame, where their derived centroids can be used to determine the best model for stellar surface $rv$ along the transit chord. After being aligned, profiles can be analyzed in the photosphere rest frame. We highlight the importance of using a Bayesian approach when fitting intrinsic profiles (\citealt{Bourrier2021}), to better explore the line model properties and their correlations, and to retrieve information from noisy or masked local stellar lines that could not be obtained through least-square minimization. Priors set on the model properties can be informed by their range of variations along the transit chord (Sect.~\ref{sec:lprof_var}) or physical arguments (e.g., local FWHM smaller than the disk-integrated value; contrast between 0 and 1; RV centroid smaller than the projected stellar rotational velocity). PDFs from the fits to intrinsic stellar lines inform on the exposures that can be exploited in time series analysis (Sect.~\ref{sec:fit_series}). Exposures with flat, non-constraining PDFs for the width, contrast, and $rv$ of the intrinsic line, typically at the limb of the star where the planet only partially occults dim regions of the stellar surface, should be excluded (Fig.~\ref{fig:Intr_CCF_map_WASP76}).

\subsubsection{Line profile variations}
\label{apn:lprof_var}

Parametric models are used to describe the variation of a line property as a function of the chosen variable. They are fitted to the properties measured over all epochs or a selection of epochs, with parameters that can be common or independent between fitted datasets to allow for variations between epochs. Disk-integrated properties measured out-of-transit can be studied as a function of ambient (e.g., airmass, seeing measured at the instrument facility), data-related (e.g., time, mean S/N over a given selection of orders), or stellar (e.g., activity indexes, retrieved from the DACE platform) variables. They are fitted directly with the chosen parametric model, which can be a polynomial, a sine function, or their combination (e.g., additive for RVs, multiplicative for FWHM and contrast). Intrinsic properties, by definition measured in-transit, are studied as a function of stellar coordinates (e.g., center-to-limb angle, sky-projected distance from star center, latitude or longitude). They are fitted with brightness-averages over the planet-occulted regions (Sect.~\ref{sec:pl_occ_grid}), tiled with the chosen parametric model. Intrinsic line centroids, which trace the surface RVs along the transit chord, are described with Eq.~\ref{eq:rv_surf}. Morphological line properties $m$ are described with absolute ($m(x) = \sum_{i\geq0}c_i x^i)$) or modulated ($m(x) = m_0 (1 + \sum_{i\geq1}c_i x^i)$) polynomial models. The latter possibility allows for a common dependence of the property with stellar coordinate $x$, with a scaling $m_0$ specific to each epoch.

\subsection{Analysis of joint intrinsic stellar profiles}

\subsubsection{Global model line profiles}
\label{apn:glob_mod}

Series of intrinsic profiles are fitted over multiple epochs with a global model for the stellar lines, defined analytically or numerically (Sect.~\ref{sec:ana_indiv_CCF}). Line profiles are linked between consecutive exposures through models describing their shape and position as a function of stellar coordinates, informed by the analysis of individual exposures (Sect.~\ref{sec:lprof_var}). Profiles can further be linked between epochs by using common model coefficients, although they can be varied to account for stellar variability. The line profiles are defined before convolution with each instrument's response, so that their coefficients remain comparable between different spectrographs. 

\subsubsection{RMR fitting}
\label{apn:RMR}

The surface $rv$ model controlling the theoretical line centroids (Eq.~\ref{eq:rv_surf}) depends at first order on the sky-projected spin-orbit angle $\lambda$ and stellar rotational velocity $v_{\rm eq} \sin i_\star$. If allowed by the data, the $rv$ model can further depend on convective blueshift and on differential rotation, the latter breaking the degeneracy on the stellar latitudes transited by the planet. The resulting measurement of the stellar inclination $i_\star$, combined with $\lambda$ and $v_{\rm eq} \sin i_\star$, respectively yields the 3D spin-orbit angle $\psi$ and the equatorial stellar rotational velocity $v_{\rm eq}$. Alternatively, $\psi$ can be derived using an independent constraint on the stellar inclination (for example through asteroseismology) or the equatorial rotation period $P_{\rm eq}$ (typically using photometric modulation or activity indexes). 

The coefficients describing the line profiles depend on the type of profile (e.g., FWHM and contrast for analytical profiles, abundances for theoretical profiles, etc), but their spatial dependence along the transit chord is always described using the polynomials described in Sect.~\ref{apn:lprof_var}. We remind that analytical models for intrinsic lines can include macroturbulence (Sect.~\ref{sec:st_grid}), which can be used to separate its contribution from that of thermal broadening. 

By construction, measured intrinsic profiles are corrected for the broadband flux scaling associated with stellar intensity variations and the area occulted by the planet. However, an intrinsic profile can be seen as the brightness-average of all elementary stellar profiles occulted by the planet during an exposure (Sect.~\ref{sec:st_surf}), so that its shape and Doppler-shift still depend on the size of the planet and on the stellar intensity properties. Model intrinsic profiles thus also depend on the planet-to-star radius ratio in the fitted band (which determines which stellar grid cells are occulted) and on the stellar gravity/limb-darkening coefficients (which determine the relative weights of the elementary profiles from each cell). In practice, intrinsic line profiles are however not measured to a sufficiently good precision, even in CCF format, to constrain these properties. On the other hand the scaled semi-major axis $a_p/R_{\star}$ and orbital inclination $i_p$ can in some cases be constrained by the data, and set as free parameters in the joint fit. Indeed, those parameters can be degenerate with either $\lambda$ (for aligned orbits) or $v\sin\,i_\mathrm{\star}$ (for polar orbits), and affect the model stellar lines by controlling the theoretical transit chord and thus the occulted stellar grid cells. Importantly, the extraction of intrinsic profiles does not require a precise knowledge of the planet-to-star radius ratio, semi-major axis, and orbital inclination, but only of the transit light curve used for broadband scaling (Sect.~\ref{sec:flux_scaling}, \ref{sec:intr_prof}).

\begin{figure*}
\begin{minipage}[tbh!]{\textwidth}
\includegraphics[trim=0cm 0cm 0cm 0cm,clip=true,width=\columnwidth]{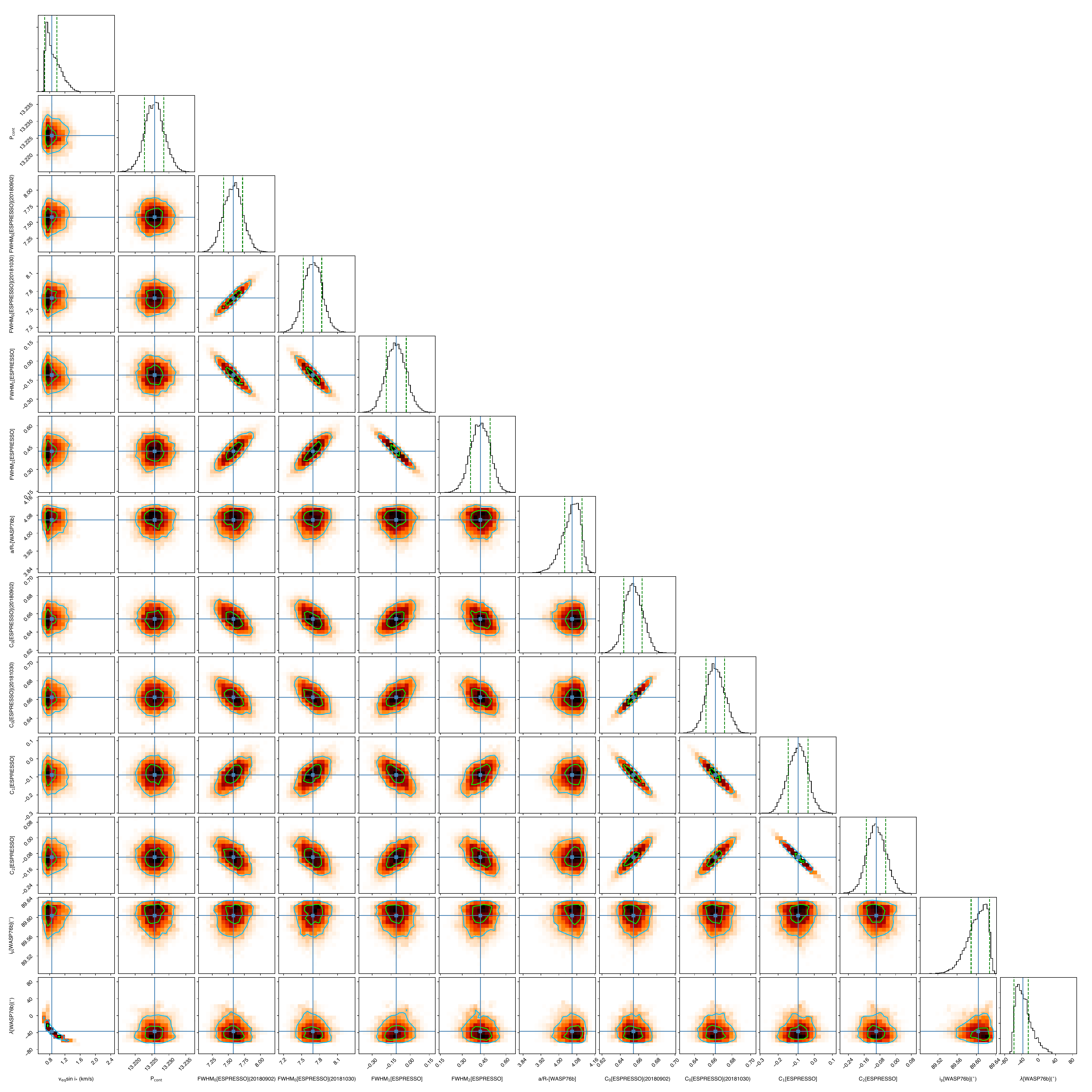}
\centering
\end{minipage}
\caption[]{Correlation diagrams for the PDFs of the RMR model parameters of the WASP-76b transits. C$i$ and FWHM$i$ indicate polynomial coefficients for the line shape properties (see text). Green and blue lines show the 1 and 2 simultaneous 2D confidence regions that contain, respectively, 39.3\% and 86.5\% of the accepted steps. 1D histograms correspond to the distributions projected on the space
of each line parameter, with the green dashed lines limiting the 68.3\% HDIs. The blue lines and squares show the median values.}
\label{fig:Corr_diag_WASP76}
\end{figure*}

\section{Performance analysis}
\label{apn:perf}

\subsection{Resampling}
\label{apn:test_resamp}

Spectra are measured in different exposures over slighly different wavelength grids, and the subsequent shifts required to align them in common rest frames (Sect.~\ref{sec:align_star}, \ref{sec:align_intr_prof}) further differentiate their spectral grids. Resampling spectra on a common grid, as is commonly done in transmission spectroscopy analyses, blurs sharp spectral features and introduces correlations between pixels (Sect.~\ref{sec:resamp}). The \textsc{antaress} workflow innovates by avoiding resampling as much as possible, and we evaluate here the benefits of this approach using the first epoch of HD\,209458 as an example.

\begin{figure}[tbh!]
\includegraphics[trim=2.5cm 0.5cm 10cm 8cm,clip=true,width=\columnwidth]{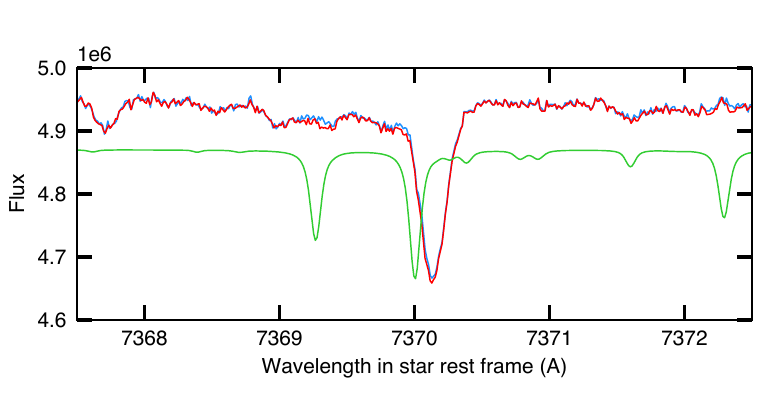}
\centering
\caption[]{Average out-of-transit spectrum of HD\,209458 (epoch 1), when resampling spectra on a common wavelength grid (red) or not (blue). In the former case, undercorrections of telluric lines (highlighted by the best-fit telluric model at mid-epoch, in green) in individual exposures add up and yield biases visible by eye.}
\label{fig:Plot_Flux_Fbal_mast_effect_tell}
\end{figure}

In our nominal approach, a model telluric spectrum is calculated on the native grid of each observed spectrum before calculating telluric CCFs of the spectra and their models (Sect.~\ref{sec:tell_corr}). If spectra are first resampled on a common grid, the best-fit telluric CCFs are shallower due to the blurring of the observed telluric lines, which are then undercorrected (Fig.~\ref{fig:Plot_Flux_Fbal_mast_effect_tell}). 

We fitted Gaussian profiles to deep and narrow stellar lines in the red, high S/N parts of the spectrum to assess the impact of resampling on the line shape. Although the lines were systematically shallower and broader when resampled, the difference remained marginal (1$\sigma$ for the constrast, 0.5$\sigma$ for the FWHM, 0.05$\sigma$ for the $rv$). We derived the same conclusion from fits to the cross-correlated stellar line averaged over out-of-transit exposures. We thus conclude that resampling spectra after they are aligned in the star rest frame has a negligible impact on individual stellar lines. This is because the Keplerian motion of HD\,209458 around the transit windows only represents a small fraction of an ESPRESSO pixel, so that the interpolation due to resampling remains limited. 

The impact of resampling intrinsic spectra is mitigated by their lower S/N. Nonetheless, the cross-correlated stellar line averaged along the transit chord shows differences on the order of 0.1\%, 10\,m\,s$^{-1}$ and 1\,m\,s$^{-1}$ for the contrast, FWHM, and $rv$ position, with typical significance of $\sim$1$\sigma$. This is because intrinsic spectra aligned in the rest frame of the stellar surface are corrected for its Doppler motion, which represents here a substantial fraction of a pixel. The RMR fit is similarly impacted, with differences at the $\sim$1$\sigma$ level for $\lambda$ and $v_\mathrm{eq} \sin i_\star$ (1.055$\pm$0.077$^{\circ}$, 4.272$\pm$0.007\,km\,s$^{-1}$ for the nominal analysis; 1.183$\pm$0.073$^{\circ}$, 4.262$\pm$0.007\,km\,s$^{-1}$ for resampled data).

\begin{figure}[tbh!]
\includegraphics[trim=2.5cm 0.5cm 10cm 9cm,clip=true,width=\columnwidth]{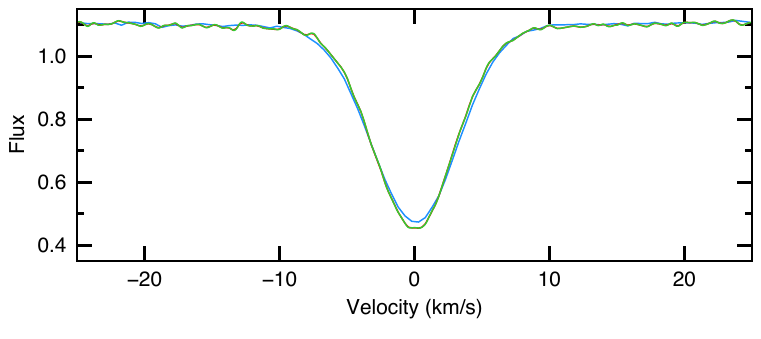}
\centering
\caption[]{Master intrinsic CCF of HD\,209458, averaged along the transit chord in epoch 1. The blue, red, and green profiles were calculated with $rv$ steps of 0.50, 0.25, and 0.10\,km\,s$^{-1}$ , respectively. }
\label{fig:Compa_HD209_osamp_CCFintr}
\end{figure}

Finally, one of the interests of the \textsc{antaress} workflow is the possibility to oversample CCFs with a $rv$ step smaller than the instrumental pixel width, thus better resolving the line profiles. The resulting correlations between pixels are propagated thanks to the use of covariance matrixes, and accounted for when fitting the CCFs (Sect.~\ref{sec:resamp}). To illustrate this point we calculated the CCFs of HD\,209458 intrinsic spectra with $rv$ steps of 0.25 and 0.10\,km\,s$^{-1}$, smaller than the nominal value of 0.50\,km\,s$^{-1}$. Fits to the master CCF along the transit chord in epoch 1 show a significant refinement of the stellar line shape from 0.5 to 0.25\,km\,s$^{-1}$, as it gets deeper (contrast from 0.5683$\pm$0.0009 to 0.58693$\pm$0.0010) and narrower (FWHM from 7.656$\pm$0.016 to 7.199$\pm$0.018\,km\,s$^{-1}$). Using even finer $rv$ steps does not resolve better the line shape (Fig.~\ref{fig:Compa_HD209_osamp_CCFintr}). Oversampling the CCFs does not change the results of the RMR fit, likely because the line profiles of HD\,209458 are symmetrical and their positions are already measured with high precision at the nominal $rv$ step.

%%%%%%%%%%%%%%%%%%%%%%%%%%%%%%%%%%%%%%%%%%%%%%%%%%%%%%%%%%%%

\subsection{Covariance matrix}

The \textsc{antaress} workflow calculates the covariance between pixels of the processed profiles, and propagates it as a banded covariance matrix (Sect.~\ref{sec:resamp}). With the optimal processing, when 2D spectra are resampled as late as possible, no covariance exists until residual profiles are extracted (Sect.~\ref{sec:res_prof}). At this stage, correlations are introduced through the difference between in-transit profiles and the master out-of-transit spectrum used as reference, since disk-integrated spectra must be aligned and resampled on a common grid before they can be averaged into a master. The master spectra of HD\,209458 and WASP-76 are associated with three covariance diagonals, whose ratios to the variance (from $\sim$3$\times$\,10$^{-3}$ to 4$\times$\,10$^{-5}$) however remain negligible.  
%NB: ratios ($\sim$3x\,10$^{-3}$, 8x\,10$^{-4}$, 4x\,10$^{-5}$) and 0.2, -5x\,10$^{-2}$, 2x\,10$^{-3}$, 10$^{-4}$, -5\,10$^{-6}$, -6\,10$^{-8}$, 9\,10$^{-9}$, 2\,10$^{-14}$
Strong correlations are then inevitably created when computing CCFs, although this is mitigated in the \textsc{antaress} workflow by accounting for the covariance in fits. The CCFs of intrinsic spectra are associated with eight covariance diagonals, with typical ratios to the variance decreasing from 0.2 to 2$\times$\,10$^{-14}$. However, covariances remain small enough that discarding them in the RMR fits does not change the derived values and their uncertainties for either HD\,209458b or WASP-76b.

%%%%%%%%%%%%%%%%%%%%%%%%%%%%%%%%%%%%%%%%%%%%%%%%%%%%%%%%%%%%

\subsection{Weighted means}
\label{apn:test_wm}

Particular care is taken in the \textsc{antaress} workflow to calculate the weighted mean of spectra (Sect~\ref{sec:spec_weights}), using estimates of the true error on individual pixels. Here we evaluate the impact of not considering various components in the weighing of disk-integrated and intrinsic spectral profiles. Our goal is not to perform an exhaustive assessment but rather to estimate the typical biases introduced by improper weighing.

\begin{figure}[tbh!]
\includegraphics[trim=0cm 0cm 0cm 0cm,clip=true,width=\columnwidth]{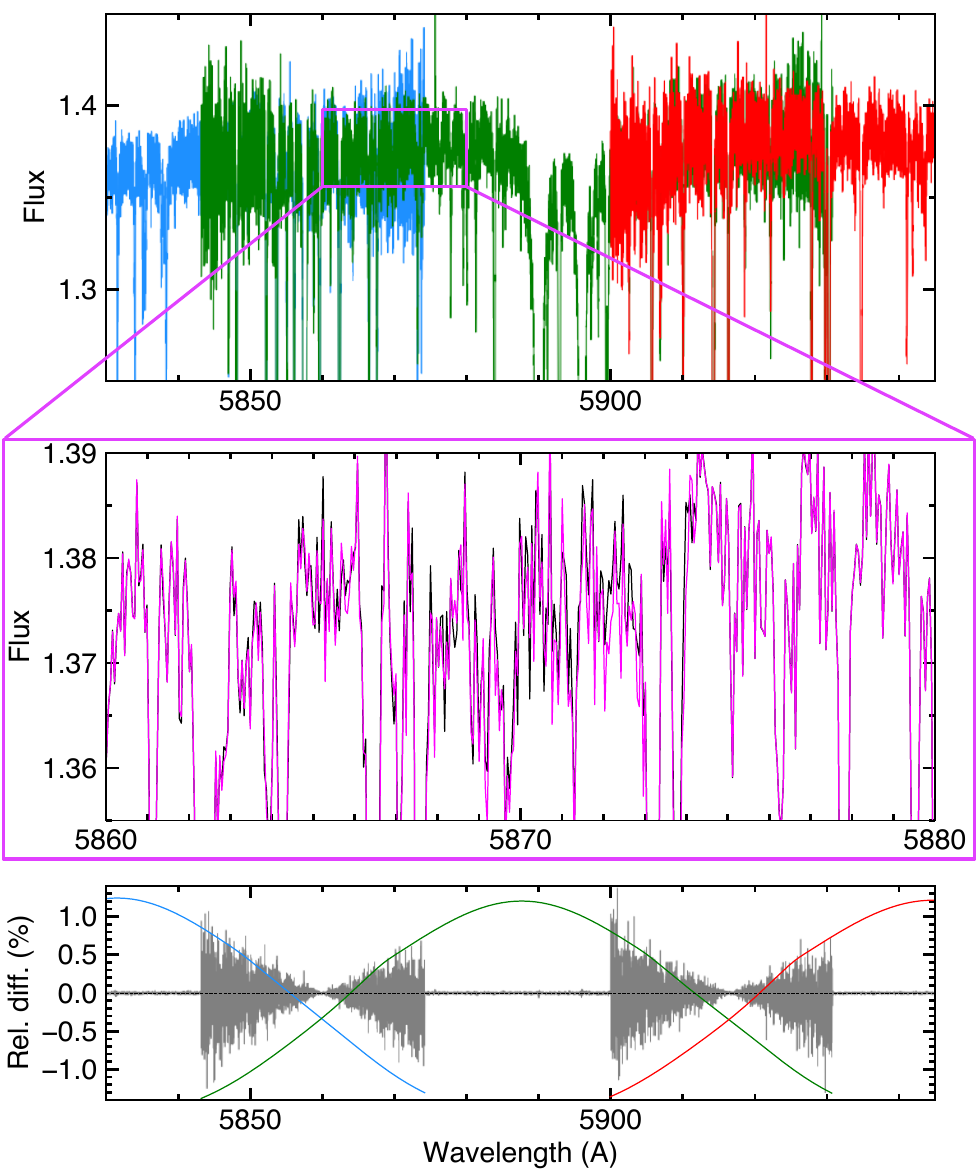}
\centering
\caption[]{Weighing performance (instrumental flux calibration) for disk-integrated spectra. \textit{Top panel:} Overlapping orders in 2D flux spectra (exposure at index 10 in HD\,209458 epoch 1). \textit{Middle panel:} Zoom on the 1D spectrum averaged from the order spectra, when accounting (black) or not (magenta) for instrumental calibration in the weight profiles. \textit{Bottom panel:} Relative difference between the magenta and black profiles, overlapped with the contribution of the instrumental calibration to the order weight profiles (arbitrary scale).}
\label{fig:Perf_weights_2D}
\end{figure}

Figure~\ref{fig:Perf_weights_2D} illustrates the role of the instrumental flux calibration profile when averaging overlapping spectral orders in a given disk-integrated spectrum (Sect.~\ref{apn:w_DI}), typically to convert 2D echelle spectra into a 1D spectrum. We observe differences up to a few percents, increasing toward the edges of the orders, when performing a simple mean that does not account for lower photon count and associated large flux errors.

\begin{figure}[tbh!]
\includegraphics[trim=0cm 0cm 0cm 0cm,clip=true,width=\columnwidth]{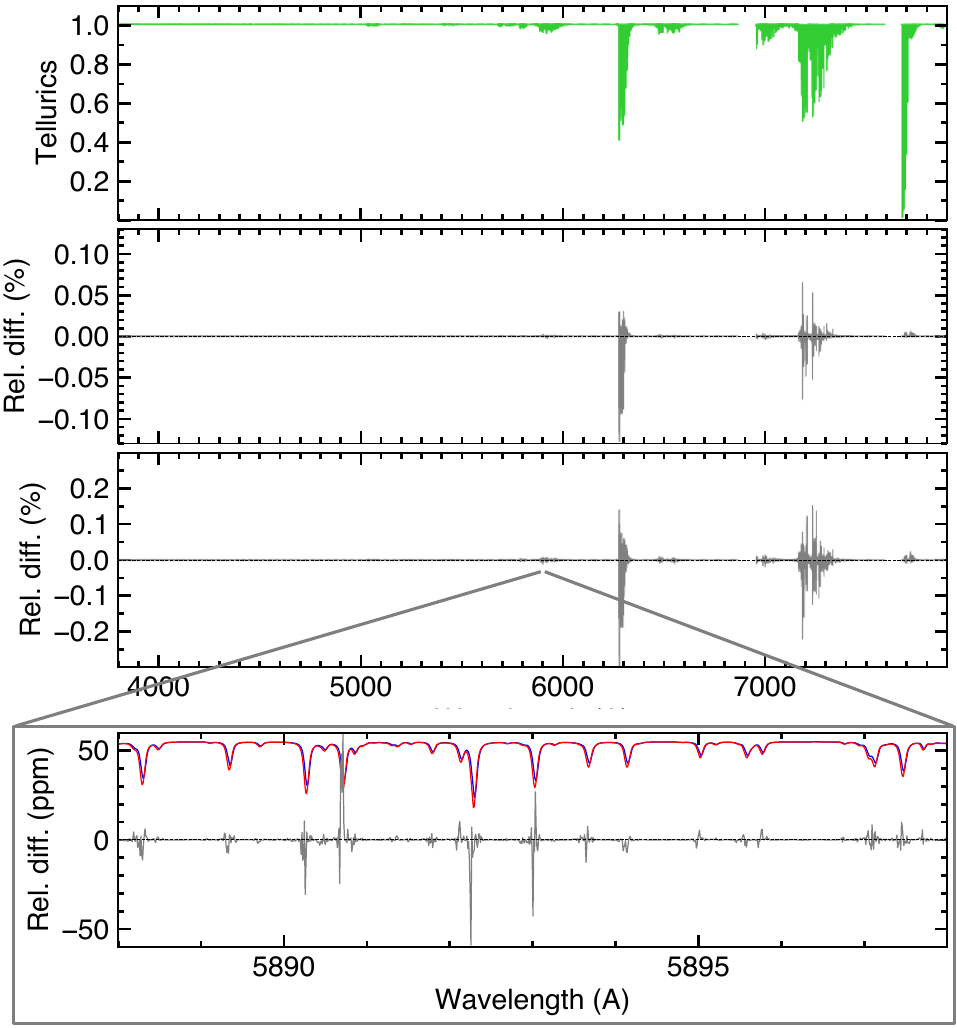}
\centering
\caption[]{Weighing performance (tellurics) for disk-integrated spectra. \textit{First (top) panel:} Typical telluric spectrum in HD\,209458 epoch 1. \textit{Second panel:} Relative difference between the master-out spectra accounting or not for telluric correction in weight profiles. \textit{Third panel:} Same as the second panel, when averaging only two exposures (index 5 and 84). \textit{Fourth (bottom) panel:} Zoom from the third panel in the region of the sodium doublet, illustrating how shifting telluric lines (reported on an arbitrary scale, in blue and red at indexes 5 and 84) bias the mean over the pixels they absorb when not included in weight profiles.}
\label{fig:Plot_DItell_weights}
\end{figure}

Figure~\ref{fig:Plot_DItell_weights} illustrates the role of telluric correction when averaging disk-integrated spectra from different exposures (Sect.~\ref{apn:w_DI}). These spectra are typically averaged in the star rest frame, in which telluric lines shift over time, so that a given pixel should be weighted differently according to its level of telluric correction. The strongest biases are observed for the deepest telluric lines when they are not included in the weight profile, but even moderately strong tellurics can induce biases up to a few tens of ppm when averaging two exposures. The bias is reduced when more exposures are averaged, as a pixel absorbed by tellurics in a given exposure will weigh less in the overall mean. Nonetheless, we observe differences up to a few hundreds of ppm in regions of deep telluric lines, even when averaging all out-of-transit exposures.

\begin{figure}[tbh!]
\includegraphics[trim=0cm 0cm 0cm 0cm,clip=true,width=\columnwidth]{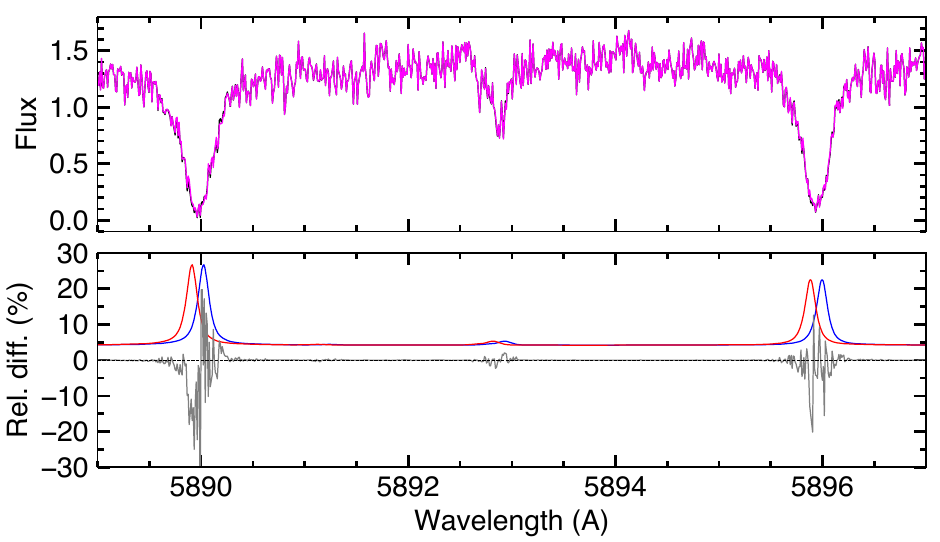}
\centering
\caption[]{Weighing performance (stellar lines) for intrinsic spectra. \textit{Top panel:} Mean of intrinsic spectra along the transit chord of HD\,209458b in the region of the sodium doublet, when accounting (black) or not (magenta) for the disk-integrated spectrum in weight profiles.\textit{Bottom panel:} Relative difference between the magenta and black profiles, overlapped with the contribution of the disk-integrated spectrum to the weight profiles (arbitrary scale, in blue and red for two exposures at ingress and egress).}
\label{fig:Plot_IntrFref_weights}
\end{figure}

Figure~\ref{fig:Plot_IntrFref_weights} illustrates the role of disk-integrated stellar lines when averaging intrinsic spectra in the photosphere rest frame (Sect.~\ref{apn:w_DI}). In that case, disk-integrated stellar lines shift between different exposures, so that a given pixel should be weighted differently according to the depth of the stellar line it overlaps with. Similarly to telluric lines, the strongest biases are observed for the deepest disk-integrated stellar lines, when they are not included in the weight profile. The sodium lines are among the deepest in the disk-integrated spectrum of HD\,209458, with nearly full absorption at the core of the line, and induce biases up to tens of percents even in the intrinsic profile averaged over all in-transit exposures. It is thus critical for the analysis of averaged planet-occulted stellar lines to account for disk-integrated stellar lines in weight profiles.

\subsection{Flux balance corrections}
\label{apn:perf_Fbal}

Correcting for spectro-temporal variations of the flux balance is necessary to avoid biases in planetary transmission spectra. While local corrections can be applied at high spectral resolution, forcing the continuum around the line of interest to a constant value, the use of a spectrum-wide correction in the \textsc{antaress} workflow may allow the retrieval of planetary signatures at medium spectral resolution. This type of correction also ensures the computation of unbiased stellar and planetary CCFs, for which it is not possible to correct the continuum of each independent line. Relative variations between the continua of different lines over time indeed translate into variations between their relative weights in the CCF, modifying its profile between exposures.

Here we evaluate how much a poor correction of the flux balance impacts the quality of disk-integrated and intrinsic stellar CCFs. We reprocessed the data from the \textsc{Flux balance} method (Sect.~\ref{sec:col_bal}) onward, using low-order polynomials instead of smoothing splines to fit the color balance variations. This approach allows capturing the overall shape of the color balance, but not its variations over a few spectral orders as finely as with splines (Fig.~\ref{fig:ColorBalanceModel}). We first assessed the quality of disk-integrated CCFs by comparing the dispersion of their out-of-transit properties with the nominal analysis. The degraded flux balance correction does not have a clear impact on WASP-76 CCFs. For HD\,209458, it increases the dispersion of the CCF contrast and FWHM by $\sim$3-8\% but has no impact on their RVs. This is likely because the residual, uncorrected color variations are broad enough that they are approximately constant over the breadth of a stellar line and thus do not modify their shape. Since the correction improves the disk-integrated CCFs of HD\,209458 we further evaluated its impact on the intrinsic CCFs and their RMR fit. The degraded correction yields $\lambda$ = 1.12$\pm$0.07$^{\circ}$ and $v_\mathrm{eq} \sin i_\star$ = 4.289$\pm$0.008\,km\,s$^{-1}$, consistent with 1.06$\pm$0.07$^{\circ}$, 4.272$\pm$0.007\,km\,s$^{-1}$ from the nominal analysis.

Our analysis suggests that an accurate flux balance correction, while important for spectral analysis and to compute high-quality CCFs, is not critical for the derivation of RVs and the interpretation of the RM effect. Biases induced by the color effect (e.g., \citealt{bourrier2014b}) mainly arise from the spectrum-wide variations reaching tens of percents, which can be captured by low-order polynomials.

%%%%%%%%%%%%%%%%%%%%%%%%%%%%%%%%%%%%%%%%%%%%%%%%%%%%%%%%%%%%

\subsection{Wiggle correction}
\label{apn:perf_wig}

\begin{figure}[tbh!]
\includegraphics[trim=0cm 0cm 0cm 0cm,clip=true,width=\columnwidth]{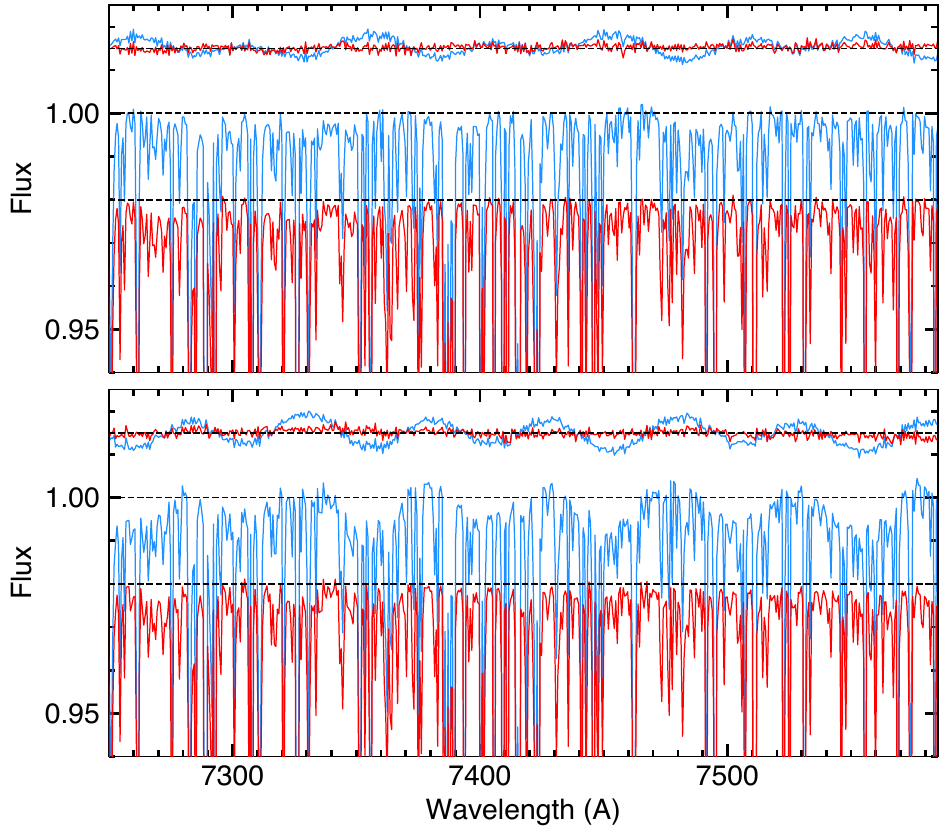}
\centering
\caption[]{Wiggle correction in epoch 1 of HD\,209458b, for exposures at index 60 (top panel) and 88 (bottom panel). Blue and red profiles show spectra before and after correction, respectively. Flux spectra are normalized by the stellar continuum (estimated from the corrected data; Sect.~\ref{sec:st_cont}) and offset for clarity. The dashed black line represents the normalized continuum unity. Transmission spectra, shown at the top of the panels and calculated with the master out-of-transit spectrum, highlight wiggle patterns strong enough to be visible in the flux spectra.}
\label{fig:Perf_wig}
\end{figure}

The same spectral region in ESPRESSO data is affected by different wiggle patterns over time (Fig.~\ref{fig:Perf_wig}). Applying a global, homogeneous correction to all spectra acquired during an epoch (Sect.~\ref{sec:wig_mod}) is thus required to compare a given spectral line between different exposures, or to combine a line assumed to be stable over several exposures. 

Here, we assess its impact on stellar CCFs and their interpretation. Similarly to color balance variations, but at shorter frequencies, wiggles indeed change the relative flux balance between different spectral regions over time and thus the relative weights between CCF lines. We deactivated the wiggle correction (Sect.~\ref{sec:wig_mod}) and reprocessed the data from the \textsc{Spectral line detrending} method (Sect.~\ref{sec:st_line_detrend}) onward. Correcting for the wiggles has no significant impact on WASP-76 disk-integrated CCFs. For HD\,209458, it improves the dispersion of their out-of-transit contrast and FWHM by $\sim$3-6\% but has no impact on their $rv$s. This is likely because the wiggle pattern is too broad to affect the line shapes and thus only changes their overall flux level, as discussed for the flux balance correction (Sect.~\ref{apn:perf_Fbal}). We then evaluated the impact of the wiggles on the intrinsic CCFs of HD\,209458 and their RMR fit. Disabling the correction yields $\lambda$ = 1.01$\pm$0.08$^{\circ}$ and $v_\mathrm{eq} \sin i_\star$ = 4.271$\pm$0.007\,km\,s$^{-1}$, to be compared to 1.06$\pm$0.07$^{\circ}$, 4.272$\pm$0.007\,km\,s$^{-1}$ for the nominal analysis. 

Although they have a strong impact on spectral data, wiggles thus appears to have a moderate to negligible impact on disk-integrated and intrinsic CCFs. This is likely because wiggles are smoothed out over the thousands of cross-correlated lines, so that temporal variations in the wiggle pattern have little impact on the CCF profiles. This suggests that wiggle correction is not critical to RMR fits on ESPRESSO data, although it may be safer to apply it to ensure unbiased results, especially if a small number of stellar lines are used in the CCFs.

%%%%%%%%%%%%%%%%%%%%%%%%%%%%%%%%%%%%%%%%%%%%%%%%%%%%%%%%%%%%

\subsection{CCF from intrinsic spectra / from disk-integrated CCF}
\label{apn:perf_intr_DI}

The \textsc{antaress} workflow offers the new possibility to apply the RMR technique (Sect.~\ref{sec:RMR}) to ``white'' CCFs calculated by cross-correlating intrinsic spectra over broad spectral ranges:
\begin{equation}
CCF[F^{\rm intr}](rv) = \sum_{l} \frac{F_{\star}^{t}(\lambda_l)-F^{\rm sc}(\lambda_l)}{1-lc(\lambda_l)} \alpha_l                                
\end{equation} 
\label{eq:CCFintr}
Where $\alpha_l$ designates the combined weights applied to local line spectra (Sect.~\ref{sec:CCFs}). It is also possible, as in past studies (e.g., \citealt{Bourrier2021}), to use the CCFs of disk-integrated spectra, readily provided by the DRS of high-resolution spectrographs or calculated with the \textsc{antaress} workflow (Sect.~\ref{sec:intr_prof}), and extract intrinsic CCFs as:
\begin{equation}
CCF^{\rm intr}(rv) = \frac{CCF_{\star}(rv)-CCF^{\rm sc}(rv)}{1 - lc_{\rm wh}}                    
\end{equation} 
Where $lc_{\rm wh}$ (Sect.~\ref{sec:flux_scaling}) is either the achromatic light curve calculated over the ``white'' band best matching the CCF linelist (when CCFs are provided as input of \textsc{antaress}), or the cross-correlation of the chromatic light curve used to scale disk-integrated spectra (when CCFs are calculated within our workflow). The $CCF[F^{\rm intr}]$ and $CCF^{\rm intr}$ are thus not equivalent when a chromatic light curve is used, as with WASP-76b, and more generally because the $CCF_{\star}$ and $CCF^{\rm sc}$ are calculated by resampling, weighing, and scaling disk-integrated CCFs in $rv$ space. 

To evaluate the bias induced by these approximations, we compared the nominal RMR fit of the $CCF[F^{\rm intr}]$ (Sect.~\ref{sec:RMR}) with the same fit applied to $CCF^{\rm intr}$ derived from disk-integrated CCFs calculated with the \textsc{antaress} workflow. The quality of the fits and the precision of the results is similar for both planets ($\chi^2_r$ = 1.1), with consistent results for HD\,209458b ($\lambda$ = $1.164\pm0.069^{\circ}$, $v_\mathrm{eq} \sin i_\star$ = 4.263$\stackrel{+0.0072}{_{-0.0083}}$\,km\,s$^{-1}$ from the $CCF^{\rm intr}$; $\lambda$ = $1.06\pm0.07^{\circ}$, $v_\mathrm{eq} \sin i_\star$ = 4.272$\pm$0.007\,km\,s$^{-1}$ from the nominal $CCF[F^{\rm intr}]$). While the derived rotational velocities are also consistent for WASP-76b (0.79$\stackrel{+0.12}{_{-0.15}}$\,km\,s$^{-1}$ from $CCF^{\rm intr}$; 0.86$\stackrel{+0.13}{_{-0.19}}$\,km\,s$^{-1}$ from $CCF[F^{\rm intr}]$), the fit to the $CCF^{\rm intr}$ prefers a positive solution for $\lambda$ (27.0$\stackrel{+29.4}{_{-18.6}}^{\circ}$, compared to the nominal -37.1$\stackrel{+12.6}{_{-20.6}}^{\circ}$). It is noteworthy that our fit to the $CCF^{\rm intr}$ yields similar results as those derived by \citet{Ehrenreich2020} without prior on the rotational velocity ($v_\mathrm{eq} \sin i_\star$ = 0.73$\stackrel{+0.21}{_{-0.03}}$\,km\,s$^{-1}$, $\lambda$ = 10.2$\stackrel{+31.0}{_{-5.7}}^{\circ}$; priv. comm.), supporting the conclusion from Sect.~\ref{sec:RMR} that their different surface $rv$ series and results arise from their use of intrinsic CCFs derived from disk-integrated CCFs. 

We remind that both types of intrinsic CCFs analyzed here are derived from the same 2D spectra processed by the \textsc{Spectral line detrending} method (Sect.~\ref{sec:st_line_detrend}). The differences we observe are thus entirely caused by cross-correlating spectra too early in the workflow. We thus recommend analyzing CCFs calculated from intrinsic spectra rather than extracted from disk-integrated CCFs, as the latter can bias the analysis of the RM effect.

%%%%%%%%%%%%%%%%%%%%%%%%%%%%%%%%%%%%%%%%%%%%%%%%%%%%%%%%%%%%

\subsection{Chromatic flux scaling}
\label{apn:perf_chrom}

The \textsc{antaress} workflow can account for broadband variations in the planetary absorption and stellar emission when rescaling disk-integrated spectra $F^{\rm sc}$ to their correct relative flux level, and when rescaling intrinsic spectra $F^{\rm intr}$ to a common flux level (Sect.~\ref{sec:flux_scaling}). As can be seen in Eq.~\ref{eq:CCFintr}, using an achromatic scaling light curve instead of the true chromatic one thus biases the continuum of the planet-occulted spectrum $F_{\star}^{t}-F^{\rm sc}$ and its intrinsic normalization by $1-lc$. This changes the relative weight given to stellar lines from different regions of the spectrum when cross-correlating intrinsic spectra, and may thus bias the resulting CCF properties.

To investigate this issue we performed a RMR fit of the $CCF[F^{\rm intr}]$ from WASP-76b, processed from the \textsc{Broadband flux scaling} method onward with the white transit depth and limb-darkening values from Table~\ref{tab:sys_prop}. The achromatic processing yields $v_\mathrm{eq} \sin i_\star$ = 0.78$\stackrel{+0.08}{_{-0.10}}$\,km\,s$^{-1}$ and $\lambda$ = -23.9$\stackrel{+16.6}{_{-24.7}}^{\circ}$, to be compared with the nominal values 0.86$\stackrel{+0.13}{_{-0.19}}$\,km\,s$^{-1}$ and -37.1$\stackrel{+12.6}{_{-20.6}}^{\circ}$. While the derived properties remain consistent, they are impacted by the chromaticity of the scaling light curve. Future studies of data measured at higher precision, or for planets with strong chromatic variations of their continuum, should thus be wary of this bias.

%%%%%%%%%%%%%%%%%%%%%%%%%%%%%%%%%%%%%%%%%%%%%%%%%%%%%%%%%%%%

%%%%%%%%%%%%%%%%%%%%%%%%%%%%%%%%%%%%%%%%%%%%%%%%%%%%%%%%%%%%

\subsection{Stellar CCF masks}
\label{apn:perf_CCF_mask}

To illustrate the use of the \textsc{Stellar mask generator} (Sect.~\ref{sec:CCF_masks}) we built custom masks from the disk-integrated spectra of HD\,209458 and WASP-76 (Fig.~\ref{fig:CCF_mask_HD209_strict}). 

We assessed the quality of disk-integrated CCFs derived with the custom masks by comparing the dispersion of their out-of-transit properties with those of CCFs derived with the standard F9 ESPRESSO DRS mask. For HD\,209458, the custom CCFs are less stable in contrast, more stable in FWHM, and comparable in RVs to the DRS CCFs. For WASP-76, the custom CCFs are less stable in contrast and FWHM, but more stable in RVs. Given that HD\,209458 and WASP\,76 have types F9 and F7, these results are not unexpected. Indeed, from a similar comparison over a wider stellar sample, \citet{Bourrier2023} noted that masks customized to a specific target star, rather than representative of a spectral type proxy as in ESPRESSO DRS, yield CCFs of comparable quality for F-type stars and CCFs of much better quality for G-type and especially K-type stars. As a complementary analysis, we performed a RMR fit of the $CCF[F^{\rm intr}]$ from HD\,209458, cross-correlating intrinsic spectra with its custom mask (Sect.~\ref{sec:CCF_intr}). The fit is of similar quality than the nominal one based on the F9 mask, and yields consistent results.

These tests show that custom masks, generated with the \textsc{antaress} workflow from input datasets of a given star, allow building CCFs of at least comparable quality to standard masks, with the interest of performing self-consistent RM analysis.

\end{appendix}

\end{document}